\documentclass[12pt,a4paper]{article}
\usepackage{amsmath,amsfonts,amsthm,graphicx,lscape}
\usepackage[dvips]{psfrag}
\usepackage{cite}

\usepackage{textcomp}

\usepackage{amsmath,amsthm,amssymb}
\usepackage{amsfonts,graphicx,lscape}

\numberwithin{equation}{section}

\def\beqa{\begin{eqnarray}}
\def\enqa{\end{eqnarray}}
\def\beq{\begin{equation}}
\def\enq{\end{equation}}

\begin{document}
\title{
%\vspace{-7mm}
\vspace{-9mm}
A systematic method for 
%finding 
constructing 
%obtaining 
%deriving 
%{\it local} \/
time 
%integrable 
discretizations of integrable lattice systems: 
{\it local} \/equations of motion
%local equations of motion
%A note on the high temperature expansion of the density matrix for 
%the isotropic Heisenberg chain 
}
\author{Takayuki \textsc{Tsuchida}\footnote{
%E-mail address: 
%zengo\_tsuboi@pref.okayama.jp
%E-mail:\ \{surname of the author\}@ms.u-tokyo.ac.jp
E-mail:\ surname at 
%atmark 
ms.u-tokyo.ac.jp
}
\\
\\
{\it Okayama Institute for Quantum Physics,
%\footnote{URL: 
%http://www.pref.okayama.jp/kikaku/kouryoushi/english/kouryoushi.htm}
}
\\
{\it Kyoyama 1-9-1, Okayama 700-0015, Japan}
}
%\date{}
\maketitle
\begin{abstract} 
%Using the zero-curvature (Lax-pair) 
%representation, 
We propose a 
%simple but powerful 
new 
%systematic 
method for 
%construction 
discretizing the 
%continuous 
time variable 
in 
%of 
integrable lattice systems 
while maintaining 
%preserving 
the {\it locality} \/of the 
%original 
equations of motion. 
The method 
%recipe 
is based on the zero-curvature (Lax
%-
pair) representation 
and the lowest-order 
%simplest 
%fundamental 
``conservation laws''. 
%of the lattices. 
%but 
In contrast to 
%By refining 
the 
%original approach due to 
pioneering work of 
%early attempt by 
Ablowitz and Ladik, 
%Ablowitz--Ladik, 
% and Suris, 
%based on the zero-curvature (Lax-pair) representation 
our 
%the proposed 
method allows 
%to express 
the auxiliary dependent variables 
%(infinite products, infinite sums, etc.) 
appearing 
in the stage of time discretization 
to be expressed 
{\it locally} \/in terms of 
the original dependent variables. 
%We illustrate our method with examples such as 
The time-discretized lattice systems 
%time discretizations 
%thus obtained 
%are 
have 
%exactly 
the same set of conserved quantities and 
%do not change 
the same structures of the 
solutions 
%as the time-continuous 
%original 
%of 
as 
%those of 
the 
%continuous-time 
%time-continuous 
continuous-time lattice systems; 
only the time 
%dependence 
evolution of 
%certain 
%some 
the parameters 
in the
solutions 
%(roughly speaking, angle variables) 
%corresponding 
that correspond 
to the angle variables 
%``angle variables'' 
%change upon time discretization. 
%are 
is discretized. 
%suffer a change. 
%
%This property possibly suggests the applicability 
%of our results in numerical simulations. 
%The constructed discrete-time evolutions 
%are, by construction, 
%%can be 
%%considered 
%regarded as
%% generalized 
%commuting flows, 
%%symmetries, 
%or equivalently, auto-B\"acklund transformations 
%of 
%%for 
%the original 
%continuous-time lattices. 
%fully discrete 
The effectiveness 
%ity 
%validity 
of our 
%the 
%wide applicability of our 
method 
is illustrated 
%with 
%by 
using examples such as 
the Toda lattice, 
the Volterra lattice, the modified Volterra lattice, 
the Ablowitz--Ladik lattice (an integrable semi-discrete 
%NLS
nonlinear Schr\"odinger system), 
%equation), 
and 
the lattice 
Heisenberg ferromagnet model. 
%lattice/chain. 
%The 
For the Volterra lattice and 
%the 
modified Volterra lattice, 
we also present their {\it ultradiscrete} \/analogues. 
\end{abstract}
%%%%%%%%%%%%%%%%%%%%%%%%%%%%%%%%%%%%%%
%{\it MSC:} 82B23; 45G15; 82B20; 82B80 \\
%{\it PACS:} 75.10.Jm, 02.30.Ik, 05.70.-a, 05.30.-d \\
%{\it Key words:}
%
\vspace{5mm}
\vspace{5mm}
\\
%{\it PACS: }75.10.Jm, 02.30.Ik, 05.70.-a, 05.30.-d \\
{\it PACS numbers: }02.30.Ik, 02.70.Bf, 05.45.Yv, 45.05.+x 
%
%02.30.Jr Partial differential equations
%02.60.-x Numerical approximation and analysis
%02.60.Cb Numerical simulation; solution of equations
%02.60.Jh Numerical differentiation and integration
%02.60.Lj Ordinary and partial differential equations; boundary value problems
%02.70.-c Computational techniques in mathematical methods in physics
%02.70.Bf Finite-difference methods
%02.70.Dh Finite-element and Galerkin methods
%05.45.-a Nonlinear dynamics, 
%05.50.+q Lattice theory and statistics, 
%42.65.Tg Optical solitons; nonlinear guided waves
%42.81.Dp Propagation, scattering, and losses; solitons
%75.10.Hk Classical spin models
%
%{\it Report-no:} {\bf OIQP-08-??}  \\
%{\bf to appear in Physica A}
%%%%%%%%%%%%%%%%%%%%%%%%%%%%%%%%%%%
\newpage
\noindent
\tableofcontents

\newpage
\section{Introduction}
%$\abs{a+b\abs{c+d}e+f}$, $\abs{a+b}c+d\abs{e+f}$
%
The quest for a 
%proper 
finite-difference 
%scheme 
analogue 
%version 
of a given differential equation 
%is supported by a natural motivation and 
can be justified for 
%many 
several 
%natural ?
sound reasons. 
%many reasons. 
%in many 
%%different 
%ways. 
%has been fully-justified. 
A suitable
%proper 
discretization
%, i.e., discretization that maintains the qualitative behavior, 
can reproduce most of the important properties 
%maintains the qualitative properties 
%behavior 
of the 
%original 
differential equation 
in 
%and/or 
%a 
the small-value range 
%region 
%or 
of the difference interval 
%and the 
%a 
%as well as its zero 
%%continuous 
%limit, 
%lattice 
%interval 
%difference step
%grid size?
%mesh size?? 
%difference interval
%step size 
%of the lattice parameters, 
%of space-time 
and 
%thus 
can be 
%is 
considered 
%as 
%giving 
%its 
a ``generalization'' 
of the 
original 
continuous
%differential 
equation. 
%of it. 
%provide 
%help 
%The difference scheme 
%It 
Such a discretization
%not only 
facilitates 
%prompts 
a better 
and 
more intuitive understanding 
of the differential equation
%, 
without 
%taking a limit 
using 
a limiting procedure, which is  
%infinitesimal calculus 
%defining 
needed to define 
%the 
differentiation, 
and 
%but also 
%provides 
%a 
%%n important 
%crucial 
%%%useful 
%idea 
%%key to 
%a candidate scheme for 
is ideal
for performing numerical 
%numerical schemes used in 
%for 
%computer 
%%numerical 
experiments. 
%

%When a given diffenrential equation is completely integrable, 
%For a discretization to be a proper one, 
%it 
%a proper discretization 
The suitable
%proper 
discretization 
of a 
%differential 
completely integrable system 
%equation 
%that is completely integrable 
is 
%preferably 
usually 
%very often 
required to 
%exhibit a qualitative behavior 
%of solution as the differential equation. 
%also integrable. 
%too. 
%be an integrable discretization; this refers to 
%the discretization of the independent variables 
%preserving the integrability. 
retain the integrability; 
%in this case 
if this is satisfied, 
%this 
it is called 
an integrable discretization. 
An integrable system 
%differential equation 
%generally 
%very 
often 
%has 
%allows 
admits more than one 
integrable discretization; in 
%this 
such a case, 
%In that case, 
%one has 
we 
%have a freedom to choose 
can consider 
%look into 
%scrutinize 
%their 
the properties of each integrable discretization 
other than integrability 
and discuss 
%decide 
which one 
%integrable discretization 
is the most favorable 
%fits his/her 
%depending on 
for our 
%the 
purpose. 
%by 
The problem of integrable discretization 
%was started in 
has been sporadically 
%extensively 
studied since the mid-1970s, 
i.e., 
%soon after 
%the early stage of the days 
%
%the early 
%%stage of the 
%days 
%%when 
%in the formation of 
%soon after 
the dawn of 
the modern theory 
of integrable systems. 
%has begun to be 
%%been 
%formulated. 
%(see 
For more than thirty years, 
various techniques 
%approaches 
%to the problem of integrable discretization 
have been 
%proposed 
developed to 
%construct 
%derive 
obtain
%provide a large number of 
integrable discretizations of 
%many 
%various 
continuous 
%integrable 
systems. 
%The 
%interested 
Readers interested in the 
%``historical'' background 
%``history'' 
history of integrable discretizations 
%is 
are 
%kindly 
referred to 
the preface 
%by 
of 
%Suris
Suris's book~\cite{Suris03}.
%)

%The 
Partial differential equations (PDEs) involve 
more than one independent variable. 
%An integrable discretization of a given PDE that is integrable 
The discretization of an 
%given 
integrable nonlinear 
PDE 
is 
%usually 
generally 
performed in two steps; 
in the first step, we discretize the spatial variable(s) 
%; 
and in the second step, the time variable is discretized. 
%However, 
Of course, 
%in 
for some PDEs 
such as 
%with the dispersion relation \mbox{$\omega \propto k^{\pm 1}$} 
%%(cf.\ 
%({\it e.g.}, 
the sine-Gordon equation 
%in the light-cone coordinate 
\mbox{$u_{xt}= \sin u$},
%), 
the roles of 
the space 
%variable and the 
and 
time variables can 
%totally 
be 
swapped 
and it is not 
%meaningless 
meaningful to discuss the order of 
discretization. 
%discretizing the variables. 
%(cf.\ the sine-Gordon equation in the light-cone 
%coordinate \mbox{$u_{xt}= \sin u$})
However, 
%in 
%all the known examples of 
%integrable discretizations of 
for an integrable nonlinear PDE
%s 
wherein the roles of 
%the 
individual 
%independent 
variables are 
essentially 
%distinct 
different 
and not 
interchangeable, 
%swappable,
%cannot be swapped, 
the 
%possible order of discretizing the independent variables 
order in which the independent variables may be discretized  
appears to be unique. 
%Anyway, 
Thus, 
%in this paper, 
we 
%avoid the inessential 
%claim 
can 
consider that the time variable 
is always 
%has to be 
discretized 
%last of all 
%in 
last, 
%lastly 
%only 
after 
%all 
the spatial variable(s) have been discretized. 
%
%
%Thus, it is not 
%%meaningless 
%meaningful to claim that the time variable has to 
%%can 
%be discretized only after all 
%the spatial variables have been discretized. 
%
%Note that no examples of an integrable PDE 
%are known that can be discretized in any order 
%of the independen variables. 
%

%This paper focuses 
In this paper we focus 
on the problem of the 
integrable 
%time 
full
%-
discretization 
of 
%integrable 
differential-difference equations 
in \mbox{$1+1$} 
%\mbox{$(1+1)$} 
dimensions. 
The 
%only 
continuous independent variable to be discretized 
is regarded as 
time, as noted 
%described
%presented 
%mentioned 
above. 
%Note that 
Most of the differential-difference 
equations 
%systems 
considered 
%do not need to have a natural continuum limit of space 
reduce to integrable PDEs 
in a 
%natural continuum 
proper 
%suitable
%appropriate
continuous limit; 
%of the 
%discrete 
%%spacial 
%variable;
%%of discrete space; 
however, this 
%information 
is not necessary and we can 
also 
%in principle 
start 
%deal 
with 
%consider 
%differential-difference equations 
integrable lattice systems 
that 
%do not 
have no 
%any 
%natural 
continuous counterpart. 
%analogue. 
The problem of 
%integrable 
time 
%full-
discretization is, by its nature, 
%rather 
%different 
%inherently 
%intrinsically
distinct 
%in nature 
from 
%has its own 
the problem of space discretization. 
In fact, 
the former problem has its own 
peculiarities 
and 
difficulties 
that 
%do not appear in the latter. 
%problem. 
the latter 
%problem 
does not 
have. 
This point 
%fact 
was 
%probably 
%revealed 
uncovered 
%for the first time 
by 
%the pioneering work of
%early attempt by
Ablowitz and Ladik~\cite{AL76,AL77} 
in their attempt to fully discretize 
%the time variable in 
%of 
the space-discretized 
nonlinear Schr\"odinger 
%(i.e., 
(semi-discrete NLS) 
%space-discrete 
%NLS 
equation~\cite{AL1}. 
%(so-called Ablowitz--Ladik lattice). 
It turned out that 
%some 
unexpected nonlocality 
%of the scheme inevitably 
%appears 
emerges 
in the stage of time discretization; 
the 
%full-discrete 
fully discrete NLS equation 
%s 
%necessarily 
involves 
%some 
infinite sums and/or 
infinite products 
(in the case of an infinite 
%lattice)  
chain) 
with respect to the discrete spatial
variable and 
%are 
is thus a 
global-in-space 
scheme. 
%Such nonlocalities can be 
%We can 
%By introducing 
%Using some auxiliary dependent 
%%additional 
%variables, 
%to rewrite 
The fully discrete NLS equation 
%s 
can superficially 
%immediately 
be 
%re
written 
in a local
%-like 
form 
%in terms of 
using additional 
%some 
%auxiliary 
dependent variables 
%that are 
called {\it auxiliary variables}, 
%~\cite{Suris97'}, 
but it 
%this 
does not provide 
any essential resolution of the nonlocality problem. 
%~\cite{Suris97'}. 
%fixing-up
%such a trick cannot provide a truly local scheme~\cite{Suris97'}, 
%on a superficial level 
%however, such a form makes it difficult to recognize it 
%as a difference analogue of the original evolutionary PDE. 
The subsequent paper 
%of 
by Taha and Ablowitz~\cite{Taha1} 
%on other integrable lattices 
%following works 
%made it clear 
%showed 
%implied 
%strongly 
%implies 
reinforces the impression 
that 
%the nonlocality 
the appearance of global 
%nonlocal 
terms 
%problem 
%is not restricted to the NLS equation, 
%rather it 
is a general feature 
of 
the problem of time discretization 
for 
%other 
integrable lattice systems. 

%More specifically, 
The pioneering 
work 
%method 
of Ablowitz 
%and Ladik~\cite{AL76,AL77} 
and coworkers~\cite{AL76,AL77,Taha1} 
%(see also Taha--Ablowitz and Suris) 
is 
based on the zero-curvature (Lax
%-
pair) representation; 
%In particular, 
%%they assumed 
%their guiding principle is that 
%used 
the 
%their 
guiding 
principle
%philosophy 
is that 
%a time-discretized lattice should 
%have the same 
the time discretization 
%should 
does not 
change 
%alter
the spatial part of the Lax pair 
%Lax matrix 
%as that for 
%%both 
%the time-continuous 
%and time-discrete 
for an integrable lattice system. 
%Their 
This 
%stance 
%approach 
automatically 
guarantees 
%results in 
the major 
%great 
advantages of 
the 
%obtained 
%time 
full
%-
discretization;
%s; 
%obtained with 
%%by 
%their approach; 
that is, 
the time-discretized lattice systems
%time discretizations
%thus obtained
%are
have
%exactly
the same 
%set of 
%conserved quantities 
integrals of motion 
and
%do not change
the same structures of the solutions
%as the time-continuous
%original
%of
as 
%those of 
the
%continuous-time
original 
%time-continuous
continuous-time
lattice systems. 
%
%Only the time
%%dependence
%evolution of
%%certain
%%some
%the parameters
%in the
%solutions
%%(roughly speaking, angle variables)
%%corresponding
%that correspond
%to the ``angle variables''
%%change upon time discretization.
%%are
%is discretized.
In 
%a more modern language,
more modern terms,
%the maps generated by the 
each of 
their 
%such 
time discretizations belongs to 
the same integrable hierarchy as the underlying 
continuous-time system~\cite{Suris97,Suris97'}. 
Despite 
%this beauty
the elegance of this result, 
%of it, 
the appearance of 
%an
infinite 
%global 
%nonlocal 
%terms 
sums/products is 
%not 
%regarded as 
a 
%serious 
shortcoming not 
%entirely 
acceptable to everybody, 
and 
%some 
new 
%teqniques are needed 
ideas 
%should be introduced in order 
are needed 
%should be proposed 
to remove 
%fix 
%resolve 
the nonlocality. 
%Since 
In this regard, 
%In the late 1990s, 
Suris 
%had 
%'s method 
%based on 
%of 
%for 
recently 
introduced 
%in the late 1990s 
%and 
%developed 
the notion of 
{\it localizing changes of 
%dependent 
variables}, 
%and 
%since then 
%used 
applied it 
%to obtain 
to a large number of 
%integrable lattices to 
%examples
integrable lattice systems, and obtained 
their 
%local 
time discretizations 
%in the form of 
written 
%as 
in local equations of motion~\cite{Suris03}. 
Note that 
some of 
%The results 
his discretizations 
%results 
%were shown to 
coincide with 
%the 
earlier results obtained 
%were obtained earlier 
%with 
using Hirota's bilinear 
%direct 
method~\cite{Hirota04}, 
as described in the bibliographical remarks 
%in the book
%of 
in his book~\cite{Suris03}. 
%in~\cite{Suris03}. 
%describe, 
%The 
Suris's idea 
%is 
was to find, {\it by guessing}, 
a change of variables 
%{\it by guessing} 
%involving one parameter corresponding to the step size of time,
%by which
\/such that 
%allows to express 
%both 
%the auxiliary variables and 
the equations 
of motion as well as the auxiliary variables 
%for time evolution 
%are 
can be 
%rewritten 
expressed 
locally in terms of the {\it new} \/dependent variables; 
the change of variables 
involves 
the step size of time as a 
%deformation 
parameter
%, 
%one parameter corresponding to the step size of time 
and is considered as a discrete 
Miura transformation giving 
%to give 
%yielding 
a one-parameter deformation 
of the original lattice 
hierarchy. 
Thus, 
%--Taha
%the time-discretized lattices
%time discretizations
%thus obtained
%are
%have
%exactly
%the same 
%set of 
%conserved quantities 
%both 
the 
integrals of motion 
and
%do not change
the solution formulas 
%same 
%structures of solutions
%as the time-continuous
%original
%of
%continuous-time
%original time-continuous lattices. 
%have to be doformed 
for the time-discretized lattice system 
%local time discretization 
are 
%have to be 
%also 
%undergo 
%a 
%the deformation 
deformed 
accordingly. 
%unlike 
%%the above-mentioned beauty of 
%%the Ablowitz--Ladik-type discretizations
%the work of Ablowitz {\it et al.}~\cite{AL76,AL77,Taha1} \
%, and 
%; 
%unfortunately, 
%Thus, 
%Unfortunately, 
%the above-mentioned beauty of the Ablowitz--Ladik-type discretizations 
%are 
%have been lost. 
% unfortunately.
%in 
%the local time discretizations based on 
%Suris's approach. 
%
%Suris's {\it localizing changes of 
%%%dependent 
%variables} approach. 
%
%On the one hand, 
Moreover, 
%The 
although Suris's 
%
%{\it localizing changes of 
%%dependent 
%variables} \/have 
approach has 
%successfully been 
%applied to 
%give 
successfully 
provided 
%a lot of 
many interesting 
examples, 
%of local 
%%%full-discrete 
%time-discrete 
%lattices, 
%full-discretizations. 
%On the other hand, 
%the 
its applicability 
%of Suris's method 
is 
rather limited. 
%
%roughly speaking, 
%to the KdV-type lattices, 
%Indeed, 
In particular, 
%it cannot be applied to NLS-type lattices 
it is not applicable to 
the time discretization
%s 
of NLS-type lattices, 
%including 
such as the semi-discrete NLS equation 
(also called the 
Ablowitz--Ladik lattice)~\cite{AL1}, 
%of Ablowitz and Ladik, 
wherein the 
two dependent variables 
%stand on an ``equal'' footing, and
can be related 
%to each other 
by a complex conjugacy 
reduction. 
%relation. 
%to form a single equation. 
%
%In particular, it is not applicable to 
%the time discretization of the semi-discrete NLS 
%equation in full generality. 
%In fact, 

The main objective of this paper is to propose a {\it systematic}
\/method for constructing 
%{\it local} \/
time 
%full-
discretizations of 
integrable lattice systems 
written
as
%in 
{\it local} \/equations of motion. 
%differential-difference equations 
%in \mbox{$(1+1)$} dimensions. 
%However, 
In contrast to the other known 
%existing 
methods, 
our method generally requires no 
%{\it a priori} \/assumption 
{\it ad hoc} \/treatment 
on a case-by-case basis 
%based on a case-by-case consideration 
and appears to have no serious limitations 
in its 
%the 
applicability; 
%of our method; 
%hopefully 
it can be applied to 
%most of the 
possibly 
%almost 
all 
%integrable 
lattice systems 
in \mbox{$1+1$} dimensions 
%having a zero-curvature representation. 
possessing a Lax
%-
pair representation. 
%
%Note that 
%%to the best of the author's knowledge, 
%%every known 
%all the existing methods 
%either have 
%%has 
%a 
%%very narrow 
%rather restricted applicability or require a human 
%interference based on lots of ad hoc process, 
%though each method 
%%it 
%%every known method 
%may have its own advantages. 
%In particular, 
%To the best of the author's knowledge, 
%no method can be applied systematically 
%to a wide spectrum of 
%%succeeds in 
%without a lot of implicit knowledge and experience 
%of the method. 
In particular, it 
%can be applied 
%is applicable to the 
can be used to obtain 
local 
%time 
full
%-
discretizations of 
%the 
NLS-type lattices, including 
the Ablowitz--Ladik lattice~\cite{AL1}.
%as well. 
Actually, 
our method 
can 
%also 
be considered 
%is 
%a refined and extended 
as 
%the 
a {\it completed version} \/of 
%both refines and extends 
the Ablowitz--Ladik approach~\cite{AL76,AL77}; 
it 
both 
refines and extends 
their 
%Ablowitz--Ladik 
work
in an essential way. 
%
%We, however, 
%make 
%%add 
%a decisive prescription 
%make decisive progress
%play a decisive role (in)
%play a critical role in
%make a decisive contribution to
%discover crucial clues to
%improvement 
%decisive breakthrough
%make decisive progress 
%To restore the locality of the equations, 
%The 
%we make 
A decisive breakthrough 
%This is a key ingredient in the recipe for success
%is made 
%can be 
has been made 
by considering
%applying 
%noting 
the lowest-order ``conservation laws'', 
derived from the zero-curvature condition 
%represenation 
written in 
%the 
matrix form. 
%rather than the component-wise form of Ablowitz--Ladik and to 
%and deriving the lowest-order ``conservation laws''. 
%of the lowest order. 
In the process, a critical role is played by 
%the existence of 
%the spectral 
an arbitrary parameter in the Lax pair, 
%in the zero-curvature representation, 
called the spectral parameter. 
The 
requirement 
%consistency of the ``conservation laws'' 
that 
%of 
all the fluxes 
corresponding to 
%associated with 
the same conserved density 
have to essentially
coincide 
%up to 
%%an overall 
%%a 
%trivial 
%%overall <- product form; note that sum form is usual
%%constant 
%factors 
%generally 
results in an ``ultralocal'' 
algebraic system for the auxiliary variables; 
the simpler 
%and trivial 
case 
%that 
where the 
%``conserved density'' 
conserved density is trivially a 
constant can be treated 
%in the same way. 
in a similar manner. 
Thus, 
by solving 
%the 
%resulting
this 
algebraic system, 
%for the auxiliary variables. 
we can restore the locality of the equations; 
that is, 
the global terms 
%having appeared 
appearing 
in the stage of 
time discretization 
can be 
%expressed locally 
replaced 
%substituted 
by {\it local} \/expressions 
in terms of the original 
dependent variables. 
%to restore the locality of the equations, 
%we have only to solve the 

This paper is organized as follows.
In section 2, 
we describe 
%explain 
%our 
the 
general method 
%recipe 
for discretizing the time variable. 
%;
%%obtaining 
%%local time discretizations; 
%%is presented 
%%described 
%%and, 
%to achieve a better understanding, 
%%of it, 
%%we 
%it 
%%the recipe 
%is illustrated 
%%it 
%with 
%%by 
%%the 
%%an 
%one 
%%a 
%basic 
%%concrete 
%example, 
%%of 
%%i.e., 
%namely, the Toda lattice in the Flaschka--Manakov 
%coordinates~\cite{Flaschka1,Flaschka2,Manakov74}. 
%variables. 
In section 3, 
%the effectiveness
%ity
%validity
%of 
%our
%the
%wide applicability of our
%method is 
%illustrated with
%by
%examples such as
%applied to 
%using our method, 
we construct 
%derive 
%the 
%local 
time discretizations of 
the Toda lattice in 
%the 
Flaschka--Manakov coordinates~\cite{Flaschka1,Flaschka2,Manakov74}, 
the Ablowitz--Ladik lattice, 
%(integrable semi-discrete NLS),
the Volterra lattice, the modified Volterra lattice,
%the Toda lattice,
%equation),
and the lattice Heisenberg ferromagnet model. 
As a spin-off, 
%by-product, 
we obtain ultradiscrete analogues~\cite{TakaMatsu95,ToTaMaSa} of 
the Volterra lattice and 
%the 
modified Volterra lattice. 
%It is also pointed out 
%
%We also 
%%mention 
%uncover 
%that 
%%our 
%the full-discrete NLS equation thus constructed 
%is closely related to the full-discrete NLS equation of 
%Quispel {\it et al.}~\cite{QNCL}
In addition, we 
uncover 
%mention 
%an 
%Some 
unexpected 
%and unnoticed 
relationships 
%overlap 
%coincidence 
with the 
%early 
%earlier 
%relevant 
work 
%full-discrete NLS equation and Heisenberg ferromagnet model 
of Nijhoff, Quispel, 
Capel
%, 
{\it et al.}~\cite{NQC,QNCL,NCWQ,NijCap}. 
%coworkers
%and van der Linden~\cite{QNCL,NCWQ,NijCap}.
%is uncovered. 
%clarified. 
%revealed. 
%commented upon. 
%The last section, section 4, 
Section 4 is devoted to concluding remarks.
%
%as will be shown in subsequent sections. 
%

%revealed that some auxiliary dependent 
%additional 
%variables have to be introduced to 
%obtain a closed 
%%system 
%time discretization of 
%%equations. 
%
%auxiliary variable 

%
%The use of the auxiliary variables changes the 
%global scheme to 
%an apparently local but 
%%or 
%``non-evolutionary'' scheme. 
%%system 
%%difficult to define ``non-evolutionary'' in the time-discrete case
%%It is better to discuss if the scheme is ``explicit/implicit'' 

%In section 2, ... 
%In addition, we explain how Suris's method 
%({\it localizing changes of 
%%dependent 
%variables}) 
%%\/
%can be characterized/interpreted using the zero-curvature 
%representation in our formulation. 

%%%%%%%%
\section{Method 
%method based on conservation laws
%the 
%general 
%The recipe 
%theory 
for 
%:\ 
%local 
time 
%integrable 
discretization}
\label{sect2}

In this section, we 
%consider
discuss 
the problem of 
%a method for
%local 
time discretization for a given 
%general 
%integrable lattice systems. 
integrable lattice system. 
We start with a 
%the 
Lax
%-
pair formulation 
in the 
%time-continuous 
continuous-time
case
%, 
%for the time-continuous lattice system, 
and then proceed to discretize the time variable. 
%the time-discrete case. 
%the 
%its time discretization based on the Lax pair. 
%In general, 
A set of auxiliary variables 
%satisfying nonlocal recursive relations 
is 
%has to be 
introduced to express
%write out 
the time-discretized lattice 
as a closed system of 
equations. 
%for the time-discretized lattice. 
%the time-discretized lattice system in closed form. 
%system. 
Using 
%some 
the fundamental ``conservation laws'' 
derived from the Lax pair, 
%Lax-pair representation, 
we 
%povide 
%present a procedure for expressing 
can obtain local expressions for 
%of 
the auxiliary variables 
%locally 
in terms of the original 
%dependent 
variables. 
%thus arrriving at 
%producing a local time discretization. 
%solving 

\subsection{Lax pair 
%auxiliary variables 
and
%%fundamental 
a 
%``conservation law'' 
conservation law in the 
%time-continuous 
continuous-time
case} 
%in the semi-discrete case} 

%Let us begin with 
%the general 
The 
%A 
Lax
%-
pair 
formulation 
%of the ISM
%for integrable systems
%in a discrete space and continuous time.
%in the space-discrete and  and time-continuous case
in a semi-discrete space-time
comprises
%i.e.,
%is given by
%We consider
a pair
%set
of linear equations, 
\begin{equation}
\Psi_{n+1} = L_n (\lambda)
\Psi_n, \hspace{5mm} \Psi_{n,t} = M_n (\lambda)\Psi_n.
\label{line}
\end{equation}
Here, $\Psi_n$ is a column-vector function
%, 
%where 
and $n$ is the discrete spatial variable. 
The subscript $t$ denotes
differentiation 
%by 
with respect to
the continuous time variable 
$t$.
The square matrices $L_n$ and $M_n$
%are square matrices which
depend on the 
%constant 
spectral parameter $\lambda$,
%that does not depend on 
%that 
which is 
%assumed to be 
an arbitrary constant 
%, namely, 
independent of $n$ and $t$. 
%a parameter.
%
The compatibility condition of the overdetermined system 
(\ref{line}) is given by~\cite{AL1,Kako,AL76,Suris97,Suris97',Suris00}
\begin{equation}
 L_{n,t} +L_n M_n - M_{n+1}L_n = O,
\label{Lax_eq}
\end{equation}
which is
(a semi-discrete version of) the zero-curvature
condition. 
%Throughout this paper, we use 
The symbol 
%italic 
$O$
%when dependent variables in the equation considered can take their 
%values in matrices.
is used to 
%imply that the left-hand side is matrix-valued. 
stress that this is a matrix equation. 
If we specify 
%choose 
the
$\lambda$-dependent
matrices $L_n$ and $M_n$ 
%properly,
appropriately,
(\ref{Lax_eq}) results in 
%gives
%the time evolution of
a 
%$\lambda$-independent 
closed 
%semi-discrete 
differential-difference 
system for some 
%elements 
$\lambda$-independent 
%entries of 
quantities in 
$L_n$ and 
%/or 
$M_n$. 
%independent of $\lambda$. 
%the parameter.
In such 
%cases, 
a case, the pair of matrices $L_n$
%, 
and $M_n$ 
%and the parameter are 
is called 
%the 
a Lax pair. 
%andthe spectral parameter, respectively. 
%We assume that 
The matrix $L_n$ is usually 
%{\it ultralocal} \/
ultralocal in the dependent variables; 
%(really necessary?); 
that is, if $L_n$ involves 
%a 
some 
%dependent 
variable, say $u_n$, then it does not 
%contain 
involve shifted variables such as $u_{n\pm 1}$ and $u_{n\pm 2}$. 
%, etc. 
In addition, the determinant of 
$L_n$ is 
%assumed 
required 
%assumed 
to be nonzero for generic $\lambda$ 
so that the spectral problem is well-posed 
on the entire 
%whole 
infinite chain. 
%line. 

The zero-curvature condition (\ref{Lax_eq}) generates 
%implies 
%the existence of 
a conservation law
%``conservation law'' 
of the following form: 
\begin{equation}
\frac{\partial}{\partial t} \log (\det L_n) 
%+ \Delta_n^+  
= \boldsymbol{\Delta}_n^+ 
%\left( \mathrm{tr} \hspace{2pt}M_n \right)
(\mathrm{tr} \hspace{1pt} M_n). 
\label{sd-cons}
\end{equation}
Here, $\boldsymbol{\Delta}_n^+$ is 
the forward difference operator in the spatial direction, i.e.,
%$\boldsymbol{\Delta}_n$ to indicate
%,
\begin{equation}
%\[
\boldsymbol{\Delta}_n^+
%f_{n+j}:=f_{n+j+1}-f_{n+j}.
f_{n}:=f_{n+1}-f_{n}.
%\]
\label{space-diff}
\end{equation}
For a {\it proper} \/integrable lattice system, 
%\mbox{$\log (\det L_n)$} 
%the quantity 
\mbox{$\det L_n(\lambda)$} 
%is 
has to be 
either 
a 
%$t$
time-independent 
%``constant'' 
%constant
%a 
%scalar 
function of $\lambda$ 
%%%some 
%a constant 
or the exponential of 
a $\lambda$-independent 
%quantity 
%and exponentiated 
conserved density 
%times 
%
%scalar 
%multiplied by it. 
%up to 
%scalar 
multiplied by 
%it. 
%with 
an overall 
%constant 
$t$-independent 
factor. 
%a scalar 
%%some 
%function of $\lambda$. 
%In fact, 
Indeed, 
if this 
%is 
was not satisfied, 
%for example, 
%{\it e.g.\ }
{\it e.g.}, \mbox{$\det L_n (\lambda)= 1+ \lambda u_n$}, 
then the expansion of \mbox{$\log \det L_n (\lambda)$} 
with respect to $\lambda$ 
%indicates that the lattice system has 
%yields 
would yield
an infinite number of 
(almost) 
ultralocal conserved densities, say 
%for example, 
\mbox{$u_n$}, \mbox{$u_n^2$}, 
%\mbox{$u_n^3$}, 
\ldots, 
%\hspace{1pt}. 
%However, this implies that 
%Thus, the lattice is not intresting 
and thus the lattice system 
%the lattice 
%is 
%must 
would be 
trivial in some sense. 

Note that 
%It should be noted that 
the zero-curvature condition (\ref{Lax_eq}) 
%there exist simple 
%implies 
has 
the following 
%its 
%``invariance'' 
invariance properties: 
%with respect to the transformations of the Lax pair:
% as follows: 
%is invariant under the following 
\begin{enumerate}
%\begin{itemize}
\item[(a)]
\mbox{$L_n \rightarrow f(\lambda) L_n$}, where $f(\lambda)$ is 
%\mbox{$L_n \rightarrow \alpha(\lambda) L_n$}, where $\alpha$ is 
%an \mbox{$(n,t)$}-independent 
a \mbox{$t$}-independent 
%n arbitrary 
%nonzero 
scalar function,
%;
%of 
%%constant that can depend on 
%$\lambda$, 
%the spectral parameter, 
\item[(b)]
%The freedom: 
%\mbox{$M_n \rightarrow M_n + \beta (\lambda)I$}, where $\beta(\lambda)$ 
\mbox{$M_n \rightarrow M_n + g(\lambda)I$}, where $g(\lambda)$ 
is an \mbox{$n$}-independent 
%\mbox{$(n,t)$}-independent 
%an arbitrary 
% nonzero 
scalar 
function and $I$ is the identity matrix,
%;
%constant that can depend on $\lambda$, 
%the spectral parameter, 
\item[(c)]
\mbox{$L_n \rightarrow 
%{\mathrm e}^
{\exp (\alpha \rho_n)} L_n$ and 
\mbox{$M_n \rightarrow M_n + \alpha j_n I$}}, 
where $\alpha$ is a 
%nonzero 
parameter, 
%constant, 
%\mbox{$\log \rho_n$} 
\mbox{$\rho_n$} is a conserved density and $j_n$ is the 
corresponding flux (up to a sign), namely, 
\mbox{$\partial_t \rho_{n}
%\partial \rho_{n}/\partial t 
= \boldsymbol{\Delta}_n^+ j_n$}. 
%Note that 
%\end{itemize}
\end{enumerate}
%Namely, the set of 
%resulting equalities does not change under any of the above 
%(almost trivial) gauge transformations. 
%In other words, 
%the Lax pair can be given up to 
%%has 
%the above indefiniteness. 
In particular, 
using (\ref{sd-cons}) and the above properties, 
%$(\mathrm{c})$, 
we can 
%always 
convert 
%``normalize'' 
the Lax pair 
%such that 
to a 
%``normalized'' 
normalized form, i.e., 
\mbox{$\det L_n=1$} and \mbox{${\mathrm{tr}}\hspace{1pt} M_n=
%0
%\mathrm{const.}$}
0$}.

\subsection{Lax pair 
in the 
%time-discrete (discrete-time?) 
discrete-time case and 
auxiliary variables}
Now, 
we 
%briefly 
discuss how to construct the time discretization of a given 
%the
%time discretizations of the
%obtained
%semi-discrete 
lattice system having 
%possessing 
%a 
the Lax pair $L_n$ and $M_n$. 
%A 
The natural 
%time discretization 
%full
%time-discrete 
discrete-time
analogue of the linear system 
(\ref{line}) is given by
%\mbox{$\Psi_{n+1} = L_n (\lambda) \Psi_n$} and 
%\mbox{$\widetilde{\Psi}_n = V_n (\lambda) \Psi_n$},
\begin{equation}
\Psi_{n+1} = L_n (\lambda) \Psi_n, \hspace{5mm}
\widetilde{\Psi}_n = V_n (\lambda) \Psi_n, 
\label{line2}
\end{equation}
where the tilde
%$(\; \widetilde{} \; )$
denotes the forward 
%step-up 
shift (\mbox{$m \to m+1$})
in the discrete time coordinate \mbox{$m \in {\mathbb Z}$}. 
Here and hereafter, 
the 
%time 
%$m$-
dependence on 
%the time variable 
$m$ is usually 
suppressed unless 
it is shifted. 
%Then we obtain a full-discrete version of the zero-curvature condition 
The compatibility condition of 
this overdetermined 
linear 
system is given by~\cite{AL76,AL77,Taha1,Orfa1,Suris97,Suris97',Suris00}
\begin{equation}
\widetilde{L}_n V_n = V_{n+1} L_n,
%\label{Lax_full}
\label{fd-Lax}
\end{equation}
which is (a fully discretized 
%full-discrete 
version of) the zero-curvature condition. 
Note that (\ref{fd-Lax}) can be rewritten as 
\begin{equation}
V_n = \widetilde{L}_n^{\hspace{1pt}-1} V_{n+1} L_n
\;\; \mathrm{or} \;\;
\widetilde{L}_n V_n L_n^{-1}= V_{n+1}.
%\label{Lax_full}
\label{fd-Lax2}
\end{equation}
%
%For a semi-discrete system possessing a Lax pair $L_n$ and $M_n$,
%we usually obtain a time discretization by choosing
%%by appropriate choices of
%$V_n$ appropriately for the same $L_n$-matrix.
Following the work of Ablowitz 
%{\it et al.}
and coworkers~\cite{AL76,AL77,Taha1}, 
%and Ladik~\cite{AL76,AL77}
%and coworkers
the matrix 
\mbox{$L_n (\lambda)$} is assumed to be the same 
as that in the semi-discrete 
case. Then, we look for a \mbox{$V_n (\lambda)$} such that 
the zero-curvature condition (\ref{fd-Lax})
results in
%provides 
a closed system of partial difference equations 
%that provides 
%giving the 
providing 
a discrete-time 
%time-discrete 
analogue 
%full-discrete analogue 
%time discretization 
of the semi-discrete system. 

For this purpose, 
%first 
we 
%first 
assume that the matrix \mbox{$V_n$} 
%(\lambda)$} 
has 
%the 
asymptotic 
%form 
behavior, 
\begin{equation}
V_n = I + h \left[ M_n + 
%{\mathcal O}(h) 
%{\cal O}(h)
O(h)
% o(1) 
\right],
% M_n + O(h^2),
\label{asymp-h}
\end{equation}
where $h$ is a sufficiently 
small (but nonzero) parameter 
independent of $\lambda$
%, 
%. 
%Here, 
and is considered 
%denotes 
the difference interval of time 
(cf.\ (\ref{Lax_eq}) and (\ref{fd-Lax})). 
%is denoted by a small parameter $h$, 
%or more precisely, 
More precisely, 
$h$ approximates 
the ``true'' 
%time step
step size of discrete 
time
%, 
%is equal to \mbox{$h\{ 1+o(1)\}$}. 
up to an \mbox{$o(h)$} error. 
%Second, 
Moreover, we assume that 
\mbox{$V_n(\lambda)$} has 
essentially the same 
$\lambda$-dependence 
%on $\lambda$ 
as \mbox{$I + h M_n (\lambda)$}. 
Note, however, that 
%the Lax matrix 
\mbox{$M_n (\lambda)$} can only 
be 
%defined 
determined 
%only 
up to 
%an 
the addition of an $n$-independent 
%any 
scalar matrix (cf.\ (b)). 
Thus, this arbitrariness has to be taken into account 
%consideration 
in determining \mbox{$V_n(\lambda)$};
this corresponds to the freedom 
(nonuniqueness) 
of 
%determining 
choosing 
the linear part 
%(i.e., dispersion relation) 
of the time-discretized lattice system 
that determines the dispersion relation~\cite{AL77}. 
%time discretization. 
%a nonzero entry 
Moreover, some $n$-independent 
%constant 
%entries 
quantities
(usually 
%taken 
set as constants) 
%terms 
in $M_n$ 
%may 
translate 
%change 
into 
%transform to
%correspond 
%will be replaced by 
%new 
%non-constant 
$n$-dependent 
%entries 
quantities
%new variables 
in $V_n$, 
%be replaced 
which typically constitute, up to 
%an inessential 
%a 
the reformulation 
%redefinition 
of the dependent variables, 
%the 
new {\it auxiliary 
%dependent 
variables}. 
Then, we
%We need to 
%should 
specify appropriate boundary 
%values 
conditions 
for these new 
%non-constant entries 
variables 
%new variables 
in $V_n$, which should retrieve the corresponding 
%constant 
$n$-independent 
%entries 
values 
in $M_n$.
%; that is, we assume constant 
%%(i.e., 
%%(time-independent) 
%boundary values of the auxiliary variables. 
%%More specifically, 
%%In addition, 
%Moreover, 
In fact, we usually 
assume 
%the 
``constant'' boundary conditions for $V_n$, 
%are assumed: 
\begin{equation}
\lim_{n \to - \infty }V_n = \lim_{n \to + \infty }V_n = 
{\textrm{finite.}}
%$n$-independent 
%const. (time-independent)}
%and 
\label{V+-}
\end{equation}
%
%This is consistent with the time independence (assumption) of $h$. 
%
%Note that 
%Still, 
However, the right-hand side 
%can 
is allowed to depend 
%, in principle, 
on the time variable \mbox{$m \in {\mathbb Z}$}. 
In the 
%actual 
application of the inverse scattering method 
based on the Lax pair, we 
need to 
%impose proper boundary conditions on 
specify the boundary conditions 
%of 
for $L_n$ as \mbox{$n \to \pm \infty$}. 
%typically, 
%\[
%\lim_{n \to - \infty }L_n = \lim_{n \to + \infty }L_n = 
%{\textrm{
%%$m$-independent 
%const. 
%%(time-independent)}
%%and finite.
%}}
%\]
In such a case,  it is
%, in general, 
{\it redundant} 
%\/in general 
\/to impose 
%assume 
%specify 
%both boundary values 
the 
%asymptotic behavior 
boundary conditions 
%of $V_n$ 
on $V_n$ 
at both spatial ends as given in (\ref{V+-}), 
and it is 
%highly 
nontrivial that the redundant boundary conditions are compatible. 
In fact, it is sufficient to know 
%specify 
only one of 
%sufices 
the two boundary values, 
%$V_{-\infty}$ and $V_{+\infty}$. 
\mbox{$\lim_{n \to - \infty }V_n$} 
%and 
or \mbox{$\lim_{n \to + \infty }V_n$}. 
In the existing literature~\cite{Suris03,AL76,AL77,Suris97,Suris97',Suris00}, 
it is 
%{\it hypothesized} \/
hypothesized that 
%the auxiliary variables have the same boundary values 
%as \mbox{$n \to - \infty$} and as \mbox{$n \to +\infty$}; 
these two limits 
%boundary values 
indeed coincide; 
%only 
a preliminary 
%partial 
%dicsussion 
consideration without using 
this hypothesis 
%assumption
is given in~\cite{Ab78}. 
%can be found in~\cite{Ab78}. 
In 
%the next 
section~\ref{Exs}, we 
%illustrate 
demonstrate 
%with 
for specific examples that this 
%``hypothesis'' 
%can be verified. 
is not a hypothesis but 
%rather 
a verifiable fact. 

%If we 
Let us decompose \mbox{$L_n (\lambda)$} into a sum of 
terms, each of which is 
the product of an \mbox{$(n,m)$}-independent 
scalar function of $\lambda$ and a 
$\lambda$-independent matrix, 
%%each multiplied by a scalar function of $\lambda$:
%%each having a $\lambda$-dependent 
%with scalar coefficients that are functions of $\lambda$ only, i.e.,
i.e.,
\[
L_n (\lambda) = \sum_{i= i_{\mathrm{min}}}^{i_{\mathrm{max}} }
 f_i (\lambda) L_n^{(i)}.
\]
The scalar 
functions  \mbox{$
%\{ 
f_i (\lambda)
%\}
\; (i_{\mathrm{min}} \le i \le i_{\mathrm{max}})
$} 
are linearly independent; 
typically, they are 
%integer 
powers of $\lambda$, {\it e.g.}, 
%namely, 
\mbox{$f_i (\lambda) = \lambda^i$}. 
The nonzero 
%entries in 
elements of 
%independent 
%quantities 
the matrices $L_n^{(i)}$ 
are classified into 
%consist of 
%three 
two types, that is, 
constants 
%, 
(or, at most, 
arbitrary 
%? 
functions 
of only one independent variable) 
%, 
%only, 
%or 
and dynamical variables depending 
%essentially 
on both 
%the 
independent variables. 
We express the entire 
set of 
%dependent variables involved 
%and 
%non-constant 
%all 
%nonzero and functionally independent 
%the 
functionally 
independent 
%non-equivalent 
dynamical variables 
%(functionally independent for fixed $n$ and $m$) 
%entries 
%elements 
in 
%all the $\lambda$-independent matrices 
\mbox{$L_n^{(i)} \; (i_{\mathrm{min}} \le i \le i_{\mathrm{max}})$} 
%and 
%those 
%that in $V_n$ 
as \mbox{$\{ \boldsymbol{l}_n \}$}. 
%and \mbox{$\{ \boldsymbol{v}_n \}$}, respectively, 
%the requirement that 
%Similarly, 
The set of 
%independent 
dynamical 
variables 
\mbox{$\{ \boldsymbol{v}_n \}$} is defined 
from 
%the decomposed form of 
$V_n(\lambda)$ in exactly the same way. 
%Thus, 
Then, 
the zero-curvature condition (\ref{fd-Lax}) 
%is satisfied for any value of $\lambda$ 
%generally 
%
%results in 
provides a 
(typically 
%usually 
bilinear 
%(at most bilinear) L_n may contain 
%higher degree terms or trignometric fundtions 
algebraic) system for 
\mbox{$\{\hspace{1pt}
\widetilde{\boldsymbol{l}}_n, \boldsymbol{l}_n \}$} 
and \mbox{$\{ \boldsymbol{v}_n, \boldsymbol{v}_{n+1} \}$}. 
%\mbox{\boldmath$l$_n}
In particular, 
%a useful subsystem of 
this 
%the whole 
%algebraic 
system contains 
a useful subsystem, that is, 
an ultralocal and 
linear system 
%for 
%%a set of ultralocal algebraic relations among 
%%various 
%some components 
%%entries 
%of $V_n$
in \mbox{$\{ \boldsymbol{v}_n \}$}, wherein the 
%some 
coefficients 
%can 
involve 
\mbox{$\{\boldsymbol{l}_n \}$} 
%$\widetilde{L}_n$ 
and 
%the entries of 
%$\widetilde{L}_n$ and 
%$L_n$ and 
its shifts; 
this subsystem 
can also 
be derived 
%understood 
%obtained 
by noting 
%requiring 
%observing 
that (\ref{fd-Lax2}) holds true 
as an identity in $\lambda$. 
%, (\ref{fd-Lax2}) 
%provides an ultralocal linear system for 
%%a set of ultralocal algebraic relations among 
%%various 
%some components 
%%entries 
%of $V_n$ wherein some coefficients can 
%involve the entries of 
%%$\widetilde{L}_n$ and 
%$L_n$ and its shifts. 
We 
%can 
solve this subsystem
%, that is, a subsystem of the whole algebraic system, 
to express a subset of 
\mbox{$\{ \boldsymbol{v}_n \}$} in terms of 
%other components 
the remaining
%rest of 
%variables in 
\mbox{$\{ \boldsymbol{v}_n \}$} 
as well as 
\mbox{$\{ \boldsymbol{l}_n \}$} and 
its shifts 
%of 
%shifted variables of 
%\mbox{$\{ \boldsymbol{l}_n \}$} 
such as
\mbox{$\{\hspace{1pt} \widetilde{\boldsymbol{l}}_n \}$}
%, 
%\mbox{$\{\hspace{1pt} \widetilde{\boldsymbol{l}}_{n-1} \}$}, 
and \mbox{$\{\hspace{1pt} {\boldsymbol{l}}_{n-1} \}$}. 
%, etc. 
%Solving it, 
Thus, 
we can reduce the number of independent 
%entries 
dynamical variables in 
%of 
$V_n$
%, 
%. Thus, we can maintain the ultralocality of $V_n$ 
while maintaining the ultralocality of $V_n$ 
%in 
with respect to 
%the variables 
\mbox{$\{ \boldsymbol{v}_n \}$}. 
It remains to 
%%is important to 
be verified 
%checked 
%It should be confirmed 
that the zero-curvature condition (\ref{fd-Lax}) 
%defines 
indeed provides 
a 
%meaningful
%{\it meaningful}
meaningful 
%``meaningful'' 
\/fully discrete system for 
the 
%this 
reduced set of dependent variables; 
%derived from 
%that is, 
it should 
%has to 
define 
a 
consistent and unique time evolution 
for 
%a 
%given 
%starting from
% any 
%a 
generic 
initial data 
%satisfying
under 
%the specified 
appropriate boundary conditions 
(cf.\ (\ref{V+-})). 
%Roughly speaking, 
%Aside from a minor reformulation and paraphrasing, 
%this is what Ablowitz and coworkers did in their work. 

\subsection{Fundamental 
``conservation laws'' in the 
%time-discrete (discrete-time?) 
discrete-time case
%derived from the Lax pair
}
\label{FCL}

%Next
%Third, 
We 
%should note the ``conservation law''
consider 
%take 
the determinant of both sides of (\ref{fd-Lax}) to obtain 
the equality 
%local 
%
\begin{equation}
(\det \widetilde{L}_n) (\det V_n) = (\det L_n) (\det V_{n+1}).
\label{fd-cons1}
\end{equation}
This relation can 
%be rewritten in the form of a 
be written more explicitly 
%as 
in the form of a 
%local 
discrete 
%``conservation law'', 
conservation law, 
\[
%\begin{equation}
\boldsymbol{\Delta}_m^+ 
\log (\det L_n) 
%+ \Delta_n^+  
= \boldsymbol{\Delta}_n^+ 
%\left( \mathrm{tr} \hspace{2pt}M_n \right)
\log (\det V_n).
%\label{fd-cons2}
%\end{equation}
\]
Here, $\boldsymbol{\Delta}_m^+$ is 
the forward difference operator in the time direction, 
i.e., 
\mbox{$\boldsymbol{\Delta}_m^+
%f_{n+j}:=f_{n+j+1}-f_{n+j}.
f_{n}:=\widetilde{f}_{n}-f_{n}$}. 
This is the discrete-time version of (\ref{sd-cons}). 
%Note again 
%Recall 
Note that 
%\mbox{$\det L_n (\lambda)$} should be uniform 
%in $\lambda$, up to an overall constant factor. 
after 
%removing 
cancelling 
%this constant factor, 
%an 
the 
%overall 
%constant 
$m$-independent 
factor 
of 
%involved 
%contained in 
\mbox{$\det L_n$}, 
% (\lambda)$}, 
the relation 
(\ref{fd-cons1}) should 
%read as 
reduce to either 
%\mbox{$
\begin{equation}
\det V_n=\det V_{n+1}
%, 
\label{Vn:2}
\end{equation}
%$}. 
%, or simply, 
or
%\mbox{$
\begin{equation}
\exp{(\widetilde{\rho}_n)} \det V_n 
	=\exp{(\rho_n)} \det V_{n+1},
\label{Vn:1}
\end{equation}
%$}, 
where \mbox{$\rho_n$} is a nontrivial 
conserved density. 
%
%Unless 
%%When 
%\mbox{$\log \det L_n (\lambda)$} 
%%does not 
%provides a nontrivial conserved density, 
%the left-hand side becomes zero and thus 
In the first 
%latter 
case (\ref{Vn:2}), 
one may speculate that 
%wonder if 
the above conservation law 
%derived in this way may be 
%is 
%might 
%be 
may become 
%a 
the meaningless 
%and fake 
relation 
%, e.g.,
%a fake one 
%such as 
\mbox{$\boldsymbol{\Delta}_m^+ (\textrm{const.})=
	\boldsymbol{\Delta}_n^+ ({\mathrm{const.}})$}. 
%; 
However, 
in all the examples that 
require the introduction of 
%required us to introduce 
%the 
auxiliary variables, 
this 
%is not 
appears 
%to be not 
not to be 
the case; 
%that is, 
%at 
%first, 
% and 
it is not 
immediately 
evident that 
the determinant of $V_n$ is 
%not apparently 
%a constant
$n$-independent, 
%of $n$, 
and thus 
%this relation is not fake and 
the relation (\ref{fd-cons1}) 
%always 
still contains meaningful 
%useful 
information. 
We 
employ a 
%the 
simplified
%, 
%and 
but still 
ultralocal (with respect to 
\mbox{$\{ \boldsymbol{v}_n \}$}) form 
of $V_n$, 
%and 
%remove an overall constant factor of 
%expand 
compute 
%the determinant of $V_n$ 
its determinant, 
and expand 
%
%in a 
%%Laurent 
%power series 
%of \mbox{$\lambda-\lambda_0$} 
%%($\lambda_0$:\ const.) 
%as 
it with respect to $\lambda$ 
in the summed form 
\begin{equation}
\det V_n (\lambda) 
%\propto 
= 
\sum_{j= j_{\mathrm{min}}}^{j_{\mathrm{max}}} 
%(\lambda-\lambda_0)^j 
g_j (\lambda) 
%\gamma
a_n^{(j)}. 
\label{expand}
\end{equation}
Here, \mbox{$g_j (\lambda) 
\;(\mbox{$j_{\mathrm{min}} \le j \le j_{\mathrm{max}}$})$} 
are 
%is a constant (typically zero), and 
linearly independent functions of $\lambda$, {\it e.g.}, 
\mbox{$g_j (\lambda) = \lambda^{2j}$}, 
and 
%$\lambda_0$ is a constant (typically zero), and 
their 
%``
coefficients
%'' 
\mbox{$a_n^{(j)}$} 
are $\lambda$-independent functions of 
\mbox{$\{ \boldsymbol{v}_n \}$}, 
%as well as 
\mbox{$\{ \boldsymbol{l}_n \}$}, and the 
space/time 
shifts of 
\mbox{$\{ \boldsymbol{l}_n \}$}. 
%
%and substitute it into 
%either (\ref{Vn:1}) or 
%(\ref{Vn:2}). 
%Then we obtain the $n$-independence of $a_n^{(j)}$ and 
%Thus, 
Substituting (\ref{expand}) into (\ref{Vn:2}), 
we obtain the $n$-independence of 
%$a_n^{(j)}$ for 
%{\it every} \/coefficient of $\lambda^j$, 
%all 
the coefficients of 
%{\it every} \/
%{\it all} 
%\/the 
\mbox{$g_j (\lambda)$} for all $j$, 
%powers
%%} \/
%of \mbox{$\lambda-\lambda_0$}, 
%\mbox{$\lambda-\lambda_0$}, 
i.e.,
\begin{equation}
a_n^{(j)} = \lim_{n \to \pm \infty} a_n^{(j)}, \hspace{5mm}
j_{\mathrm{min}} \le j \le j_{\mathrm{max}}.
\label{alge1}
\end{equation}
In the 
%former 
second 
%and major 
case (\ref{Vn:1}), 
%that 
which is more common 
%``usual'' 
%frequent
%popular 
%often 
than 
%the first case 
(\ref{Vn:2}), 
%substituting (\ref{expand}) into 
the substitution of (\ref{expand}) gives 
%provides 
%into 
%or (\ref{Vn:2}). 
%we have 
the set of relations 
%\begin{equation}
\[
\exp{(\widetilde{\rho}_n)} \hspace{1pt}a_{n}^{(j)}
	=\exp{(\rho_n)} \hspace{1pt}a_{n+1}^{(j)},\hspace{5mm}
j_{\mathrm{min}} \le j \le j_{\mathrm{max}}.
%\label{}
%\end{equation}
\]
%The crucial point is that 
Thus, there 
%generally 
exist seemingly 
%, in general, 
more than one 
%fluxes 
flux \mbox{$\log a_{n}^{(j)}$} 
associated with the same conserved density $\rho_n$, 
%\mbox{$\log \det L_n$}. 
but the \mbox{$a_{n}^{(j)}$} 
%they 
should coincide up to trivial 
proportionality 
%coefficients/
factors. 
%a trivial factor. 
%Thus
Indeed, calculating 
%taking 
the ratio of 
%both sides of 
the above 
%equation 
equality for 
different values of $j$ 
on both sides, we obtain 
\begin{equation}
\frac{a_n^{(j_1)}}{a_n^{(j_2)}}
= \lim_{n \to \pm \infty} \frac{a_n^{(j_1)}}{a_n^{(j_2)}}, \hspace{5mm}
j_{\mathrm{min}} \le j_1 \neq  j_2 \le j_{\mathrm{max}}.
\label{alge2}
\end{equation}
That is, the ratio 
\mbox{$a_n^{(j_\mathrm{min})}: \cdots : a_n^{(j_\mathrm{max})}$} 
is independent of $n$. 
%In either case, 
In both the above cases, 
the right-hand side of (\ref{alge1}) or (\ref{alge2}) 
is 
%can be 
determined 
by the boundary conditions 
%of 
for $V_n$, in particular, 
the boundary 
%conditions 
values of the dependent variables 
%appearing 
contained 
in $V_n$;
%and 
%it 
%in this paper, 
%the 
each right-hand side is 
%is assumed to be 
%can be 
set as a definite value 
%a constant
%%, namely, 
%independent of time (cf.\ (\ref{V+-})). 
independent of $n$ (cf.\ (\ref{V+-})).
This 
%generally 
results in an ``ultralocal'' algebraic system 
for 
%the remaining 
%
a subset of 
%%dependent variables
\mbox{$\{ \boldsymbol{v}_n \}$}
%, namely, 
that essentially constitutes 
%form 
the auxiliary variables; 
%and 
the number of independent 
unknowns is usually 
%empirically 
equal to that 
%the number 
of the 
independent equalities 
%relations. 
%equations. 
so that 
%the 
this algebraic system is neither overdetermined nor underdetermined. 

\subsection{Algebraic system for the 
%Determination of 
auxiliary variables
%}
%\subsection{
and 
%local time discretization
%local full discretization
local equations of motion
%time discretization
}
\label{sec2.4}

%Finally, 
%Fourth, 
%and finally, 
We 
%have only to 
solve the 
%thus 
obtained algebraic system 
for the auxiliary 
%dependent 
variables 
appearing in $V_n$ 
%and not in $M_n$. 
%that first appeared in $V_n$ 
%but 
(and not 
%present 
in $M_n$). 
By eliminating all but one of the 
auxiliary 
%dependent 
variables, 
%we arrive at a 
%The algebraic system for \mbox{$\{ \boldsymbol{v}_n \}$} 
this 
system 
%can be reduced 
%amounts to 
becomes 
a scalar algebraic equation 
%for 
in the remaining 
%single 
%one 
auxiliary 
variable; the other auxiliary variables are expressible in terms of 
the solution of this 
%scalar 
equation. 
%this auxiliary variable. 
The degree and 
%the 
complexity of this algebraic equation 
%system 
%depends 
depend 
%not only 
%on the number of the auxiliary variables 
%and the functional form of \mbox{$\det V_n (\lambda)$} 
%%, but also 
%that 
%%depend 
%%%critically 
%susceptibly
%%on 
%are determined by 
on the boundary 
%values 
conditions for $V_n$, 
%that 
which determine 
%in particular, 
%In particular, the degree of the scalar algebraic 
%equation in 
%%a specific lattice 
%each model depends essentially 
%on 
the dispersion relation of the time-discretized lattice system. 
%time discretization. 
%
%%dependent variables \mbox{$\{ \boldsymbol{v}_n \}$}. 
%The degree of the algebraic equation depends on 
%the boundary conditions, 
%or equivalently, 
%%namely, 
%the 
%%linear 
%dispersion relation of the time-discretized lattice. 
In this paper, we mainly 
%focus on at most the quadratic 
consider 
%the degree 
%one or 
%two 
the case for a degree of two 
%of the degree 
so that the equation is solvable 
%can be solved 
%at most 
by the quadratic formula. 
For sufficiently small $h$, 
%owing to 
referring to the prescribed 
%asymptotic 
behavior (\ref{asymp-h}) 
of 
%the matrix 
$V_n$, 
we can discard one of the two solutions as improper, 
%, 
%turns out to be improper, and 
%can be discarded 
%for sufficiently small $h$, 
and 
%we 
obtain 
%the 
a
%unique 
unique proper 
solution of the quadratic equation. 
%uniquely. 
%properly. 
%under the assumption that $h$ is sufficiently small. 
In general, the 
%more 
larger the number of 
%is the number of 
grid points 
%lattice sites 
%points 
defining 
%are needed to define 
the lattice system as well as 
the dispersion relation, the higher 
%is the 
the degree of 
the algebraic equation determining the auxiliary variables.
%variable(s). 
%becomes. 
%The time evolution in 
The higher-degree case can be interpreted as 
%composite mapping
a composition of 
%time evolutions in 
%the 
%some 
lower-degree cases; 
that is, the time evolution in the higher-degree case 
can be 
%factored 
factorized into 
%an 
%``
sequential 
%consecutive
%alternate 
%(?) ``product'' 
applications
%'' 
of 
%different 
more 
%fundamental 
elementary 
time evolutions. 
%in the lower degree cases.
This point will be 
illustrated in 
%the next 
section~\ref{Exs}. 
Once all the auxiliary variables 
have been 
%are 
expressed 
%{\it locally} \/
locally in terms of 
the 
%{\it original} \/
original dependent variables 
%appearing 
that 
have 
already appeared 
%in $L_n$ and $M_n$ 
in the semi-discrete case, we 
only have to substitute them into 
%some 
%a proper 
an appropriate 
%minimal 
subset of equations arising from the zero-curvature 
condition (\ref{fd-Lax}). 
%; 
%so that 
%Note that 
Because of the 
%derivation and 
use of the ``conservation laws'' 
(cf.\ (\ref{Vn:2}) or (\ref{Vn:1})), 
not all of the 
equations 
%relations 
arising from 
%the zero-curvature condition 
(\ref{fd-Lax}) are 
independent 
and necessary 
any longer. 
%The results define
%%a
%consistent and unique time evolution for a
%%given
%%starting from
%% any
%%a
%generic initial data
%satisfying the specified 
%%appropriate 
%boundary conditions. 
%(cf.\ (\ref{V+-})). 
%Now that ..., 
%more. 
We 
%have to 
choose a
%n independent 
%necessary 
minimal 
%minimum 
subset of 
%them 
these equations so that 
%the substitution of the auxiliary variables by their local expressions 
the substitution of the local expressions 
%for 
in the auxiliary variables 
%can 
%uniquely 
%determines 
%provides 
%defines 
%describes 
produces
the 
%consistent and 
{\it unique} \/discrete-time evolution of the lattice system. 
%under the specified boundary conditions. 

Last but not least, 
%Finally, 
%It should be noted that 
%we comment that 
the above 
set of 
%``conservation laws'' (cf.\ (\ref{Vn:2}) or (\ref{Vn:1})) 
conservation laws 
%(cf.\ (\ref{Vn:2}) or (\ref{Vn:1})) 
used to determine the auxiliary variables 
can, in principle, 
%of course, 
be derived 
%directly using 
from the original 
%initial 
system of partial difference 
%difference-difference 
equations 
%implied by 
resulting from the zero-curvature condition (\ref{fd-Lax}). 
%
%it is 
%%at least 
%in principle possible to verify 
%derive 
%the above set of 
%%the lowest-order 
%``conservation laws'' 
%%obtained above 
%%derived in 
%%this way 
%%are 
%%can of course (in principle) be derived 
%directly using 
%%from 
%the equations of motion, i.e., 
%the set of partial difference 
%%difference-difference 
%equations 
%%implied by 
%resulting from the zero-curvature condition (\ref{fd-Lax}). 
However, 
%as a matter of reality, 
in practice, 
this is an extremely difficult task; 
%// as a practical matter // in practice
%The fact of the matter is 
%the latter type of 
%such 
any 
%``component-wise'' 
computation 
%are 
%performed 
conducted 
at the component level 
%in a blind way
without following a set procedure
%usually leads to a 
%shotgunning 
%in a blind way
%surely 
%will stray into a labyrinth and one can hardly 
is highly 
%very
%quite 
unlikely to 
arrive at 
%all 
%a
the nontrivial 
%``conservation laws'' 
conservation laws
to be derived. 
%found. 
Thus, our derivation 
performed at the matrix 
%determinant 
level using 
%, especially, 
a determinantal formula 
and 
%the existence of 
the 
%arbitrary 
spectral parameter 
%in it 
%can be 
%appears to be 
is possibly 
the 
%one and 
only 
%best 
way 
of obtaining them. 
%the nontrivial conservation laws. 

\subsection{Remarks on nonautonomous extensions}
\label{nonauto}

%Note that 
Actually, 
the 
%``time step'' 
``step size'' 
parameter $h$ introduced in (\ref{asymp-h}) 
need not be a constant
%, 
and can 
%be an arbitrary (nonzero) function of 
depend 
arbitrarily 
on 
the discrete 
time coordinate \mbox{$m \in {\mathbb Z}$}; 
%in an arbitrary manner;
%way; 
in other words, the 
%one-parameter 
discrete-time flow involving an arbitrary parameter 
$h$ 
belongs to the same hierarchy, 
%regardless of the value of $h$, 
and we can 
specify 
%choose 
any value of $h$ at 
%each 
every step of the time evolution. 
Indeed, this does not result in 
%make 
any essential difference in the 
%resulting 
subsequent computations, 
%of the zero-curvature condition, 
%as 
because the zero-curvature condition (\ref{fd-Lax}) involves 
equal-time 
$V_n$
%'s 
%$V_n$-matrices 
only. 
In this way, we can obtain 
%a 
nonautonomous extensions of 
%every 
time-discretized lattice systems 
%systems 
involving one arbitrary function 
of the discrete time (see, 
%{\it e.g.}
for example, \cite{Kaji05} 
and references therein). 
%note that if 
%when 
%Actually
Moreover, if the time discretization considered 
%not an elemenray one and is factored into a 
%a 
%is equivalent to 
can be factorized into a 
%the 
composition of $M$ 
%two 
elementary time evolutions (cf.~\cite{Suris97,Suris97',Suris00}), 
%discretizations, 
then it 
%the discretization 
%can be generalized 
%extended 
%to 
generally 
%may 
allows an 
%non-autonomous 
extension 
%involve 
involving 
%two 
$M$ arbitrary functions of time; however, we do not 
proactively discuss 
%consider 
%mention 
this possibility to avoid unnecessary confusion. 
%in this section. 
%different 
%variable. 
%Note that 
From the modern point of view (cf.~\cite{Orfa1,Orfa2,Date1,Sada
%Date4
}), 
a 
%one-step 
%discrete time evolution 
discrete-time flow 
can be identified with 
the spatial part of an auto-B\"acklund transformation 
of the original continuous-time
%lattice 
flow
%, 
%system, 
or
%, 
%in a more unified point of view, 
an auto-B\"acklund transformation of the whole
hierarchy of continuous-time flows. 
Thus, the consistency of 
this nonautonomous (``nonuniform 
%inhomogeneous 
in time'') 
%equally-spaced intervals
%equal distance,  equal interval
%unequally-spaced discrete point
%nonuniform grids 
%nonuniform meshes 
%
%(grid) with variable/non-uniform spacing 
%
%at regular intervals
%generalization 
extension 
%fact 
%possibility 
can 
%easily 
%also 
be 
%interpreted as 
understood as 
%through 
the commutativity 
%and associativity 
of 
%the 
auto-B\"acklund transformations 
%corresponding to 
%with 
%at 
%two 
for 
different values of the 
%``spectral'' 
B\"acklund parameter~\cite{Lamb,Scott}, 
%; the simplest pairwise commutativity is 
known as 
%the Bianchi permutability. 
Bianchi's permutability theorem. 
%it should be recalled that a time-discrete flow 
%
%the Bianchi identity. 
%
%and 
%If 
%we can express 
%When 
%For more details of B\"acklund transformations, 
%The 
Readers 
interested in B\"acklund transformations 
%is 
are referred to
%can refer to 
the Proceedings~\cite{Miura}. 
%NSF Research Workshop on Contact Transformations 
%For details, refer to ...

%Sometimes, 
Bianchi's permutability often 
%generally 
%theorem 
%provides 
implies 
% leads to 
%can produce 
%be expressed in 
%an ultralocal relation 
an 
%simple 
ultralocal 
%four-point 
relation among 
%that connects 
the 
%corresponding 
four 
%different 
solutions 
of the same system, 
%and is ultralocal 
%form 
%such a way that four different solutions of the 
%can relate 
%in 
%with respect to the original independent variables, 
called a nonlinear superposition formula; 
%the nonlinear superposition formula can be derived 
its derivation is based on the compatibility condition 
of 
%can be performed using 
%most systematically 
%preparing and using two 
%from 
two 
%``independent''
``different'' 
%discrete 
time 
evolutions: 
%equations 
%\mbox{$\widetilde{\Psi}_n = V_n (\lambda, h_1) \Psi_n$}, 
\mbox{$\widetilde{\Psi} = V (\lambda, h_1) \Psi$} 
%, 
and 
\mbox{$\widehat{\Psi} = V (\lambda, h_2) \Psi$}. 
%\mbox{$\widehat{\Psi}_n = V_n (\lambda, h_2) \Psi_n$}. 
%Here, we only consider the simpler case of \mbox{$W_n \equiv V_n$}
%In case 
%When 
If there exist 
two or more distinct 
%more than one 
%different 
one-parameter
%one-parametric 
%more than one 
%independent 
auto-B\"acklund transformations, 
%with a free parameter $h$, 
then 
the 
%evolution 
matrix 
\mbox{$V
%_n 
(\lambda, h_2)$} may 
%can 
be 
%replaced by 
generalized to \mbox{$W
%_n 
(\lambda, h_2)$}. 
%In such a 
%%that 
%case, 
%Anyway, 
The nonlinear superposition formula
%we can 
%allows 
motivates us to 
place the four solutions appropriately
at the four vertices 
of a rectangle~\cite{QNCL,NCWQ}
%, 
%quadrangle 
and to assign 
each 
%of the the two 
%arbitrary
value of the B\"acklund 
%``spectral'' 
parameter to 
%the two 
each pair of parallel 
%opposite 
sides; this is often 
%sometimes 
referred to as the Lamb diagram 
(see Fig.~3 
%of
in~\cite{Lamb}). 
It is convenient 
%for intuitive understanding 
to identify 
each value of the B\"acklund parameter with 
the 
%``
length
%'' 
of each side, although 
%it is 
values are not 
restricted to 
%real 
positive numbers. 
%Then, 
We can 
%now consider 
use this rectangle as an elementary cell 
%building block
defining 
%that will generate 
two directions 
%{\it new directions} \/
of {\it new} \/independent variables~\cite{QNCL,NCWQ,NijCap}. 
%of {\it new} \/independent variables. 
%, and 
%Then
%Indeed, 
In fact, 
%we consider 
repeated 
%so that 
%multiple 
applications of the 
auto-B\"acklund transformation at 
(generally) 
distinct 
%different 
values of the 
%``spectral'' 
B\"acklund parameter 
%can 
%will 
%that 
%tile 
%fill 
%make up 
generate a two-dimensional 
%inhomogeneous 
%non-uniform
unequally
%-
spaced 
(in both directions) 
lattice 
in 
%on 
%span 
the quadrant (cf.\ 
Fig.~12 in~\cite{Scott} and 
the main figure in~\cite{Kono82}), or even in the 
%whole 
plane,  
%by 
with 
%in terms of 
%the 
elementary 
%rectangles. 
cells of various sizes. 
With this understanding, we can 
%regard 
reinterpret 
%the 
every nonlinear superposition formula  
as a fully discretized 
%time-discretized 
lattice system 
involving 
%two arbitrary functions:\ 
%two arbitrary functions each of which depends on each independent variable. 
one arbitrary function of one independent variable 
%the discrete space variable 
and 
another 
%one 
arbitrary function of the other independent variable. 

Thus, if 
we encounter 
%obtain 
%a derived time discretization of an integrable 
a time discretization of an integrable 
%time-discretized 
%``new'' 
lattice system 
%turns out to have 
that has the 
%``
same
%'' 
form as a nonlinear superposition formula for some continuous/discrete 
integrable system, then it allows a 
%further 
%``natural'' 
natural
%we can directly 
%%replace the two values of the ``spectral'' parameter with 
%obtain its 
nonautonomous extension involving 
%{\it two} \/
two arbitrary functions 
originating from two values of the B\"acklund parameter. 
For example, the nonlinear superposition formula for the 
potential 
KdV hierarchy (cf.\ (14) or (16) in~\cite{WahEst}) 
%can be reinterpreted as 
suggests 
a nonautonomous extension 
of a discrete potential KdV equation (cf.\ (5) and (6) 
in~\cite{PGR}; 
%see also
(5.16) in~\cite{STZ}; (68) in~\cite{KajiOh}), 
while the nonlinear superposition formula for the sine-Gordon 
equation (or
%, 
the potential mKdV hierarchy) 
%(6.12) in~\cite{Lamb}
with simple sign handling 
implies
%, after simple sign handling,  
%that 
%to 
%can 
%motivates us to extend 
a nonautonomous extension of 
the fully discretized sine-Gordon equation~\cite{Hiro77,Orfa1} 
%can be 
%allows 
%to its 
%a 
%natural 
%non-autonomous version, 
%non-autonomous extension, 
given by 
\begin{equation}
\tan \left( \frac{{u}_{n+1,m+1} + u_{n,m}
%\frac{{u}_{n+1,m+1} - u_{n,m}
%\frac{\widetilde{u}_{n+1} - u_{n}
}{4}
\right) = 
%\frac{\delta(n)+h (m)}{ \delta(n)- h (m)}
\frac{f(n)+g(m)}{f(n) - g (m)}
\tan \left( \frac{ u_{n+1,m}+{u}_{n,m+1} 
%\frac{ u_{n+1,m}-{u}_{n,m+1} 
}{4} \right), 
%
%\sin \Bigl( \frac{a_n + \tilde{a}_{n+1} - \tilde{a}_n - a_{n+1}}{4}
%\Bigr) = \frac{\frac{h \eta}{2}}{\sqrt{1+ (\frac{h\eta}{2} )^2 }}
%\sin \Bigl( \frac{a_n + \tilde{a}_{n+1} + \tilde{a}_n
% + a_{n+1}}{4} \Bigr).
%\label{fdSG}
\end{equation}
or equivalently, 
%\begin{equation}
\begin{align}
&
\frac{f
%\delta
(n)}{g (m)}
\sin \left( \frac{{u}_{n+1,m+1} - u_{n+1,m}-{u}_{n,m+1} + u_{n,m}}{4} \right) 
\nonumber \\
&
= 
\sin \left( \frac{{u}_{n+1,m+1} + u_{n+1,m}+{u}_{n,m+1} + u_{n,m}}{4}
\right). 
\label{nonauSG}
\end{align}
%\end{equation}
%
If we rescale/redefine the arbitrary functions and 
%the 
%independent 
variables as 
\mbox{$f(n)=
%: 
(4/\varDelta) F(n \varDelta
) $}
%, 
and \mbox{$g(m)=
%: 
h \hspace{1pt}G(m h) $}, 
and \mbox{$u_{n,m}=u(n \varDelta,m h)$}, 
\mbox{$n \varDelta =: x$}, and \mbox{$m h =: t$}, 
respectively, 
then the continuous limit \mbox{$\varDelta , h \to 0$} 
reduces (\ref{nonauSG}) to the variable-coefficient 
sine-Gordon 
equation 
%for \mbox{$u(x,t)= u_{n,m}$}
\[
\frac{F(x)} {G(t)} u_{xt} = \sin u , 
\]
which 
%can be transformed 
is obviously equivalent 
to the constant-coefficient sine-Gordon equation 
\mbox{$u_{XT} = \sin u$}. 
%by a ``conformal'' transformation. 
Thus, 
%each 
this type of nonautonomous extension 
in the discrete case 
%involving two arbitray functions. 
%an arbitrary function 
%%in a discrete system 
%of 
%one 
%a 
%discrete coordinate 
%variable 
is thought to be a 
%surely 
%corresponds to 
%originates from 
%can be considered 
%a 
%the discrete analogue 
%the 
vestige 
of 
%the freedom of 
%an arbitrary 
%change of 
coordinate 
transformations 
%of one continuous coordinate. 
%coordinate transformation 
%of a 
%one 
%each 
%continuous variable. 
in the continuous case 
that do 
not mix the two independent variables. 

It is still 
%not yet clear 
unclear 
%at 
%%present 
%the present moment 
%time 
whether 
%some 
any 
of the time-discretized lattice systems 
obtained in 
the next section 
%allows an interpretation as 
in their 
%its 
present form 
can be 
identified with a 
%some 
%a 
nonlinear superposition formula. 
%either in its original form 
%%directly 
%or in its transformed form via a Miura map. 
%and thus 
%However, 
Alternatively, 
%Instead, 
we 
%can 
propose 
an 
%alternative 
intriguing
%interesting 
%constructive 
%n alternate 
%method 
procedure 
%for producing 
%that applies 
%can apply 
%to 
%most of the 
%a large number 
%a lot of 
%many examples; indeed it can 
%successfully 
%to 
%provide 
%allows a non-autonomous extension involving 
%a 
for constructing 
%time-discretized 
fully discrete 
%lattice 
nonautonomous 
systems involving 
two arbitrary functions; 
%it can be applied to many examples, though not all. 
this procedure is 
closely 
%connected with the concept 
related 
%with 
to the property 
of three-dimensional consistency~\cite{NijWal,ABS03,ABS09}. 
%in the following steps. 
%We assume that the temporal Lax matrix 
It is assumed that \mbox{$L_n (\lambda)$} is ultralocal 
in the 
set of 
dynamical 
%dependent 
variables 
\mbox{$\{ \boldsymbol{l}_n \}$}, while 
%and 
\mbox{$V_n (\lambda,h
)$} can be written in terms of 
%only contains 
%either 
%the pair of 
%the paired sets of variables 
\mbox{$\{ \boldsymbol{l}_{n-1}, \widetilde{\boldsymbol{l}}_{n} \}$} 
or 
%the pair of 
\mbox{$\{ \hspace{1pt}\widetilde{\boldsymbol{l}}_{n-1},\boldsymbol{l}_n \}$}. 
%, in addition to 
%%, as well as 
%some constants and arbitrary function(s) of time. 
We only consider the former 
%first 
case, 
%as 
because the 
%second 
latter case 
can be dealt with 
%treated 
in a similar
%the same 
manner. 
Then, the linear problem 
(\ref{line2}) with the 
%``time step'' 
``step size'' 
$h$ set as 
% two values 
%``values'' 
%of $h$, 
\mbox{$h=h_1$} and {$h=h_2$}
%, 
respectively
%,
%The pair of linear equations 
implies the relations 
%the 
%temporal 
%Lax matrix \mbox{$V_n (\lambda, h)$}
%
\begin{subequations}
\begin{equation}
\widetilde{\Psi}_{n} = N_n^{(1)} \Psi_{n-1}, \hspace{5mm} 
	N_n^{(1)} := V_n (\lambda, h_1) L_{n-1}(\lambda),
%\widetilde{\Psi}_{n-1} = N_n^{(1)} \Psi_{n}, \hspace{5mm} 
%	N_n^{(1)} := (\widetilde{L}_{n-1})^{-1} V_n (\lambda, h_1),
%\label{}
\end{equation}
and 
%while the 
%%at 
%choice {$h=h_2$} gives 
%
\begin{equation}
\widehat{\Psi}_{n} = N_n^{(2)} \Psi_{n-1}, \hspace{5mm} 
	N_n^{(2)} := V_n (\lambda, h_2) L_{n-1}(\lambda).
%	N_n^{(2)} := (\widetilde{L}_{n-1})^{-1} V_n (\lambda, h_2), 
\end{equation}
\label{h2-evol}
\end{subequations}
Here, 
the newly introduced 
%Lax 
matrix 
% Evolution Matrices 
\mbox{$N_n^{(1)}$} 
connects the values of $\Psi$ 
%eigenfunctions 
at 
%the 
two 
lattice points 
%\mbox{$\{ \boldsymbol{l}_{n-1} \}$}
\mbox{$(n-1,m)$} 
and 
\mbox{$(n,m+1)$}, 
and 
%indeed 
%and \mbox{$N_n^{(2)}$} 
%contain
%are expressed is 
depends only on the dynamical variables at 
%on 
these 
%two 
%lattice 
points, namely,  
\mbox{$\{ \boldsymbol{l}_{n-1}, \widetilde{\boldsymbol{l}}_{n} \}$}. 
%The same applies for \mbox{$N_n^{(2)}$}, except that 
%%now 
%the forward time evolution now reaches 
%connects to 
%arrives at 
%leads to 
%the new point \mbox{$(n,m+1')$} instead of \mbox{$(n,m+1)$}; 
Similarly, 
\mbox{$N_n^{(2)}$} connects the 
two points 
\mbox{$(n-1,m)$} 
and 
\mbox{$(n,m+1')$}, 
and depends on 
the dynamical variables at 
these points:\ 
\mbox{$\{ \boldsymbol{l}_{n-1}, \widehat{\boldsymbol{l}}_{n} \}$}. 
%;
%\mbox{$(n,m+1')$} instead of \mbox{$(n,m+1)$}; 
Note that \mbox{$(n,m+1')$} can be identified 
with \mbox{$(n,m+1)$} only if \mbox{$h_1 = 
%\equiv 
h_2$}. 
%Note that 
The compatibility 
condition 
of 
%the overdetermined system 
%the pair of equations 
(\ref{h2-evol}) is given by 
%is given by
%provides 
a ``new'' zero-curvature equation, 
%condition 
%\mbox{$
\[
\widetilde{N}_{n+1}^{(2)} N_{n}^{(1)} 
= \widehat{N}_{n+1}^{(1)} N_{n}^{(2)}. 
\]
%$}. 
Thus, the substitution of the 
explicit forms 
%expressions 
of the matrices 
\mbox{$N_n^{(1)} 
(\{ \boldsymbol{l}_{n-1}, \widetilde{\boldsymbol{l}}_{n} \}, h_1)$} 
and \mbox{$N_n^{(2)} 
(\{ \boldsymbol{l}_{n-1}, \widehat{\boldsymbol{l}}_{n} \}, h_2)$} 
%\mbox{$N_n^{(2)}$} 
into this equation 
should 
%result in 
provide 
%yield 
%a four-point relation among 
a closed system for the dynamical variables 
%defined on 
at the four lattice points 
%\mbox{$ 
\[
\boldsymbol{l}_{n-1}, \; \widetilde{\boldsymbol{l}}_{n}, \;
\widehat{\boldsymbol{l}}_{n}, \; \widehat{\widetilde{\boldsymbol{l}}}_{n+1}
(= 
%\equiv 
\widetilde{\widehat{\boldsymbol{l}}}_{n+1}). 
\]
%$} 
%involving 
We can regard 
the corresponding 
%thus formed 
%this 
parallelogram 
%formed by 
%these four points 
as 
defining two directions of {\it new} \/independent variables, 
%an elementary cell, 
so that
%which 
the system now involves two arbitrary functions 
%of each independent variable 
originating from 
$h_1$ and $h_2$. 

%This generalization can easily be noticed 
%%seen 
%and 
%%thus 
%it is 
%%too 
%%difficult to 
%%we do not know who first 
%%identify 
%trace back to the 
%%original 
%first references; 
%%that pointed out 
%%noticed 
%%this possibility; 
%we only refer to the recent papers 
%Hirota('97) and Kajiwara--Mukaihira ('05) 
%%and 
%as well as 
%references cited therein (see also Miwa variables '82 ?). 
%

%{\it Example) Toda lattice.} 
%\subsection{Illustration 
%with the Toda lattice in the Flaschka--Manakov coordinates}
%%variables}

\section{Examples}
%Applications}
\label{Exs}
In this section, we apply the general 
%recipe 
method for 
%local 
time discretization
% presented 
in section 2 to 
%several 
five important 
%interesting 
examples:\ 
%that is, 
% of integrable lattices. 
the Toda lattice, 
%in the Flaschka--Manakov coordinates
the Ablowitz--Ladik lattice (semi-discrete NLS), 
the Volterra lattice, 
the modified Volterra lattice, 
and the lattice Heisenberg ferromagnet model. 
%Moreover, 
%
%In the first two 
%%three 
%examples, 
%%subsections 3.1--3.4, 
%we also describe how Suris's 
%{\it localizing changes of variables} \/can be characterized 
%using the zero-curvature condition (\ref{fd-Lax}). 
%%Lax pair. 
%In particular, his results can be {\it systematically} reproduced, 
%if we think of 
%%the 
%%some 
%%non-constant entries 
%dynamical variables 
%in $V_n$ 
%($\{{\boldsymbol{v}_n}\}$ in section 2)
%%the previous section) 
%as playing the leading 
%%a major 
%role in (\ref{fd-Lax}) 
%%is played by 
%%instead of 
%and those 
%%the non-constant entries 
%in $L_n$ 
%($\{{\boldsymbol{l}_n}\}$ in 
%%the previous 
%section 2) 
%%in (\ref{fd-Lax})
%%play a subordinate
%%
%as playing the supporting role,
%%central player
%%leading player
%%contrary to what most people think. 
%%contrary to popular belief. 
%contrary to common conception. 
%%contrary to common belief. 
%%contrary to accepted wisdom. 

We are 
only 
%mainly
%plays a crucial role
%particularly 
interested in time discretizations 
that can accurately approximate the 
%corresponding 
continuous-time dynamics
%flows. 
%for sufficiently small 
%value 
in the 
%a 
small-value range of the 
``step size'' parameter
%of discrete time 
%``time step'' 
%interval'' 
$h$. 
%At the level of the Lax pair, 
This implies 
%at the level of the Lax pair 
that the discrete-time Lax 
matrix $V_n$ 
%in the Lax pair 
allows 
the asymptotic expansion with respect to $h$ 
%as 
given in 
%by 
(\ref{asymp-h}); 
%, 
%and that the coefficients in this expansion are sufficiently small; 
%small enough; 
% and 
in particular, \mbox{$(V_n-I)/h$} does not 
%{\it e.g.}, they do not 
involve \mbox{$O(1/h)$} terms. 
This requirement 
%is useful 
%%for 
plays a crucial role 
in obtaining 
%determining 
%the proper 
suitable 
local expressions 
%of 
for the auxiliary variables appearing 
%that appear 
in 
%the discrete-time Lax matrix 
$V_n$. 
%, but not in 
%%the continuous-time Lax matrix 
%$M_n$. 

The parameter $h$ is generally assumed to be nonzero; 
%throughout
%, though it is too bothersome to state it 
alternatively, 
%one can regard 
one can allow 
%and understand 
the case \mbox{$h=0$} 
%can be trivially understood 
as the trivial identity mapping
%, 
%{\it e.g.}, \mbox{$\widetilde{u}_n =u_n$}. 
\mbox{$\widetilde{\boldsymbol{l}}_n = \boldsymbol{l}_n $}. 

\subsection{The Toda lattice in 
%the 
Flaschka--Manakov coordinates}

We consider the Toda lattice written 
%represented 
in 
%terms of 
%the 
Flaschka--Manakov coordinates~\cite{Flaschka1,Flaschka2,Manakov74}: 
\begin{equation}
u_{n,t} = u_n (v_{n} -v_{n-1}), \hspace{5mm} 
v_{n,t} = u_{n+1} -u_{n}.
\label{Toda-F}
\end{equation}
%The change of variables 
%Indeed, 
The parametrization 
%\[
\begin{equation}
%a_n 
u_n = \mathrm{e}^{x_{n}-x_{n-1}}
%\exp ()
, \hspace{5mm} 
%b_n 
v_n = x_{n,t}, 
\label{FlaMa}
\end{equation}
%changes 
%recasts 
enables 
the system (\ref{Toda-F}) 
%to 
%in 
to be rewritten as the 
%usual 
%form 
Newtonian equations of motion 
%of 
for the Toda lattice, 
\[
x_{n,tt} = \mathrm{e}^{x_{n+1}-x_{n}} - \mathrm{e}^{x_{n}-x_{n-1}}. 
\]
The Lax pair for the Toda lattice (\ref{Toda-F}) 
in 
%the 
Flaschka--Manakov coordinates
%, system (\ref{Toda-F}), 
is given by~\cite{Suris03} 
\begin{subequations}
\begin{align}
L_n &= \left[
\begin{array}{cc}
\lambda + v_n & u_{n}\\
 -1 & 0 \\
\end{array}
\right], 
\label{Toda-L}
\\[3mm]
M_n &= \left[
\begin{array}{cc}
0 & -u_{n}\\
1 & \lambda + v_{n-1}\\
\end{array}
\right].
\label{Toda-M}
\end{align}
\label{Toda-LM}
\end{subequations}
Indeed, the substitution of (\ref{Toda-LM}) into
the zero-curvature condition (\ref{Lax_eq})
results in (\ref{Toda-F}).

%In the discrete-time case, 
Let us move to the discrete-time case. 
A comprehensive overview of the existing results 
%papers 
on time discretizations of the Toda lattice 
is given in \S3.22 and \S5.11 of \cite{Suris03}, 
%so 
thus we do not repeat it here. 
%
%Cite Wadati--Toda (1975 \& 1975), Hirota
%
In view of the zero-curvature condition (\ref{fd-Lax}), 
we assume the Lax matrix $V_n$
of the following form:
\begin{align}
V_n &= I + h \left[
\begin{array}{cc}
-\lambda a + \alpha_n & -a u_{n} - \widetilde{u}_n b_n \\
a+b_n & \lambda b_n + \alpha_{n-1} + a {v}_{n-1} 
%+ \alpha_{n} + a \widetilde{v}_{n-1} 
+ b_n v_{n-1} \\
\end{array}
\right]. 
\label{Toda-V}
\end{align}
Here, $\alpha_n$ 
and $b_n$ are auxiliary variables. 
%
%Substituting (\ref{Toda-L}) and (\ref{Toda-V}) into (\ref{fd-Lax}), 
%we obtain 
%
The zero-curvature condition
(\ref{fd-Lax}) for the Lax pair (\ref{Toda-L}) and (\ref{Toda-V}) 
amounts to the following 
%difference-difference 
system 
of partial difference equations: 
\begin{equation}
\left\{
\hspace{2pt}
\begin{split}
& \frac{1}{h}\left(\widetilde{u}_n -u_n \right) = 
 \alpha_{n+1} u_n - \widetilde{u}_n \alpha_{n-1}
 + \widetilde{v}_n \left( a u_n + \widetilde{u}_n b_{n}\right) 
       - \left( a \widetilde{u}_n + \widetilde{u}_n b_n \right) v_{n-1},
% \left( \alpha_{n+1} + a \widetilde{v}_n + \widetilde{v}_{n}b_{n+1}\right) u_n
%        - \widetilde{u}_n \left( \alpha_{n}
%\beta_n
\\[0.5mm]
%[1mm]
& \frac{1}{h}\left(\widetilde{v}_n -v_n \right) = 
 \alpha_{n+1} v_n - \widetilde{v}_n \alpha_n 
 + a \left( u_{n+1}-\widetilde{u}_{n}\right) 
        + \widetilde{u}_{n+1}  b_{n+1} - \widetilde{u}_{n}b_n,
\\
& \alpha_n - \alpha_{n+1} = a \left(\widetilde{v}_{n} -v_{n} \right),
\\
& \widetilde{u}_n b_n = b_{n+1} u_n.
\end{split}
\right.
\label{fd-Toda1}
\end{equation}
The general form (\ref{fd-Toda1}) of the 
%fully 
time-discretized Toda lattice
is integrable for matrix-valued dependent variables, but in the following,
we consider only the case of scalar dependent variables. 
In view of (\ref{V+-}) and (\ref{FlaMa}),
%If
we impose the
%constant
%rapidly decaying
following boundary conditions for $u_n$, $v_n$, 
$\alpha_n$, and $b_n$:
\begin{equation}
\lim_{n \to \pm \infty}
% \left(
u_n =1, \hspace{5mm}
\lim_{n \to \pm \infty}
v_n=0, \hspace{5mm}
\lim_{n \to \pm \infty}
\alpha_n=0, \hspace{5mm}
\lim_{n \to \pm \infty} b_n
=b.
\label{Todabc}
\end{equation}
The boundary value of $\alpha_n$ is set as zero
by redefining
%rescaling
$h$. 
To be precise, the boundary conditions 
(\ref{Todabc}) contain {\it redundant} \/information. 
%it is {\it redundant} 
%and nontrivial 
%to hypothesize 
%assume 
Indeed, it can be shown that 
%As is shown below, 
%it can be verified that 
the auxiliary variables 
%can 
%should
%/can 
have the same limit 
%boundary 
values 
%as 
for \mbox{$n \to -\infty$} and \mbox{$n \to +\infty$}. 
%It will be briefly described how 
%In fact, 
%this 
%%fact 
%%can be verified 
%is a verifiable fact 
%by assuming 
%using 
%by 
%%only 
%imposing 
%Thus
Therefore, it is sufficient to assume 
either 
\mbox{$\lim_{n \to - \infty}(\alpha_n,b_n)=(0,b)$}
%\mbox{$\lim_{n \to - \infty} b_n=b$} 
or \mbox{$\lim_{n \to + \infty}(\alpha_n,b_n)=(0,b)$}. 
%\mbox{$\lim_{n \to + \infty} b_n=b$} 
%as is shown below. 
%
In the continuum limit of time \mbox{$h \to 0$}, 
the auxiliary variables reduce to constants, i.e., 
%typically 
%%unity, 
%$1$, 
%and the locality of the semi-discrete lattice system 
%%equations 
%in the original dependent variables 
%is recovered. 
%we obtain
%\[
\mbox{$\alpha_n \to 0$} and
%\hspace{5mm}
\mbox{$b_n \to b$}.
Thus,
in this
%the
limit,
%of time
%\mbox{$h \to 0$}
the time discretization
(\ref{fd-Toda1}) 
%certainly 
reduces to 
%the Toda lattice
\[
u_{n,t} = (a+b) u_n (v_{n} -v_{n-1}), \hspace{5mm}
v_{n,t} = (a+b) (u_{n+1} -u_{n}), 
\]
which
is, if \mbox{$a+b \neq 0$},
equivalent to
%conincides
the Toda lattice (\ref{Toda-F})
up to a 
%scaling 
%``scaling'' 
redefinition 
of
%$\partial_t$
$t$. Note that
in the case \mbox{$a+b =0$}, (\ref{fd-Toda1}) has the trivial
solution \mbox{$\widetilde{u}_n=u_n$}, \mbox{$\widetilde{v}_n=v_n$}, 
\mbox{$\alpha_n=0$}, \mbox{$b_n=b$}. 

The determinant of the \mbox{$2 \times 2$}
Lax matrix $L_n$ (\ref{Toda-L})
%is 
can be immediately computed as \mbox{$ \det L_n = u_n$}.
Using the recurrence formula for $\alpha_n$ 
%fully discretized system
in (\ref{fd-Toda1}),
we can rewrite the Lax matrix $V_n$ (\ref{Toda-V}) 
%to
in 
%the form
an 
%``ultralocal'' 
ultralocal form with respect to
%in
the auxiliary variables $\alpha_n$ and $b_n$.
Thus, its
%compute the
\mbox{$2 \times 2$} determinant
%of the \mbox{$2 \times 2$} matrix $V_n$ (\ref{LV-Vn})
is computed as
\begin{align}
& \hphantom{= \mbox{}}
\det V_n (\lambda)
\nonumber \\
&= \left[ 1+ h \left( -\lambda a + \alpha_n \right)
  \right] \left[ 1 + h \left(\lambda b_n
        + \alpha_n + a\widetilde{v}_{n-1} + b_n v_{n-1} \right) \right]
\nonumber \\
& \hphantom{=\mbox{}}
+ h^2 \left( a u_n + \widetilde{u}_n b_n \right) \left( a+ b_n \right)
\nonumber \\
&=
-\lambda^2 h^2 a b_n -\lambda h \left[
\left( a - b_n \right) \left( 1+ h \alpha_n \right)
+ h a^2 \hspace{1pt}\widetilde{v}_{n-1} + ha b_n v_{n-1}
\right]
\nonumber \\
& \hphantom{=\mbox{}}
+ \left( 1+ h \alpha_n \right)^2 + \left( 1+ h \alpha_n \right)
  \left( ha \widetilde{v}_{n-1} + hb_n v_{n-1} \right) 
  + h^2 \left( a u_n + \widetilde{u}_{n}b_n \right) \left( a+b_n \right).
\nonumber
\end{align}
Therefore,
the equality
(\ref{fd-cons1})
%together
combined with the boundary conditions (\ref{Todabc}) 
(or
%, 
%their 
the streamlined version as stated 
%described 
%mentioned 
%remarked
above)
implies the set of relations
\begin{equation}
\left\{
\hspace{2pt}
\begin{split}
& b_n = b \Lambda_n,
\\[1mm]
&  \left( a - b_n \right) \left( 1+ h \alpha_n \right)
+ h a^2 \hspace{1pt}\widetilde{v}_{n-1} + ha b_n v_{n-1}
%\\[-1mm] &
%\mbox{}
= \left( a - b \right) \Lambda_n,
\\[1mm]
& \left( 1+ h \alpha_n \right)^2 + \left( 1+ h \alpha_n \right)
  \left( ha \widetilde{v}_{n-1} + hb_n v_{n-1} \right) 
\\[-1mm] & \mbox{}
  + h^2 \left( a u_n + \widetilde{u}_{n}b_n \right) \left( a+b_n \right)
= \left[ 1+ h^2 \left( a+b \right)^2 \right] \Lambda_n.
\end{split}
\right.
\label{T-alg1}
\end{equation}
Here, \mbox{$\Lambda_n$} is the quantity that satisfies
%satisfying
the recurrence formula
%recursive relation
%\mbox{$
\begin{equation}
 \widetilde{u}_n \Lambda_n = u_n \Lambda_{n+1},
\label{T-Lam}
\end{equation}
and has the {\it normalized}
%``normalized''
\/boundary value,
\mbox{$\lim_{n \to \pm \infty} \Lambda_n =1$}. 
More precisely, 
%the normalization at the left end 
%\mbox{$\lim_{n \to - \infty} \Lambda_n =1$} 
%guarantees 
%implies 
%the normalization at the right end 
%\mbox{$\lim_{n \to + \infty} \Lambda_n =1$} 
%and vice versa. 
%we have only to assume 
%we require 
%either 
only 
one of the two conditions 
\mbox{$\lim_{n \to - \infty} \Lambda_n =1$} and 
%or 
\mbox{$\lim_{n \to + \infty} \Lambda_n =1$} 
is required, 
%assumed, 
in accordance 
%compliance
with the boundary conditions for $\alpha_n$ and $b_n$. 
Then, the other condition can be confirmed. 
%deduced. 
Thus, $\Lambda_n$ can be written explicitly (but globally)
%(globally)
as
\[
\Lambda_n = \prod_{j=-\infty}^{n-1}
        \frac{\widetilde{u}_j}{u_j}
= \prod_{j=n}^{+\infty}
        \frac{{u}_j}{\widetilde{u}_j}.
\]
Note that
the first equality in (\ref{T-alg1}) is valid
even if \mbox{$a=0$} (cf.\ (\ref{fd-Toda1})).
The algebraic system (\ref{T-alg1})
essentially comprises two nontrivial relations 
%for 
containing 
the two unknowns, $\alpha_n$ and $\Lambda_n$.

%Because of the nonzero boundary
%%conditions
%value of $u_n$ (\ref{Todabc}), 
It is rather cumbersome to identify
%discuss
%the linear leading part
%of
%,and
%as well as
%consequently,
the 
%corresponding 
dispersion relation 
%for 
of (\ref{fd-Toda1}) from 
%the 
its linear leading part 
%in (\ref{fd-Toda1})
% (cite Manakov?),
%, as we did in
(cf.\ subsection~\ref{secAL}).
Instead,
%following Suris's approach/fashion~\cite{Suris03,Suris97'},
%Alternatively,
we can compute the asymptotic form of the Lax matrix $V_n$ (\ref{Toda-V})
as
%a function of $\lambda$ in the limit
%\mbox{$n \to \pm \infty$},
\mbox{$n \to - \infty$} or \mbox{$n \to + \infty$}
%, 
and consider
%discuss
its factorization 
%problem 
(cf.~\cite{Suris03,Suris97, Suris97',Suris00}).
The results
%implies
imply that 
%the case \mbox{$a=0$} and the case \mbox{$b=0$} 
the cases \mbox{$a=0$} and \mbox{$b=0$} 
form
%a
the basis for the general case; that is,
the
%discrete-
time evolution
for
%with
general
%values of
$a$ and $b$
is equivalent to
%a proper
the order-independent
composition of
%the two time evolutions corresponding to \mbox{$a=0$} and \mbox{$b=0$},
%the
two
%discrete-
time evolutions
%for
corresponding to \mbox{$a=0$} and
%for
\mbox{$b=0$}, respectively.
In addition, these two
%fundamental
cases are related
%with
to
each other through 
%space/
%a 
the time reflection, as
we will see below.
%All these
%%The same
%results
%%conclusion
%can also be inferred
%%drawn
%from
%the results on
%%the decomposition
%%%factorization
%%of the composite mapping into simpler ones
%the discrete-time Ablowitz--Ladik lattice
%given in subsection~\ref{secAL}.
%

Now, 
%Thus, 
we consider 
%only 
the two
%special
fundamental cases:\
\mbox{$a=0$} or \mbox{$b=0$}.
\\
\\
$\bullet${\it The case \mbox{$a=0$}}, {$b \neq 0$}.

In this case, the
%two
auxiliary variable \mbox{$\alpha_n$}
vanishes. 
%(cf.~(\ref{Todabc})). 
Thus, the discrete-time system (\ref{fd-Toda1}) reduces to 
(cf.\ (3.6) in~\cite{Dodd})
\begin{equation}
\left\{
\hspace{2pt}
\begin{split}
& \frac{1}{h}\left(\widetilde{u}_n -u_n \right) =
 b \Lambda_{n+1} {u}_n \left( \widetilde{v}_n - v_{n-1} \right),
% \left( \alpha_{n+1} + a \widetilde{v}_n + \widetilde{v}_{n}b_{n+1}\right) u_n
%        - \widetilde{u}_n \left( \alpha_{n}
%\beta_n
\\[0.5mm]
%[1mm]
& \frac{1}{h}\left(\widetilde{v}_n -v_n \right) =
 b \Lambda_{n+1} \left( \widetilde{u}_{n+1} - {u}_{n}\right),
\\[0.5mm]
& \widetilde{u}_n \Lambda_n = u_n\Lambda_{n+1}, \hspace{5mm} 
%\lim_{n \to \pm \infty} \Lambda_n =1.  
\lim_{n \to - \infty} \Lambda_n =1 \;\, \mathrm{or} \;\,
\lim_{n \to + \infty} \Lambda_n =1.
\end{split}
\right.
\label{fd-Toda2}
\end{equation}
%
%Note that in the ``critical'' case \mbox{$hb =\pm 1$},
%(\ref{fd-Toda2}) has the trivial solution \mbox{$\widetilde{u}_n = u_{n-1}$},
%\mbox{$\widetilde{v}_n = v_{n}$}, \mbox{$\Lambda_n = 1/u_{n-1}$}.
%
The algebraic system (\ref{T-alg1}) 
%boils down 
simplifies to
a quadratic equation 
%for 
in $\Lambda_n$,
\begin{equation}
  (hb)^2 \hspace{1pt}\widetilde{u}_{n}\Lambda_n^2 
 -\left[ 1 + (h b)^2 - hb v_{n-1} \right] \Lambda_n + 1 = 0.
\label{T-alg2}
\end{equation}
The asymptotic behavior (\ref{asymp-h})
of the Lax matrix $V_n$ implies that
the proper solution of (\ref{T-alg2}) is given by
\begin{equation}
 \left[ 1 + (h b)^2 \right] \Lambda_n =
\frac{2}{1- \epsilon \hspace{1pt}v_{n-1} + \sqrt{
 \left(1 -\epsilon \hspace{1pt}v_{n-1} \right)^2
  -4 \epsilon^2 \hspace{1pt}\widetilde{u}_n}},
\label{T-Lam1}
\end{equation}
where \mbox{$ \epsilon := hb/ \left[ 1 + (h b)^2 \right]$}.
Thus, if \mbox{$hb \in \mathbb{R}$}, then \mbox{$-1/2 \le \epsilon \le 1/2$}.
%; in this case,
%and
In this case, the local expression (\ref{T-Lam1}) is valid
%for
only if \mbox{$-1 < hb <1$},
%\mbox{$-1 \le hb \le 1$},
which 
%still
%corresponds to
covers
the 
%entire
%whole
range
%of
%\mbox{$-1/2 \le \epsilon \le 1/2$}. 
\mbox{$-1/2 < \epsilon < 1/2$}. 
%
%To be precise, there exist some subtleties for 
%in the cases 
%\mbox{$hb =\pm 1$}, 
%is delicate, and should be 
%but we do not discuss 
%%consider 
%these exceptional cases. 
%Indeed, 
The borderline
%delicate 
%exceptional 
cases \mbox{$hb =\pm 1$} are excluded from our consideration. 
If \mbox{$(hb)^2 > 1$},
%\mbox{$|hb| > 1$},
(\ref{T-Lam1}) is inconsistent
with the boundary conditions, and the other solution
of (\ref{T-alg2}) should be
%used.
adopted.
%gives the proper expression.
%employed.
In any case, the boundary conditions for $u_n$ and $v_n$ imply 
that 
\mbox{$\lim_{n \to - \infty} \Lambda_n =
 \lim_{n \to + \infty} \Lambda_n =1$}. 

Substituting (\ref{T-Lam1}) into the first and second equations 
in (\ref{fd-Toda2}), we obtain a
%n integrable
%local 
time discretization
of the Toda lattice (\ref{Toda-F}) in the local form,
\begin{equation}
\left\{
\hspace{2pt}
\begin{split}
& \frac{\widetilde{u}_n -u_n }{\epsilon}
%\frac{1}{\epsilon}\left(\widetilde{u}_n -u_n \right)
 =
\frac{2 {u}_n \left( \widetilde{v}_n - v_{n-1} \right)}
 {1- \epsilon \hspace{1pt}v_{n} + \sqrt{
 \left(1 -\epsilon \hspace{1pt}v_{n} \right)^2
  -4 \epsilon^2 \hspace{1pt}\widetilde{u}_{n+1}}},
\\[0.5mm]
%[1mm]
& \frac{\widetilde{v}_n -v_n }{\epsilon} =
%\frac{1}{\epsilon}\left(\widetilde{v}_n -v_n \right) =
\frac{2 \left( \widetilde{u}_{n+1} - {u}_{n}\right)}
{1- \epsilon \hspace{1pt}v_{n} + \sqrt{
 \left(1 -\epsilon \hspace{1pt}v_{n} \right)^2
  -4 \epsilon^2 \hspace{1pt}\widetilde{u}_{n+1}}}.
\end{split}
\right.
\label{loToda1}
\end{equation}
%
%This is an 
%%{\it 
%explicit scheme 
Under the boundary conditions
\mbox{$\lim_{n \to \pm \infty} u_n =1$} and 
\mbox{$\lim_{n \to \pm \infty} v_n =0$},
%\hspace{1pt}; 
%,as the value of 
$\widetilde{v}_n$ is 
%uniquely 
determined from 
%by
%from
%given in terms of
$u_n$, $v_{n}$, and $\widetilde{u}_{n+1}$, and subsequently, 
%the value of 
$\widetilde{u}_n$ is 
%uniquely 
determined from 
%by
%given in terms of
$v_{n-1}$, $u_n$, $v_n$, $\widetilde{v}_n$, and 
$\widetilde{u}_{n+1}$.
%(\ref{LVbc}).
%Note that (\ref{loToda1}) can be rewritten
%
If \mbox{$hb \in \mathbb{R}$}, 
%and
%$u_n$'s 
the $u_n$ are nonzero
and
real-valued, and 
%$v_n$'s 
the $v_n$
are real-valued at the initial time,
%\mbox{$m=m_0$},
then
%\mbox{$\epsilon \le 1/4$}.
(\ref{fd-Toda2}) implies that
%the reality of
the auxiliary variable $\Lambda_n$ is always real-valued.
%As the result,
Consequently, the discriminant of the quadratic equation (\ref{T-alg2})
%is
must be 
nonnegative.
%takes values in $\mathbb{R}$.
%\mbox{$\Lambda_n \in \mathbb{R}$}.
%Therefore
Thus, 
%if \mbox{$-1/2 < \epsilon < 1/2$} 
%and
%\mbox{$u_n \to 1$} sufficiently smoothly 
%and fast as \mbox{$n \to \pm \infty$},
%%\mbox{$0 < \epsilon \le 1/4$} or \mbox{$\epsilon < 0$},
%and \mbox{$u_n$}'s stay in
%%the
%proximity to $1$ 
%and 
as long as 
%\mbox{$u_n$} and $v_n$ can be regarded as a perturbation around 
%from 
%their boundary values, 
the amplitudes of 
%effects of 
\mbox{$u_n-1$} and $v_n$ are sufficiently small 
%small enough 
and 
their effects can be regarded as 
%a
%weak 
perturbations, 
%to 
%%from 
%their boundary values $1$ and $0$, 
%at the initial time,
%then 
the real-valuedness of $u_n$ and $v_n$ 
is
%maintainedand 
preserved
under the time evolution
of the discrete-time Toda lattice (\ref{loToda1}) 
for \mbox{$-1/2 < \epsilon < 1/2$}. 
%The case \mbox{$\epsilon =0$} is trivially understood
%as \mbox{$\widetilde{u}_n =u_n$}.

%In order to
To express
%explicitly obtain
%write out 
the backward time evolution explicitly,
we only have to replace \mbox{$\Lambda_{n+1} $} in
the first and second equations of (\ref{fd-Toda2})
%by 
with \mbox{$\Lambda_n \widetilde{u}_n/u_n$}
%,
and then substitute the local expression (\ref{T-Lam1}).
The resulting system 
%equation 
%reads as
is 
\begin{equation}
\left\{
\hspace{2pt}
\begin{split}
& \frac{\widetilde{u}_n -u_n }{\epsilon}
%\frac{1}{\epsilon}\left(\widetilde{u}_n -u_n \right)
 =
\frac{2 \widetilde{u}_n \left( \widetilde{v}_n - v_{n-1} \right)}
 {1- \epsilon \hspace{1pt}v_{n-1} + \sqrt{
 \left(1 -\epsilon \hspace{1pt}v_{n-1} \right)^2
  -4 \epsilon^2 \hspace{1pt}\widetilde{u}_{n}}},
\\[0.5mm]
%[1mm]
& \frac{\widetilde{v}_n -v_n }{\epsilon} =
%\frac{1}{\epsilon}\left(\widetilde{v}_n -v_n \right) =
\frac{\widetilde{u}_n}{u_n} \times
\frac{2 \left( \widetilde{u}_{n+1} -{u}_{n}\right) 
%\widetilde{u}_n/u_n
}
{1- \epsilon \hspace{1pt}v_{n-1} + \sqrt{
 \left(1 -\epsilon \hspace{1pt}v_{n-1} \right)^2
  -4 \epsilon^2 \hspace{1pt}\widetilde{u}_{n}}}.
\end{split}
\right.
\label{loToda2}
\end{equation}
%Note that 
Using the first equation, the second equation in (\ref{loToda2}) 
can also be 
%re
written as 
\[
\frac{\widetilde{v}_n -v_n }{\epsilon} =
%\frac{1}{\epsilon}\left(\widetilde{v}_n -v_n \right) =
\frac{2 \left( \widetilde{u}_{n+1} -{u}_{n}\right) 
%\widetilde{u}_n/u_n
}
{1- \epsilon \left( 2\widetilde{v}_n -v_{n-1} \right) + \sqrt{
 \left(1 -\epsilon \hspace{1pt}v_{n-1} \right)^2
  -4 \epsilon^2 \hspace{1pt}\widetilde{u}_{n}}}.
\]
\\
$\bullet${\it The case \mbox{$b=0$}}, \mbox{$a \neq 0$}. 

In this case, the auxiliary variable $b_n$ vanishes.
%\mbox{$b_n =0$}.
Thus, the discrete-time system (\ref{fd-Toda1}) reduces to
\begin{equation}
\left\{
\hspace{2pt}
\begin{split}
& \frac{1}{h} \left( 1 + h\alpha_n + ha \widetilde{v}_{n-1} \right)
\left(\widetilde{u}_n -u_n \right) 
= a u_n \left( v_n - \widetilde{v}_{n-1} \right),
\\[0.5mm]
%[1mm]
& \frac{1}{h} \left( 1 
%\left(\widetilde{v}_n -v_n \right) 
 + h\alpha_n 
%\left(\widetilde{v}_n -v_n \right) 
 + ha v_n \right) \left(\widetilde{v}_n -v_n \right) = 
 a \left( u_{n+1}-\widetilde{u}_{n}\right),
\\
& \alpha_n - \alpha_{n+1} = a \left(\widetilde{v}_{n} -v_{n} \right), 
\hspace{5mm} \lim_{n \to - \infty} \alpha_n = 0 \;\, 
\mathrm{or} \;\, \lim_{n \to + \infty} \alpha_n =0. 
\end{split}
\right.
\label{fd-Toda3}
\end{equation}
%Note that in the ``critical'' case \mbox{$ha =-1$},
%(\ref{fd-Toda3}) has the trivial solution
%%\mbox{$\widetilde{u}_{n-1} = u_{n+1}$},
%%\mbox{$1+h \alpha_n = \widetilde{u}_{n-1}$},
%\mbox{$1+h \alpha_n = {u}_{n+1}$}, \mbox{$\widetilde{u}_{n} = u_{n+2}$}.
%Thus, we can assume \mbox{$ha \neq -1$}.
The algebraic system (\ref{T-alg1}) 
%boils down to
simplifies to
\begin{equation}
\left\{
\hspace{2pt}
\begin{split}
&  1+ h \alpha_n + h a \hspace{1pt}\widetilde{v}_{n-1} = \Lambda_n,
\\
%[1mm]
& \left( 1+ h \alpha_n \right)^2 + \left( 1+ h \alpha_n \right)
  ha \widetilde{v}_{n-1} + (h a)^2 u_n 
 = \left[ 1+ (h a)^2 \right] \Lambda_n.
%\\[1mm] & 
% \widetilde{u}_n \Lambda_n = u_n \Lambda_{n+1},
\end{split}
\right.
\label{T-alg3}
\end{equation}
%It should be commented 
%noted 
Note that, using $\Lambda_n$ instead of $\alpha_n$, 
system (\ref{fd-Toda3}) 
%using 
with (\ref{T-Lam}) and (\ref{T-alg3}) 
can be identified 
%rewritten 
with the discrete-time Toda lattice given in \S3.8 of \cite{Suris03} 
(also see \cite{Suris95}). 

By eliminating $\Lambda_n$, (\ref{T-alg3}) reduces to
a quadratic equation 
%for 
in \mbox{$1+ h \alpha_n $},
\[
\left( 1+ h \alpha_n \right)^2 
 - \left[ 1+ (h a)^2 - ha \widetilde{v}_{n-1}\right] 
 \left( 1+ h \alpha_n \right)
  + (h a)^2 u_n -\left[ 1+ (h a)^2 \right] ha \widetilde{v}_{n-1} =0.
\]
The proper solution of this quadratic equation is given by
\begin{equation}
\frac{1+ h \alpha_n}{1+ (h a)^2} = 
\frac{ 1 - \delta \hspace{1pt}\widetilde{v}_{n-1} 
 + \sqrt{\left( 1+ \delta \hspace{1pt}\widetilde{v}_{n-1} \right)^2 
 - 4\delta^2 u_n }
%1+ (h a)^2 - ha \widetilde{v}_{n-1} + \sqrt{\left[1+ (h a)^2 
%+ ha \widetilde{v}_{n-1} \right]^2 -4 (h a)^2 u_n }
}{2},
\label{T-alp}
\end{equation}
where \mbox{$\delta := ha/ \left[ 1 + (h a)^2 \right]$}. 
Thus, if \mbox{$ha \in \mathbb{R}$}, then \mbox{$-1/2 \le \delta \le 1/2$}.
In this case, the local expression (\ref{T-alp}) is valid
%for
only if \mbox{$-1 < ha <1$}, which
%still
%corresponds to
covers
the
%entire
%whole
range
%of
%\mbox{$-1/2 \le \epsilon \le 1/2$}.
\mbox{$-1/2 < \delta < 1/2$}.
The borderline
%delicate
%exceptional
cases \mbox{$ha =\pm 1$} are excluded from our consideration.
If \mbox{$(ha)^2 > 1$},
%\mbox{$|hb| > 1$},
(\ref{T-alp}) is inconsistent
with the boundary conditions, 
%(cf.~(\ref{Todabc})), 
and the other solution of the quadratic equation
should be employed. 
%(\ref{T-alg2}) 
%should be used.
%adopted.
%otherwise,
%Unless \mbox{$ha = -1/2$},
%a unified expression for \mbox{$1+ h \alpha_n$},
%can resolve the sign problem of the square root,
%but it looks
%%too
%unwieldy and unattractive. 
In any case, the boundary conditions for $u_n$ and $v_n$ imply
that 
\mbox{$\lim_{n \to - \infty} \alpha_n = \lim_{n \to + \infty} \alpha_n =0$},
%\mbox{$\lim_{n \to \pm \infty} \alpha_n =0$}, 
and consequently, 
\mbox{$\lim_{n \to - \infty} \Lambda_n =\lim_{n \to + \infty} \Lambda_n =1$}.
%\mbox{$\lim_{n \to \pm \infty} \Lambda_n =1$}.

Substituting the local expression (\ref{T-alp}) 
%for \mbox{$1+ h \alpha_n$}
%into \mbox{$1+ h \alpha_n$}
into the first and second equations 
%relation
%in
of (\ref{fd-Toda3}),
%results in ...
we obtain a 
%local 
time discretization of the Toda lattice (\ref{Toda-F}), 
%When $ha$ is a constant independent of the discrete time $m$,
%i.e., a constant,
%In terms of the new
%%single
%parameter
%\mbox{$\delta := 
%%=:\delta
% ha/ \left[ 1 + (h a)^2 \right]$}, 
%we can write this time discretization as
\begin{equation}
\left\{
\hspace{2pt}
\begin{split}
& \frac{\widetilde{u}_n -u_n }{\delta} = 
\frac{2 u_n \left( v_n - \widetilde{v}_{n-1} \right)}
 { 1 + \delta \hspace{1pt}\widetilde{v}_{n-1} 
 + \sqrt{\left( 1+ \delta \hspace{1pt}\widetilde{v}_{n-1} \right)^2 
 - 4\delta^2 u_n }},
\\[0.5mm]
%[1mm]
& \frac{\widetilde{v}_n -v_n }{\delta} = 
 \frac{2 \left( u_{n+1}-\widetilde{u}_{n}\right)}
 { 1 + \delta \left( 2 v_n -\widetilde{v}_{n-1} \right)
 + \sqrt{\left( 1+ \delta \hspace{1pt}\widetilde{v}_{n-1} \right)^2 
 - 4\delta^2 u_n }}. 
\end{split}
\right.
\label{loToda3}
\end{equation}
Under the boundary conditions \mbox{$\lim_{n \to \pm \infty} u_n =1$} 
and \mbox{$\lim_{n \to \pm \infty} v_n =0$}, 
%this is an 
%%{\it 
%explicit scheme
%%with respect to the forward time evolution
%%toward 
%in the forward time direction; 
%Indeed, 
%the value of
$\widetilde{u}_n$ is
%uniquely
determined from
%by
%given in terms of
$\widetilde{v}_{n-1}$, $u_n$, and $v_n$, 
%$\widetilde{v}_n$, and ${u}_{n+1}$, 
and subsequently,
$\widetilde{v}_n$ is
%uniquely
determined from
%by
%from
%given in terms of
$\widetilde{v}_{n-1}$, $\widetilde{u}_{n}$, 
$u_n$, $v_{n}$, and ${u}_{n+1}$. 
%If \mbox{$ha \in \mathbb{R}$}, and 
%%and
%$u_n$'s 
%%are nonzero and real-valued 
%and
%$v_n$'s are real-valued at the initial time,
%%\mbox{$m=m_0$},
%then
%%\mbox{$\epsilon \le 1/4$}.
%(\ref{fd-Toda3}) implies that
%%the reality of
%the auxiliary variable $\alpha_n$ is always real-valued.
%%As the result,
%Consequently, the discriminant of the quadratic equation (\ref{T-alg3})
%%is
%must be non-negative. 
%Thus,
As long as 
\mbox{$u_n$} and $v_n$ can be regarded as 
%a 
perturbations around 
%from 
%their 
the boundary values $1$ and $0$, 
%the amplitudes of
%%effects of
%\mbox{$u_n-1$} and $v_n$ are small enough and
%their effects can be regarded as a
%%weak
%perturbation,
%the 
their real-valuedness 
%of $u_n$ and $v_n$
is
%maintainedand
preserved
under the time evolution
of the discrete-time Toda lattice (\ref{loToda3}) 
for \mbox{$-1/2 < \delta < 1/2$}. 
%
%If this inequality is satisfied,
%%and
%\mbox{$u_n \to 1$} sufficiently smoothly
%%\mbox{$u_n$} stays a proper proximity of $1$
%and fast as \mbox{$n \to \pm \infty$},
%and \mbox{$u_n$}'s stay in
%%the
%proximity to $1$ at the initial time,
%then the real-valuedness of $u_n$ is
%preserved
%%maintained
%under the time evolution (\ref{loToda3}).

%In order 
To
%write
obtain
the backward time evolution explicitly,
%in the case \mbox{$b=0$}
%can be given explicitly by
%can be written by
we rewrite \mbox{$\alpha_{n}$} in the first and second equations 
of (\ref{fd-Toda3})
%in terms of $\alpha_{n+1}$ using the second equation
%to 
as \mbox{$\alpha_{n+1} + a \left( \widetilde{v}_{n} - v_{n}\right)$}
%,
and then substitute the local expression (\ref{T-alp}).
The resulting system 
%equation 
%reads as
is 
\begin{equation}
\left\{
\hspace{2pt}
\begin{split}
& \frac{\widetilde{u}_n -u_n }{\delta} = 
\frac{2 \widetilde{u}_n \left( v_n - \widetilde{v}_{n-1} \right)}
 { 1 + \delta \hspace{1pt}\widetilde{v}_{n} 
 + \sqrt{\left( 1+ \delta \hspace{1pt}\widetilde{v}_{n} \right)^2 
 - 4\delta^2 u_{n+1} }},
\\[0.5mm]
%[1mm]
& \frac{\widetilde{v}_n -v_n }{\delta} = 
 \frac{2 \left( u_{n+1}-\widetilde{u}_{n}\right)}
 { 1 + \delta \hspace{1pt}\widetilde{v}_{n} 
   + \sqrt{\left( 1+ \delta \hspace{1pt}\widetilde{v}_{n} \right)^2 
 - 4\delta^2 u_{n+1} }}. 
\end{split}
\right.
\label{loToda4}
\end{equation}
It is easy to see that
%This equation
(\ref{loToda3}) is equivalent to 
%(\ref{loToda1}) (or 
(\ref{loToda2}) 
through 
%space (or 
%a 
the time reflection and the identification
%\mbox{$\epsilon \leftrightarrow -\delta$}.
\mbox{$\delta \leftrightarrow -\epsilon$}.
In the same manner, (\ref{loToda4}) can be identified with (\ref{loToda1})
through 
%a 
the time reflection.
%is equivalent to
Thus, the forward/backward time evolution in the case \mbox{$b=0$}
corresponds to the backward/forward
time evolution in the case \mbox{$a=0$},
%\mbox{$=0$},
up to a redefinition of the
%time step
parameters.

\subsection{The Ablowitz--Ladik lattice}
\label{secAL}

%We 
%start with 
%%first discuss 
In this subsection, 
we 
%consider 
discuss the time discretization of 
the Ablowitz--Ladik lattice, 
%since it
which we consider 
%appears 
%seems 
to be 
%is considered 
the 
%optimal 
%best
most instructive 
%and optimum 
example to illustrate 
%explain 
our general method. 
%recipe. 
%educational educative instructive pedagogic pedagogical
%In this paper, we refer to the following system as well as its as 
%The equations of motion 
%%and the Lax pair 
%for 
The (nonreduced form of the) 
%the 
Ablowitz--Ladik lattice is 
%reads as 
\begin{equation}
\left\{
\hspace{1pt}
\begin{split}
& q_{n,t}- a q_{n+1} + b q_{n-1} + (a-b)q_n 
	+ a q_{n+1} r_n q_n - b q_n r_n q_{n-1} =0, \\
& r_{n,t}- b r_{n+1} + a r_{n-1} + (b-a)r_n 
	+ b r_{n+1} q_n r_n - a r_n q_n r_{n-1} =0, 
\end{split}
\right.
\label{sd-AL}
\end{equation}
%while 
and its Lax pair (cf.~(\ref{line}) and (\ref{Lax_eq})) 
is given by 
\begin{subequations}
\begin{align}
L_n &
= 
\lambda \left[
\begin{array}{cc}
1 & \\
 & 0 \\
\end{array}
\right] + 
\left[
\begin{array}{cc}
0 & q_n\\
r_n  & 0 \\
\end{array}
\right]
+ \frac{1}{\lambda}
\left[
\begin{array}{cc}
0 & \\
 & 1 \\
\end{array}
\right]
=\left[
\begin{array}{cc}
 \lambda & q_n \\
 r_n & \frac{1}{\lambda} \\
\end{array}
\right], 
\label{AL-Ln}
%\nonumber 
\\[3mm]
M_n &= 
%\lambda^2 \left[
%\begin{array}{cc}
%a & \\
% & 0 \\
%\end{array}
%\right] + \lambda \left[
%\begin{array}{cc}
% & a q_n\\
% a r_{n-1} & \\
%\end{array}
%\right]
%+ \left[
%\begin{array}{cc}
%-a (1+q_n r_{n-1}) & \\
% & -b (1+ r_n q_{n-1})\\
%\end{array}
%\right]
%+ \frac{1}{\lambda} \left[
%\begin{array}{cc}
% & b q_{n-1}\\
% b r_n & \\
%\end{array}
%\right] +
%\frac{1}{\lambda^2} \left[
%\begin{array}{cc}
%0 & \\
% & b \\
%\end{array}
%\right] 
%\\
%&=
\left[
\begin{array}{cc}
 \lambda^2 a -a (1+q_n r_{n-1}) & 
	\lambda a q_n + \frac{b}{\lambda} q_{n-1}\\
 \lambda a r_{n-1} + \frac{b}{\lambda} r_{n} & 
	\frac{b}{\lambda^2} -b (1+ r_n q_{n-1})\\
\end{array}
\right].
\label{AL-Mn}
%\nonumber
\end{align}
\end{subequations}
%where 
Here, the free parameters $a$ and $b$ are 
usually 
set as 
%considered 
%regarded as 
%(arbitrary) 
constants, but 
%Note that $a$ and $b$ in the semi-discrete case 
they can 
%actually 
depend 
on the time variable $t$ in an arbitrary manner, namely, 
%:\ 
\mbox{$a := a(t)$}, \mbox{$b := b(t)$}. 
%This is evident if one notices that the flow is actually 
%a linear combination of two (three?) commutative 
%%independent 
%flows. 
The 
%more 
familiar form of the Ablowitz--Ladik lattice 
is obtained from 
%the general form 
(\ref{sd-AL}) through the reduction of the complex conjugate: 
\mbox{$b=a^\ast$}, \mbox{$r_n=\sigma q_n^\ast$}, 
where $\sigma$ is a real constant. 
%In addition, 
The case where $a$ and $b$ are purely imaginary 
%{\it e.g.}, \mbox{$a=-b=\mathrm{i}$}, 
leads to 
%provides 
%corresponds to 
an integrable semi-discretization of the NLS equation. 
Taking into account 
the $\lambda$-dependence of the Lax 
matrix $M_n$ (\ref{AL-Mn}), 
%in the continuous-time case 
%motives us to 
we look for a
%the 
Lax matrix $V_n$ 
%of the following form: 
with the following 
%$\lambda$-
dependence on 
%the spectral parameter 
$\lambda$:
%assume that 
%
\begin{equation}
V_n 
(\lambda) 
= I + h 
%\sum_{j=-2}^2 \lambda^{j} V_n^{(j)}
\left\{ \lambda^{2} V_n^{(2)}+\lambda V_n^{(1)} 
+ V_n^{(0)} + \frac{1}{\lambda} V_n^{(-1)} + \frac{1}{\lambda^2} V_n^{(-2)} 
\right\}. 
\label{AL-Vn1}
\end{equation}
Here, 
%$\lambda$-independent 
\mbox{$V_n^{(j)}\,(j=2,\, 0,\, -2)$} 
are 
%is a 
diagonal matrices, 
%if $j$ is even
%for even $j$ 
%and 
while \mbox{$V_n^{(j)}\,(j=1,\, -1)$} 
%is 
are off-diagonal matrices. 
The nonzero parameter $h$ may depend on 
%are allowed to depend on 
the discrete time coordinate \mbox{$m \in {\mathbb Z}$}. 
%, respectively. 
%%if $j$ is odd. 
%for odd $j$. 
Substituting 
(\ref{AL-Ln}) and (\ref{AL-Vn1}) into (\ref{fd-Lax}), 
we find that the matrix $V_n$ should assume the form 
\begin{equation}
V_n = I + h 
\left[
\begin{array}{c|c}
\begin{array}{l}
\lambda^2 a + \widetilde{q}_n c_n r_{n-1}
\\ \mbox{}
+ \alpha_n +\frac{1}{\lambda^2} d_n 
\end{array}
 & 
\begin{array}{l} 
\lambda (a q_n - \widetilde{q}_n c_n) 
\\ \mbox{}
	+\frac{1}{\lambda_{\vphantom \sum}}(b\widetilde{q}_{n-1} -d_n q_{n-1}) 
\end{array}
\\
\hline
\begin{array}{l}
 \lambda (a\widetilde{r}_{n-1} -c_n r_{n-1})  
\\ \mbox{}
	+ \frac{1}{\lambda}(b r_n - \widetilde{r}_n d_n) 
\end{array}
& 
\begin{array}{l}
\lambda^{2^{\vphantom \sum}} c_n + \widetilde{r}_n d_n q_{n-1}
\\ \mbox{}
+ \beta_n +\frac{b}{\lambda^2} 
\end{array}
\\
\end{array}
\right].
\label{AL-Vn2}
\end{equation}
Here,
%where 
$\alpha_n$, $\beta_n$, $c_n$, and $d_n$ are 
auxiliary variables. 
%Then, 
The zero-curvature condition 
(\ref{fd-Lax}) 
%amounts to 
%produces
is equivalent to 
the following 
%difference-difference 
system of partial difference equations:
%of difference-difference equations
%
%\begin{subequations}
\begin{equation}
\left\{
\hspace{2pt}
\begin{split}
&
\begin{split}
\frac{1}{h}\left(\widetilde{q}_n -q_n \right) & - a q_{n+1} 
 + b \widetilde{q}_{n-1} - \alpha_{n+1} q_n + \widetilde{q}_{n} \beta_n 
%\nonumber 
\\[-1mm]
& \mbox{}
 + \widetilde{q}_{n+1} c_{n+1} (1-r_n q_n) 
 -(1-\widetilde{q}_n \widetilde{r}_n) d_n q_{n-1}=0, 
\\[1mm]
\frac{1}{h}\left(\widetilde{r}_n -r_n \right) & - b r_{n+1} 
 + a \widetilde{r}_{n-1} - \beta_{n+1} r_n + \widetilde{r}_{n} \alpha_n 
%\nonumber 
\\[-1mm]
& \mbox{}
 + \widetilde{r}_{n+1} d_{n+1} (1-q_n r_n) 
 -(1-\widetilde{r}_n \widetilde{q}_n) c_n r_{n-1}=0, 
\end{split}
%\right.
\\[1mm] &
%\end{equation}
%\begin{equation}
%\left\{
\begin{split}
 & \alpha_{n+1} - \alpha_n  = 
	a (\widetilde{q}_n \widetilde{r}_{n-1} -q_{n+1} r_n), 
\\[1mm]
 & \beta_{n+1} - \beta_n = 
	b (\widetilde{r}_n \widetilde{q}_{n-1} -r_{n+1} q_n), 
\\[1mm]
 & (1-\widetilde{r}_n \widetilde{q}_n) c_n =
  c_{n+1} (1-{r}_n {q}_n),
\\[1mm]
 & (1-\widetilde{q}_n \widetilde{r}_n) d_n =
  d_{n+1} (1-{q}_n {r}_n).
\end{split}
%\label{}
\end{split}
\right.
\label{fd-AL1}
\end{equation}
%\label{fd-AL1}
%\end{subequations}
Actually, 
%this 
the general 
%and 
nonreduced form 
(\ref{fd-AL1}) 
of the 
%fully 
time-discretized Ablowitz--Ladik lattice 
is integrable 
%even 
for 
%the case of 
matrix-valued dependent 
variables. 
%In the following, 
%In what follows, 
However, 
%in the following, 
in this paper, 
%for simplicity, 
we consider only 
%focus on 
the case of scalar 
%-valued 
dependent variables. 
%for simplicity. 
%Note that 
%Moreover, 
In addition to (\ref{V+-}), 
%If 
we impose 
%the 
rapidly decaying 
boundary conditions for 
%the 
%original 
%dependent variables 
$q_n$ and $r_n$, that is, 
%i.e., 
\begin{equation}
\lim_{n \to \pm \infty}
% \left( 
q_n = \lim_{n \to \pm \infty} r_n 
%\right) = \boldsymbol{0}, 
=0, \hspace{5mm}
\lim_{n \to \pm \infty} \left( \alpha_n, \beta_n, c_n, d_n \right)
%\hspace{5mm}
%\lim_{n \to \pm \infty} \beta_n 
= \left( \alpha, \beta, c,d \right). 
\label{ALbc}
\end{equation}
%
%Note that 
Similarly to the semi-discrete 
%continuous-time 
case, 
the parameters $a$, $b$, 
$\alpha$, $\beta$, $c$, and $d$ 
%can 
are allowed to depend on 
the discrete time coordinate \mbox{$m \in {\mathbb Z}$}. 
Thus, the time dependence of $h$ can be absorbed 
%into 
by $a$, $b$, $\alpha_n$, $\beta_n$, $c_n$, and $d_n$. 
For the same reason, 
it is 
%in principle 
possible to set $h$ as unity, 
%$h$ can be set as unity in principle, 
but we prefer 
%it is much more convenient 
%for the following discussion 
to leave it 
%$h$ 
as a small 
%expansion 
parameter, usually 
\mbox{$0 < |h| \ll 1$}. 
%these parameters. 
%are allowed to depend on 
%
To be precise, the boundary conditions
(\ref{ALbc}) contain {\it redundant} \/information. 
Indeed, it can be shown that
the auxiliary variables
have the same limit
%boundary
values for 
%as 
\mbox{$n \to -\infty$} and \mbox{$n \to +\infty$}.
%
%a sketchy description of how to verify this fact 
%
Therefore, it is sufficient 
%at first
to assume
either
\mbox{$\lim_{n \to -\infty} \left( \alpha_n, \beta_n, c_n, d_n \right)
= \left( \alpha, \beta, c,d \right)$}
or 
\mbox{$\lim_{n \to +\infty} \left( \alpha_n, \beta_n, c_n, d_n \right)
= \left( \alpha, \beta, c,d \right)$}. 
%Thus, 
The last four 
%equalities 
relations in (\ref{fd-AL1}) 
imply the following 
%{\it 
global
%} \/
expressions 
%of 
for
the auxiliary variables $\alpha_n$, $\beta_n$, 
$c_n$, and $d_n$ 
in terms of 
%\mbox{$q_n\textrm{'s}$} and \mbox{$r_n\textrm{'s}$}
\mbox{$q_n$} and 
\mbox{$r_n$}~\cite{AL77,Suris03,Taha1,Suris97',Suris00}: 
\begin{equation}
\begin{split}
& \begin{split}
 \alpha_n &= \alpha 
%+ a \left[ 
- a q_n r_{n-1} + a \sum_{j= -\infty}^{n-1} 
	(\widetilde{q}_j \widetilde{r}_{j-1} -q_{j} r_{j-1})
%\right]
\\
&= \alpha -  a \widetilde{q}_n \widetilde{r}_{n-1} 
  - a \sum_{j= n+1}^{+\infty} 
	(\widetilde{q}_{j} \widetilde{r}_{j-1} -q_{j} r_{j-1}), 
\end{split}
\\[3mm]
& \begin{split}
 \beta_n &= \beta - b r_n q_{n-1} + b \sum_{j= -\infty}^{n-1} 
	(\widetilde{r}_j \widetilde{q}_{j-1} -r_{j} q_{j-1}) 
\\
 &= \beta -  b \widetilde{r}_n \widetilde{q}_{n-1} 
  - b \sum_{j= n+1}^{+\infty} 
	(\widetilde{r}_{j} \widetilde{q}_{j-1} -r_{j} q_{j-1}), 
\end{split}
\\[3mm]
%& \begin{split}
& c_n = c \prod_{j=-\infty}^{n-1} 
	\frac{1-\widetilde{q}_j \widetilde{r}_j}{1-{q}_j {r}_j}
= c \prod_{j=n}^{+\infty} 
	\frac{1-{q}_j {r}_j}{1-\widetilde{q}_j \widetilde{r}_j},
\\[3mm]
& d_n = d \prod_{j=-\infty}^{n-1} 
	\frac{1-\widetilde{q}_j \widetilde{r}_j}{1-{q}_j {r}_j}
 = d \prod_{j=n}^{+\infty} 
	\frac{1-{q}_j {r}_j}{1-\widetilde{q}_j \widetilde{r}_j}. 
%\end{split}
\end{split}
\nonumber
%\label{}
\end{equation}
The equivalence of the two expressions for each auxiliary 
variable 
%is guaranteed by the boundary conditions 
%%((\ref{V+-}) and) 
%(\ref{ALbc}); in other words, 
%guarantees 
is related to the fact 
that 
%indicates that 
the three quantities 
%thus 
\mbox{$\sum_{j= -\infty}^{+\infty}q_{j} r_{j-1}$}, 
\mbox{$\sum_{j= -\infty}^{+\infty} r_{j} q_{j-1}$}, and 
\mbox{$\prod_{j=-\infty}^{+\infty} \left( 1-{q}_j {r}_j \right)$} 
are 
%indeed 
conserved 
%in time 
%quantities 
as in the continuous-time case. 
These conserved quantities are assumed to be finite, 
%; 
and the last one should be nonzero. 
The substitution of these expressions 
%in 
into the first two 
%relations 
equations in (\ref{fd-AL1}) provides 
%the 
a global-in-space  
time discretization of the nonreduced 
Ablowitz--Ladik 
lattice (\ref{sd-AL}). 
%~\cite{AL76,AL77}. 
%compatibility 
In the continuum 
%continuous 
limit of time \mbox{$h \to 0$}, we obtain 
\[
\alpha_n \to \alpha - a q_n r_{n-1}, \hspace{5mm}
\beta_n \to \beta - b r_n q_{n-1}, \hspace{5mm}
c_n \to c, \hspace{5mm} d_n \to d. 
\]
Thus, with an additional 
%the 
but unessential constraint on 
%%condition among 
the parameters, 
%and a proper redefinition of the parameters 
%\mbox{$$}, 
the time discretization 
%full-discrete Ablowitz--Ladik lattice 
(\ref{fd-AL1}) reduces to 
the nonreduced 
Ablowitz--Ladik lattice (\ref{sd-AL}) 
in the 
%continuous 
limit 
%of time 
\mbox{$h \to 0$}. 

%Note that 
The determinant of the \mbox{$2 \times 2$} 
Lax matrix $L_n$ (\ref{AL-Ln}) 
%is 
can be immediately 
%easily 
computed as 
%given as 
%by 
\mbox{$ \det L_n = 1 - q_n r_n$}. 
We compute the determinant of the \mbox{$2 \times 2$} Lax matrix 
$V_n$ (\ref{AL-Vn2}) as
\begin{align}
&\det V_n (\lambda) 
%\nonumber \\=& 
= \lambda^4 \hspace{1pt}h^2 a c_n 
 + \frac{1}{\lambda^4} \hspace{1pt}h^2 b d_n 
\nonumber \\
& \mbox{}+\lambda^2 \left\{ ha  - h^2 a^2 q_n \widetilde{r}_{n-1} 
 +h^2 a \beta_n 
+ h^2 a q_{n-1} \widetilde{r}_n d_n 
%+h c_n +h^2 \alpha_n c_n + h^2 a \widetilde{q}_n \widetilde{r}_{n-1} c_n 
%+ h^2 a q_n r_{n-1} c_n 
+h c_n \left[ 1 + h \alpha_n 
+h a \left(\widetilde{q}_n \widetilde{r}_{n-1} +  q_n r_{n-1} \right)\right] 
\right\}
\nonumber \\
& \mbox{}+\frac{1}{\lambda^2} \left\{ hb  - h^2 b^2 \widetilde{q}_{n-1} r_n 
 +h^2 b \alpha_n 
+ h^2 b \widetilde{q}_n r_{n-1} c_n 
+h d_n \left[ 1 + h \beta_n 
+h b \left(\widetilde{q}_{n-1} \widetilde{r}_n + q_{n-1} r_n \right)\right] 
\right\}
\nonumber \\
& \mbox{} + h^2 ab + h^2 c_n d_n  
 + \left( 1+ h \widetilde{q}_{n} r_{n-1}c_n + h \alpha_n \right)
   \left( 1+ h q_{n-1} \widetilde{r}_{n} d_n + h \beta_n \right)
\nonumber \\
& \mbox{} - h^2 \left( a q_n - \widetilde{q}_{n} c_n \right) 
  \left( b r_n - \widetilde{r}_{n} d_n \right) 
  - h^2 \left( b \widetilde{q}_{n-1} - q_{n-1} d_n \right) 
  \left( a \widetilde{r}_{n-1} - r_{n-1} c_n \right). 
\nonumber 
\end{align}
Thus, the equality 
(\ref{fd-cons1}) 
%together 
combined with the boundary conditions (\ref{ALbc}) 
(or
%, 
the streamlined version as stated above) 
implies the set of relations 
\begin{equation}
\left\{
\hspace{2pt}
\begin{split}
& c_n = c \Lambda_n, 
\\[1mm]
& d_n = d \Lambda_n, 
\\[1mm]
&  ha  - h^2 a^2 q_n \widetilde{r}_{n-1} 
 +h^2 a \beta_n 
+ h^2 a q_{n-1} \widetilde{r}_n d_n 
\\[-1mm] &
\mbox{}+h c_n \left[ 1 + h \alpha_n 
+h a \left(\widetilde{q}_n \widetilde{r}_{n-1} +  q_n r_{n-1} \right)\right] 
= \left[ ha + h^2 a \beta  + hc (1+h \alpha) \right] \Lambda_n,
\\[1mm]
& hb  - h^2 b^2 \widetilde{q}_{n-1} r_n 
 +h^2 b \alpha_n 
+ h^2 b \widetilde{q}_n r_{n-1} c_n 
\\[-1mm] &
\mbox{}+h d_n \left[ 1 + h \beta_n 
+h b \left(\widetilde{q}_{n-1} \widetilde{r}_n + q_{n-1} r_n \right)\right] 
= \left[ hb + h^2 b \alpha  + hd (1+h \beta) \right] \Lambda_n,
\\[1mm]
& h^2 ab + h^2 c_n d_n  
 + \left( 1+ h \widetilde{q}_{n} r_{n-1}c_n + h \alpha_n \right)
   \left( 1+ h q_{n-1} \widetilde{r}_{n} d_n + h \beta_n \right)
\\[-1mm] 
& \mbox{}- h^2 \left( a q_n - \widetilde{q}_{n} c_n \right) 
  \left( b r_n - \widetilde{r}_{n} d_n \right) 
  - h^2 \left( b \widetilde{q}_{n-1} - q_{n-1} d_n \right) 
  \left( a \widetilde{r}_{n-1} - r_{n-1} c_n \right)
\\[-1mm] 
& = \left[ h^2 ab + h^2 c d + (1+h \alpha)(1+h \beta) \right] \Lambda_n.
%\nonumber 
\end{split}
\right.
\label{ALalg1}
\end{equation}
%where 
Here, \mbox{$\Lambda_n$} is the quantity that satisfies 
%satisfying 
the recurrence formula
%recursive relation 
%\mbox{$ 
\begin{equation}
\left( 1-\widetilde{q}_n \widetilde{r}_n \right) \Lambda_n 
= (1-q_n r_n) \Lambda_{n+1}, 
\label{Lambda}
\end{equation}
%$}, 
%, \hspace{5mm} 
%and 
%as well as 
and has the {\it normalized} 
%``normalized'' 
\/boundary value
%, 
\mbox{$\lim_{n \to \pm \infty} \Lambda_n =1$}. 
%or, equivalently, 
More precisely, only one of the two conditions
\mbox{$\lim_{n \to - \infty} \Lambda_n =1$} and
\mbox{$\lim_{n \to + \infty} \Lambda_n =1$} is required, in accordance
with the boundary conditions for 
%$\alpha_n$ and $b_n$
\mbox{$\left( \alpha_n, \beta_n, c_n, d_n \right)$}. 
Then, the other condition can be confirmed. 
Thus, $\Lambda_n$ can be written explicitly as 
%\[
\begin{equation}
\Lambda_n = \prod_{j=-\infty}^{n-1} 
	\frac{1-\widetilde{q}_j \widetilde{r}_j}{1-{q}_j {r}_j}
= \prod_{j=n}^{+\infty} 
	\frac{1-{q}_j {r}_j}{1-\widetilde{q}_j \widetilde{r}_j}.
%\]
\label{la-rep}
\end{equation}
Note that 
the first two equalities in (\ref{ALalg1}) are valid 
%hold true 
even if \mbox{$a=0$} or \mbox{$b=0$} 
(cf.\ 
%see equations of motion 
(\ref{fd-AL1})). 
The algebraic system 
%of five equations 
(\ref{ALalg1})
%in the general case 
%contains 
%is 
essentially 
comprises 
%five 
three nontrivial relations 
%equalities 
for 
the 
%five 
three unknowns:\ 
%$c_n$, $d_n$, 
$\alpha_n$, $\beta_n$, 
and $\Lambda_n$. 
Before discussing the 
%most 
general case, 
%\mbox{$(a,b) \neq {\boldsymbol{0}}$}, 
we consider two special 
%simpler 
cases:\ 
%the case 
\mbox{$a=b=0$}
%; 
or 
%, \hspace{1pt}
%and the case 
\mbox{$c=d=0$}. 
%wherein the algebraic system can be solved through the quadratic formula. 
\\
\\
$\bullet$
%$\circ$ 
%(1) 
{\it 
%(i) 
The case \mbox{$a=b=0$}}. 

In this case, the two auxiliary variables, 
\mbox{$\alpha_n$} 
and 
\mbox{$\beta_n$}, 
%are $n$-independent and 
become $n$-independent, 
%%thus determined by 
%and their 
%%boundary 
%values are determined by the boundary conditions (\ref{ALbc}), 
i.e., \mbox{$\alpha_n=\alpha$}, 
%and 
\mbox{$\beta_n=\beta$}. 
Thus, 
%while 
%and 
the discrete-time system 
(\ref{fd-AL1}) reduces to 
\begin{equation}
\left\{
\hspace{2pt}
\begin{split}
& \frac{1}{h}\left(\widetilde{q}_n -q_n \right) 
 + (1-q_n r_n) \Lambda_{n+1} \left( c \widetilde{q}_{n+1} 
 - d 
%(1-\widetilde{q}_n \widetilde{r}_n) \Lambda_n 
q_{n-1}\right) 
 - \alpha q_n + \beta \widetilde{q}_{n} =0, 
\\[1mm]
& \frac{1}{h}\left(\widetilde{r}_n -r_n \right)  
 + (1-q_n r_n) \Lambda_{n+1} \left( d \widetilde{r}_{n+1} 
 - c 
%(1-\widetilde{r}_n \widetilde{q}_n) \Lambda_n 
 r_{n-1} \right) - \beta r_n + \alpha \widetilde{r}_{n} =0, 
\\[1mm]
& 
\left( 1-\widetilde{q}_n \widetilde{r}_n \right) \Lambda_n
= (1-q_n r_n) \Lambda_{n+1}, \hspace{5mm} 
%	\lim_{n \to \pm \infty} \Lambda_n =1,
	\lim_{n \to - \infty} \Lambda_n =1 \;\,\mathrm{or} \;\,
	\lim_{n \to + \infty} \Lambda_n =1,
\end{split}
\right.
\label{fd-AL3}
\end{equation}
and 
the algebraic system (\ref{ALalg1}) 
%boils down 
simplifies to 
%a quadratic equation in $\Lambda_n$, 
%
%\begin{equation}
%\left\{
%\hspace{2pt}
%\begin{split}
%& c_n = c \Lambda_n, 
%\\[1mm]
%& d_n = d \Lambda_n, 
%\\[1mm]
%& h^2 \left(1-\widetilde{q}_{n} \widetilde{r}_{n} - q_{n-1}r_{n-1}\right) 
%  c_n d_n  
% + \left( 1+ h \widetilde{q}_{n} r_{n-1}c_n + h \alpha \right)
%   \left( 1+ h q_{n-1} \widetilde{r}_{n} d_n + h \beta \right)
%\\[-1mm] 
%& = \left[ h^2 c d + (1+h \alpha)(1+h \beta) \right] \Lambda_n.
%%\nonumber 
%\end{split}
%\right.
%\label{ALalg3}
%\end{equation}
%
\begin{align}
& h^2 cd \left( 1- \widetilde{q}_{n} \widetilde{r}_{n}\right) 
\left( 1 - q_{n-1} r_{n-1} \right)\Lambda_n^2
-\left[
\left( 1 + h \alpha \right) \left( 1 + h \beta \right) + h^2 cd
\right.
\nonumber \\
& \left.
\mbox{} 
- \left( 1 + h \alpha \right)h d q_{n-1} \widetilde{r}_{n}
- \left( 1 + h \beta \right)h c \widetilde{q}_{n} r_{n-1}
\right] \Lambda_n 
+ \left( 1 + h \alpha \right) \left( 1 + h \beta \right) 
=0.
\label{quadra3}
\end{align}
System (\ref{fd-AL3}) is invariant 
%with respect to 
%has 
%the 
%invariance 
under 
%with respect to 
the following transformation: 
\mbox{$q_n \mapsto \mu^n q_n$}, \mbox{$r_n \mapsto \mu^{-n} r_n$}, 
\mbox{$hc \mapsto \mu^{-1} hc$}, \mbox{$hd \mapsto \mu hd$},
\mbox{$h\alpha \mapsto h\alpha$}, \mbox{$h\beta \mapsto h\beta$}, 
where $\mu$ is 
%any 
a nonzero constant. 
Using a similar transformation 
%involving 
with respect to the time direction, 
%discrete time $m$, 
it is 
%also 
possible to remove the parameters 
$\alpha$ and $\beta$, 
but we prefer to 
%leave 
retain them. 
%do not present it here. 

%In (\ref{ALalg3}), 
%using the first and second equalities, the third equality
%results in a
%%(generally)
%quadratic equation in 
%%for 
%$\Lambda_n$, 
In terms of the normalized parameters
\begin{equation}
%\widehat{a}
\hat{c} := \frac{hc}{1 + h \alpha}, 
\hspace{5mm} \hat{d} :=\frac{h d}{1 + h \beta}, 
\label{cdhat}
\end{equation}
(\ref{quadra3}) can be rewritten as 
\begin{align}
& \hat{c} \hat{d} \left( 1- \widetilde{q}_{n} \widetilde{r}_{n}\right) 
\left( 1 - q_{n-1} r_{n-1} \right)\Lambda_n^2
 -\left[ 1 + \hat{c}\hat{d}
- \hat{d} q_{n-1} \widetilde{r}_{n} - \hat{c} \hspace{1pt}
	\widetilde{q}_{n} r_{n-1}
\right] \Lambda_n 
+ 1 =0.
\label{quadra4}
%\nonumber
\end{align}
When \mbox{$cd \neq 0$}
%, 
(and thus \mbox{$\hat{c}\hat{d} \neq 0$}), 
this is a quadratic equation in 
%for 
$\Lambda_n$, 
%that 
which has 
%with 
%and has 
%having 
%has 
two solutions. 
%Let us first consider 
%In particular, in the simplest case where
%In order to discard 
The simplest way to reject
%repulse
%discard 
%exclude 
the improper solution is to recall 
%resort to
the 
asymptotic behavior of the Lax 
matrix $V_n$ for
%in 
small $h$ 
%asymptotic expansion of $V_n$ with respect to $h$ 
(\ref{asymp-h}), but here we 
take a different route. 
%provide a more intuitive rationale. 
%we 
Let us first consider the 
%simplest 
``trivial'' 
case where
%\mbox{$q_n=r_n=0$}, 
\mbox{$q_n$} and \mbox{$r_n$} are 
%identically 
zero 
for all $n$; 
if this is satisfied at 
%some 
%the initial time 
%some 
%a certain time 
some instant
%and 
%at any time $m$.  
\mbox{$m=m_0$}, 
%\mbox{$m=0$}, 
then it 
%should 
holds true 
identically for 
%at 
any time $m$. 
%(cf.~(\ref{ALbc}) and (\ref{fd-AL3})).  
Thus, the two solutions of (\ref{quadra4}) are 
%they are 
given by \mbox{$\Lambda_n=1, 
\hspace{2pt}1/(\hspace{1pt}\hat{c}\hat{d}\hspace{1pt})$}. 
% or 
%\mbox{$\Lambda_n=1/(\hspace{1pt}\hat{c}\hat{d}\hspace{1pt})$}. 
%\left(\hat{c}\hat{d} \right)$}. 
The recurrence formula for $\Lambda_n$ in 
%equations of motion 
(\ref{fd-AL3}) 
implies that these two solutions are unconnected, 
%``unconnected'', 
i.e., \mbox{$\Lambda_n$} takes the same value for all $n$. 
%; 
%and that 
%Moreover, 
The solution 
\mbox{$\Lambda_n=1/(\hspace{1pt}\hat{c}\hat{d}\hspace{1pt})$} 
%should 
can 
be discarded if \mbox{$\hat{c}\hat{d} \neq 1$}, 
%(if \mbox{$\hat{c}\hat{d} \neq 1$}), 
%as 
because 
it is inconsistent with the boundary condition
%s 
for $\Lambda_n$. 
Next, 
in the general case where 
\mbox{$q_n$} and \mbox{$r_n$} are not identically zero, 
we assume that 
%consider 
their amplitudes are always so small that 
%enough 
%much smaller than $1$ 
%they 
their effects can be regarded as 
%a 
%an \mbox{$O(h)$} 
weak perturbations 
%to 
of the identically zero case. 
In particular, 
%we have either 
the 
%possible 
value of 
%the auxiliary variable 
\mbox{$\Lambda_n$} 
is restricted 
%assumed 
to 
%take values in 
%the 
%\mbox{$O(h)$} 
%vicinity
%proximity 
a neighborhood 
of unity. 
%%$1$, 
%i.e., 
%%quantity 
%\mbox{$\Lambda_n \sim 1 + O(h)$}. 
%for all $n$ and $m$. 
Thus,
%%taking into account
%recalling the prescribed
% asymptotic behavior
%of $\Lambda_n$ as \mbox{$h \to 0$} (or \mbox{$n \to \pm \infty$}),
%and
we obtain the proper
%unique
solution of the quadratic equation (\ref{quadra4}) 
%,
%given by 
as
\begin{equation}
%\frac{1}{\Lambda_n} =
\Lambda_n = \frac{2}{1 + C_n + \sqrt{\left( 1 + C_n \right)^2
-4 D_n}},
\label{Lambda3}
\end{equation}
with
\begin{equation}
\begin{split}
C_n & := \hat{c}\hat{d}
- \hat{d} q_{n-1} \widetilde{r}_{n} - \hat{c} \hspace{1pt}
	\widetilde{q}_{n} r_{n-1}, 
\\
D_n & := \hat{c} \hat{d} 
\left( 1- \widetilde{q}_{n} \widetilde{r}_{n}\right) 
\left( 1 - q_{n-1} r_{n-1} \right). 
\end{split}
%\nonumber
\label{CnDn}
\end{equation}
%
%Thus, 
Note that this 
%expression 
solution 
is also 
valid for the linear case \mbox{$\hat{c}\hat{d}=0$}. 
%In addition, 
The 
%rapidly 
decaying 
%zero 
boundary conditions for $q_n$ and $r_n$ imply that 
\mbox{$\lim_{n \to - \infty} \Lambda_n =
\lim_{n \to + \infty} \Lambda_n =1$}. 
When \mbox{$\hat{c}\hat{d} \in \mathbb{R}$},
%this
the local expression (\ref{Lambda3}) is
valid only if \mbox{$\hat{c}\hat{d} \le 1$}. 
%(cf.~(\ref{ALbc}) and (\ref{la-rep})). 
If \mbox{$\hat{c}\hat{d} > 1$}, 
the other solution of (\ref{quadra4})
should be
adopted; alternatively, 
%otherwise,
one can understand 
the right-hand side of (\ref{Lambda3}) as being defined by 
%through 
%its 
%the 
a Taylor series for small $\hat{c}$ and $\hat{d}$, 
and 
its 
%proper 
analytic continuation. 
%for large $\hat{c}$ and $\hat{d}$. 
%used.
%displace
%supersede
At the ``threshold'' value 
%of $\hat{c}\hat{d}$, 
%in the case 
\mbox{$\hat{c}\hat{d} =1$},
the discrete-time system (\ref{fd-AL3})
%under the boundary conditions (\ref{ALbc})
%allows
has the
%``trivial''
trivial solution \mbox{$\widetilde{q}_n = \hat{d} q_{n-1}$},
\mbox{$\widetilde{r}_n = \hat{c} r_{n-1}$},
\mbox{$\Lambda_n = 1/(1- q_{n-1}r_{n-1})$}. 
Thus,
the discriminant
%in the square root
of the quadratic equation (\ref{quadra4})
vanishes at \mbox{$\hat{c}\hat{d}=1$},
%at \mbox{$h =\pm 1$},
and the two solutions
indeed intersect. 
Unless \mbox{$\hat{c}\hat{d} =1$}, 
a unified expression for \mbox{$\Lambda_n$}, 
\[
%\frac{1}{\Lambda_n} =
\Lambda_n = \frac{2}{1 + C_n + \bigl(1-\hat{c}\hat{d}\hspace{1pt}\bigr)
\sqrt{\left( \frac{1 + C_n}{1-\hat{c}\hat{d} }\right)^2
-\frac{4 D_n}{( 1-\hat{c}\hat{d} \hspace{1pt})^2}}},
\]
can resolve the sign problem of the square root. 
%where 
%Note, however, that 
%Moreover, 
In addition, 
the above square root 
%in the above expression 
%formula 
%which 
allows the Taylor expansion 
%for small 
with respect to 
%perturbative 
\mbox{$ \{ \widetilde{q}_{n}, \widetilde{r}_{n}, q_{n-1}, r_{n-1} \}$}. 
%; 
However, 
%but 
we do not use 
%it looks 
%%too 
this unwieldy 
%and unattractive 
%form. 
formula. 
%for 

%Now, 
The first and second equations in 
%system 
(\ref{fd-AL3}) are linear in \mbox{$\widetilde{q}_n$} and
\mbox{$\widetilde{r}_n$}, respectively. 
Thus, 
the
%{\it explicit}
forward
time evolution 
%
%that is
%%{\it 
%explicit under the decaying 
%boundary conditions 
%%(\ref{ALbc})
can be expressed
%written
%given
as follows (cf.\ (\ref{cdhat})):
\begin{equation}
\left\{
\hspace{2pt}
\begin{split}
& \widetilde{q}_n = \frac{1+h \alpha}{1+h \beta}\hspace{1pt}q_n
 + \left( 1-q_n r_n \right) \Lambda_{n+1}
 \left( - \frac{1+h \alpha}{1+h \beta} \hspace{1pt}
	\hat{c}\hspace{1pt} \widetilde{q}_{n+1} +
 \hat{d} {q}_{n-1}\right),
\\[1mm]
&
\widetilde{r}_n = \frac{1+h \beta}{1+h \alpha}\hspace{1pt}r_n
 + \left( 1-q_n r_n \right) \Lambda_{n+1}
 \left( -\frac{1+h \beta}{1+h \alpha} \hspace{1pt}\hat{d} 
	\hspace{1pt}\widetilde{r}_{n+1}
 +\hat{c} r_{n-1} 
\right),
\end{split}
\right.
\label{ALmap2}
\end{equation}
where $\Lambda_n$ is given by
(\ref{Lambda3}) with (\ref{CnDn}). 
In the simplest case of \mbox{$c=\hat{c}=0$} or \mbox{$d=\hat{d}=0$},
%\mbox{$ab= \hat{a} \hat{b}=0$},
%of \mbox{$a=0$} or \mbox{$b=0$},
%the quadratic
%%algebraic
%equation (\ref{quadra4}) for $\Lambda_n$ reduces to a linear equation,
%and
%%consequently,
the forward time evolution (\ref{ALmap2})
is
%written
%expressed as
%becomes
given by
a simple rational mapping.
%the mapping becomes rational
This fact
%result
was
%simple case recovers the result of S
%discovered 
%uncovered 
disclosed 
%unveiled 
by Suris~\cite{Suris97',Suris00}. 
%It is worthwhile to mention that 
In the general case, 
%of \mbox{$-1< \hat{c}\hat{d} \le 1$}, 
%in terms of 
%by introducing 
using the new parameters 
\mbox{$\check{c}:=\hat{c}\hspace{1pt}
	\bigl(1+\hat{c}\hat{d}\hspace{1pt}\bigr)^{-1}$} and
%, 
\mbox{$\check{d}:=\hat{d}\hspace{1pt}
	\bigl(1+\hat{c}\hat{d}\hspace{1pt}\bigr)^{-1}$}, 
the mapping (\ref{ALmap2}) 
can be rewritten in a slightly simpler form: 
%\begin{equation}
\[
\left\{
\hspace{2pt}
\begin{split}
& \widetilde{q}_n = \frac{1+h \alpha}{1+h \beta}\hspace{1pt}q_n
 + \frac{2 \left( 1-q_n r_n \right)}{F_{n}}
 \left( - \frac{1+h \alpha}{1+h \beta} \hspace{1pt}
	\check{c}\hspace{1pt} \widetilde{q}_{n+1} +
 \check{d} {q}_{n-1}\right),
\\[1mm]
&
\widetilde{r}_n = \frac{1+h \beta}{1+h \alpha}\hspace{1pt}r_n
 + \frac{2 \left( 1-q_n r_n \right)}{F_{n}}
 \left( -\frac{1+h \beta}{1+h \alpha} \hspace{1pt}\check{d} 
	\hspace{1pt}\widetilde{r}_{n+1}
 +\check{c} r_{n-1} 
\right),
\end{split}
\right.
%\label{ALmap3}
%\end{equation}
\]
%Here, 
with 
\[
\begin{split}
F_n & := 1 - \check{d} q_{n} \widetilde{r}_{n+1} - \check{c} 
	\hspace{1pt}\widetilde{q}_{n+1} r_{n} 
\\[1mm]
& \hphantom{=}\;\; \mbox{} + \sqrt{\left( 1 - \check{d} q_{n} \widetilde{r}_{n+1} 
	- \check{c} \hspace{1pt}
	\widetilde{q}_{n+1} r_{n}  \right)^2
-4 \check{c} \check{d} 
\left( 1- \widetilde{q}_{n+1} \widetilde{r}_{n+1}\right) 
\left( 1 - q_{n} r_{n} \right)} \hspace{1pt}. 
\end{split}
\]
That is, the first 
%``constant'' 
term \mbox{$\hat{c}\hat{d}$} 
of $C_{n+1}$ 
%can be 
is removed. When \mbox{$\hat{c}\hat{d} \in \mathbb{R}$}, 
%the additional condition 
%the above 
this expression for $F_n$ is valid only if 
\mbox{$-1< \hat{c}\hat{d} \le 1$}; this corresponds to
the range \mbox{$\check{c} \check{d} \le 1/4$}. 
%Otherwise, 
If \mbox{$\hat{c}\hat{d} <-1$} or \mbox{$\hat{c}\hat{d} >1$}, 
the sign in front of the square root has to be changed. 

When $h$ is real, \mbox{$\beta= \alpha^\ast$}, and
\mbox{$d = c^\ast$}, we can impose the 
%reality (
complex conjugacy 
reduction \mbox{$r_n
%\pm
%\propto
=\sigma q_n^\ast$} with a real constant $\sigma$.
%is allowed.
%is possible.
In particular, setting 
%as
\mbox{$\alpha= -\beta=-\mathrm{i} \gamma /\varDelta^2$},
\mbox{$c = -d=-\mathrm{i}/\varDelta^2$},
and \mbox{$r_n= - \varDelta^2 q_n^\ast$}, we obtain
%\mbox{$$}
the
%a 
fully discretized NLS equation
\begin{equation}
\hspace{2pt}
\begin{split}
 \frac{\mathrm{i}}{h}\left(\widetilde{q}_n -q_n \right)
 & + \frac{2 \left( 1 + \varDelta^2\left| q_n \right|^2 \right)}{
 1 + C_{n+1} + \sqrt{\left( 1 + C_{n+1} \right)^2 -4 D_{n+1}}}
 \frac{\widetilde{q}_{n+1} + q_{n-1} }{\varDelta^2}
 -\frac{\gamma}{\varDelta^2} \left(\widetilde{q}_{n} + q_n \right)
%\\[-1mm] & \mbox{}
 =0,
\end{split}
\label{fdNLS2}
\end{equation}
%with
where
\begin{equation}
\begin{split}
%h
C_{n+1} & = \frac{h^2}{
%\left( \frac{\varDelta^2}{h} - \mathrm{i}\gamma \right)
\left(\varDelta^2  + \mathrm{i}\gamma h \right)
\left( \varDelta^2 - \mathrm{i}\gamma h \right)}
 + \frac{\mathrm{i} \varDelta^2 h
}{\varDelta^2 + \mathrm{i}\gamma h }
q_{n} \widetilde{q}_{n+1}^{\hspace{2pt}\ast}
 - \frac{\mathrm{i} \varDelta^2 h}{ \varDelta^2 - \mathrm{i}\gamma h}
\widetilde{q}_{n+1} q_{n}^{\ast},
\\
%h^2
D_{n+1} & = \frac{h^2}{
\left(\varDelta^2  + \mathrm{i}\gamma h \right)
\left( \varDelta^2 - \mathrm{i}\gamma h \right)}
\left( 1 + \varDelta^2 \left| \widetilde{q}_{n+1} \right|^2 \right)
\left( 1 + \varDelta^2 \left| q_{n} \right|^2
\right).
\end{split}
\nonumber
\end{equation}
The choice
%case
of \mbox{$\gamma=1$} 
%of 
for the real
%free
parameter $\gamma$
is the most natural when 
%for
%directly
taking
%considering
the continuum limit,
while the choice of 
%case
\mbox{$\gamma=0$} simplifies the equation considerably.
The aforementioned 
condition \mbox{$\hat{c}\hat{d} \le 1$} implies 
that 
%\mbox{$h^2 \le \varDelta^4 + (\gamma h)^2$}; 
\mbox{$ (1-\gamma^2) h^2 \le \varDelta^4 $}; 
this is automatically satisfied if \mbox{$\gamma^2 \ge 1$}. 
%At first sight, 
At first glance, it is far from evident that the 
quantity 
%inside 
in the square root is 
%non-negative: 
nonnegative: 
\mbox{$ \left( 1 + C_{n+1} \right)^2 -4 D_{n+1} \ge 0 $}. 
However, this inequality should hold true 
%as 
because the left-hand side represents 
%is 
the discriminant of the quadratic equation 
%for 
in $\Lambda_{n+1}$ 
(cf.\ (\ref{quadra4})). 
%is non-negative. 
%Indeed, 
In fact, 
%because of 
the reduction \mbox{$r_n= - \varDelta^2 q_n^\ast$} 
with \mbox{$\varDelta^2 >0$} 
guarantees 
%the values of 
the auxiliary variable 
$\Lambda_n$ 
%is 
%restricted to
%guaranteed 
to be 
%real-valued 
positive 
(cf.\ (\ref{la-rep})). 
%%In addition
%In fact, \mbox{$\Lambda_n$} 
%%\mbox{$\Lambda_n > 0$} 
%%is restricted to a positive number 
%is always 
%positive so that 
Thus, we also 
%obtain 
have 
the inequality \mbox{$ 1 + C_{n+1} > 0 $}. 
To 
%sum up
summarize, 
%the two quantities 
$\widetilde{q}_{n+1}$ and $q_n$ are not 
fully 
independent 
%; rather, they are 
and 
%somehow correlated
%and satisfies the inequality 
%In a nutshell, 
%to 
satisfy the inequality \mbox{$ 1 + C_{n+1} \ge 2 \sqrt{D_{n+1}} $}. 
Note that 
%using (\ref{fdNLS2}), 
(\ref{fdNLS2}) with \mbox{$n \to n+1$} determines 
$\widetilde{q}_{n+1}$ 
%is determined 
from $q_{n+1}$, 
$\widetilde{q}_{n+2}$, and $q_n$. 

%It is interesting to note
%%should be noted
%that, like 
Similarly to the continuous NLS equation,
this full discretization is homogeneous under the following weighting scheme:
\mbox{weight\hspace{1pt}$(\varDelta)=-1$},
\mbox{weight\hspace{1pt}$(h)=-2$},
\mbox{weight\hspace{1pt}$(q_n)=1$}.
%invariant possesses
%under the recaling
At present,
it is unclear
%at the present time
whether
the corresponding one-parameter group of scaling symmetries,
\mbox{$(\varDelta, h, q_n) \mapsto
(\varDelta/k, h/k^2,
%\mu
k q_n)$},
can define
%a meaningful class of self-similar
meaningful ``self-similar'' solutions
%of
to (\ref{fdNLS2}).

As mentioned previously, 
%above, 
%from an aesthetic point of view, 
%aesthetically pleasing 
%It is worthwhile to mention that 
it might be aesthetically pleasing 
%better 
to eliminate 
%remove 
%can be 
the first ``constant'' term of $C_{n+1}$ 
by a suitable
%proper 
%appropriate
redefinition of 
%redefining 
%scaling 
the parameters. 
For example, 
in the simple case of \mbox{$\gamma=0$} and 
\mbox{$\varDelta^2=1$}, (\ref{fdNLS2}) can be rewritten as 
\begin{equation}
\hspace{2pt}
\begin{split}
%{\mathcal{C}\cal{C}\mathcal{D}\cal{D}}{\mathcal C}
 \frac{\mathrm{i}}{\delta}\left(\widetilde{q}_n -q_n \right)
 & + \frac{2 \left( 1 + \left| q_n \right|^2 \right)
  \left( \widetilde{q}_{n+1} + q_{n-1} \right)}{
 1 + \mathrm{i} \delta {\mathcal C}_n
%\widehat{C}_{n+1} 
+ \sqrt{\left( 1 + \mathrm{i} \delta {\mathcal C}_n
%\widehat{C}_{n+1} 
	\right)^2 -4 \delta^2 
%\widehat{D}_{n+1}
{\mathcal D}_n}}
 =0,
\end{split}
\label{fdNLS2'}
\end{equation}
with
%where
%\begin{equation}
\[
%h
%\widehat{C}_{n+1}
{\mathcal C}_n :=  
%\mathrm{i} \delta 
%\left( 
q_{n} \widetilde{q}_{n+1}^{\hspace{2pt}\ast}
	- \widetilde{q}_{n+1} q_{n}^{\ast},
%\right),
\hspace{5mm}
%\widehat{D}_{n+1}
{\mathcal D}_n:= 
%\delta^2  
\left( 1 + \left| \widetilde{q}_{n+1} \right|^2 \right)
\left( 1 + \left| q_{n} \right|^2
\right).
%\nonumber
%\end{equation}
\]
Here, \mbox{$\delta := h/(1+h^2)$} is a new parameter. 
%Note that 
%If \mbox{$h \in \mathbb{R}$} (or, \mbox{$-1 \le h \le 1$}), then
%As 
Because \mbox{$-1 \le h \le 1$}, the range of $\delta$ is 
%given by 
%we have 
%\mbox{$-\frac{1}{2} \le \delta \le \frac{1}{2}$}.
\mbox{$-1/2 \le \delta \le 1/2$}; 
at the end points \mbox{$\delta =\pm 1/2$}, 
the time evolution is trivial 
%that corresponds to 
because 
%of the relation 
\mbox{$\hat{c}\hat{d} =1$}. 
%It should be noted 
Note that (\ref{fdNLS2'}) is invariant 
under the transformation
\mbox{$q_n \mapsto (-1)^n q_n$}, \mbox{$\delta \mapsto -\delta$}. 

%In order 
To
%explicitly obtain
%write out 
express the backward time evolution explicitly, 
%which
%%is necessary for
%%should be done for
%%apears to be especially
%is highly desired for
%%in
%%necessary in
%the case \mbox{$h < 0$},
we only have
%need
to replace 
%rewrite 
\mbox{$(1-q_n r_n) \Lambda_{n+1}$} in 
the first 
%two 
and second equalities 
%equations 
of (\ref{fd-AL3}) 
%by 
with \mbox{$\left( 1-\widetilde{q}_n \widetilde{r}_n \right) \Lambda_n$}, 
%$\Lambda_n$ in (\ref{fd-AL3}) to \mbox{$\Lambda_{n+1}$}
%using
%with the help of (\ref{Lambda}),
and then
%\mbox{$\alpha_{n}$} and \mbox{$\beta_{n}$}
%to \mbox{$\alpha_{n+1}$} and \mbox{$\beta_{n+1}$}
%before substituting the local expressions for
%\mbox{$\alpha_{n}$} and \mbox{$\beta_{n}$}.
substitute
%before substituting
the local expression
%for $\Lambda_{n}$
(\ref{Lambda3}) with (\ref{CnDn}).
%
%Note that in order to explicitly obtain the backward time evolution 
%(or, equivalently, \mbox{$h < 0$}),  
%we need to rewrite \mbox{$(1-q_n r_n) \Lambda_{n+1}$} 
%as \mbox{$\left( 1-\widetilde{q}_n \widetilde{r}_n \right) \Lambda_n$} 
%before substituting the local expression for \mbox{$\Lambda_n$}. 

In the 
complex conjugacy reduction 
wherein  
%When 
$h$ is real, \mbox{$\beta= \alpha^\ast$}, and
\mbox{$d = c^\ast$}, 
we can ``normalize'' 
the scaling of 
%can change the scaling of 
%rescale 
the dependent variable 
%and set as 
by setting 
%as 
\mbox{$r_n =- q_n^\ast$}. 
%, we leave the free parameters unspecified and change 
%the scaling of the dependent variable 
%impose the 
%%reality (complex conjugacy) 
%reduction \mbox{$r_n =- q_n^\ast$} 
In this case, 
%We can 
we rewrite the first equation for $q_n$ in (\ref{fd-AL3}) 
as 
\begin{equation}
\left(1 + h\alpha^\ast \right)\widetilde{q}_n 
  - \left(1 + h\alpha \right)q_n 
 + (1 + \left| \widetilde{q}_n \right|^2 ) \Lambda_{n} 
  \left( h c \widetilde{q}_{n+1} - h c^\ast q_{n-1}\right) =0.
\label{rational1}
\end{equation}
Moreover, (\ref{quadra3}) can be rewritten as 
\begin{align}
& \left| 1 + h \alpha \right|^2 
%\left( 1 + h \beta \right) 
+ h^2 |c|^2
%\left| c \right|^2
+ \left( 1 + h \alpha \right)h c^\ast q_{n-1} 
	\widetilde{q}_{n}^{\hspace{2pt}\ast}
+ \left( 1 + h \alpha^\ast \right)h c \widetilde{q}_{n} 
 q_{n-1}^{\hspace{1pt}\ast}
\nonumber \\
&= \left| 1 + h \alpha \right|^2
%\left( 1 + h \beta \right) 
\frac{1}{\Lambda_n}
%\right. \nonumber \\ & \left. \mbox{}
+ 
h^2 | c |^2 \left( 1+\left| \widetilde{q}_{n} \right|^2 \right)
\left( 1 + \left|q_{n-1}\right|^2 \right)\Lambda_n.
\label{rational2}
\end{align}
Using (\ref{rational1}), 
%to 
we can replace \mbox{$1/\Lambda_n$} and \mbox{$\Lambda_n$}
%\mbox{$\Lambda_n\textrm{'s}$} 
in 
%and 
(\ref{rational2}) 
%by 
with rational expressions 
in $q_n$ 
%and 
as well as its shifts and complex conjugate. 
Thus, we obtain a rational form of the fully discrete NLS equation, 
\begin{align}
& \left| 1 + h \alpha \right|^2 
%\left( 1 + h \beta \right) 
+ h^2 |c|^2
%\left| c \right|^2
+ 2 \hspace{1pt}\mathrm{Re}\left[ \left( 1 + h \alpha \right)h c^\ast q_{n-1}
 \widetilde{q}_{n}^{\hspace{2pt}\ast} \right]
%+ \left( 1 + h \alpha^\ast \right)h c \widetilde{q}_{n} 
%q_{n-1}^{\hspace{1pt}\ast}
\nonumber \\
& = 
\left| 1 + h \alpha \right|^2
%\left( 1 + h \beta \right) 
\left( 1+\left| \widetilde{q}_{n} \right|^2 \right) 
\frac{ h c \widetilde{q}_{n+1} - h c^\ast q_{n-1} }
{\left(1 + h\alpha \right)q_n - \left(1 + h\alpha^\ast \right)\widetilde{q}_n}
%\right. 
\nonumber \\ 
& \hphantom{=}
\mbox{}
+ h^2 | c |^2 
\left( 1 + \left|q_{n-1}\right|^2 \right)
%\Lambda_n
\frac{ \left(1 + h\alpha \right)q_n 
- \left(1 + h\alpha^\ast \right)\widetilde{q}_n }
{h c \widetilde{q}_{n+1} - h c^\ast q_{n-1}}.
\label{rational3}
\end{align}
Surprisingly, this coincides with 
%rational form 
%expression 
%form 
%of 
the 
%full
double-discrete NLS equation 
%was 
%obtained 
proposed by Quispel, Nijhoff, Capel, 
and van der Linden~\cite{QNCL,NijCap}, 
up to a minor 
%reparametrization and 
change of coordinates and parameters; 
%it is seemingly 
despite its 
%``elegant''
%beautiful 
%expression
``elegance'',
%~\cite{NijCap}, 
%it 
this rational version 
has the drawback that the 
forward/backward time evolution cannot be uniquely determined. 
%in a unique way. 
In addition, 
one 
%it 
cannot immediately recognize that 
(\ref{rational3}) 
%allows 
reduces to the NLS equation in a continuous limit. 
Thus, we prefer our version, 
%involving the square root 
%that 
which is seemingly less elegant
%``less elegant'' 
because of the existence of 
the square root
%, 
%(use /application of the quadratic formula), 
but can 
%determine the time evolution uniquely. 
define the unique 
time evolution properly 
and allow an easy-to-follow continuous limit. 

Actually, using a 
%simple 
coordinate transformation, 
it is 
%in principle
possible to ``identify'' 
our time discretization 
%(
%[\hspace{1pt}
[(\ref{ALmap2}) with 
\mbox{$\hat{d} = \hat{c}^\ast$}, \mbox{$\alpha=\beta=0$}, 
and \mbox{$r_n=-q_n^\ast$}\hspace{1pt}] 
%can be identified 
with the auto-B\"acklund transformation of 
the Ablowitz--Ladik 
%lattice 
hierarchy 
derived by Nijhoff, Quispel, and Capel~\cite{NQC}. 
%(see (19) in~\cite{NQC}). 
However, 
%unlike our time discretization, 
their expression (see (19) in~\cite{NQC}) 
involves the indefinite sign $\pm$ 
%sign indefiniteness
in front of the square root
%, 
and
% so that 
%because they were unaware of this coordinate transformation 
%and interpretation. 
%no explanation was given there on how to was given there 
it is not clear how to 
%determine the sign.
understand and determine
%read 
it. 

%Write the explicit form of the normalized Lax pair, 
%which is expected to provide the equations of motion without any constraint. 
Once 
the local expressions for 
%all 
the auxiliary variables 
%in the original dependent variables 
have been derived, 
%obtained, 
%are replaced by 
%their 
we can 
%, in principle, 
%``normalize'' 
normalize
%convert 
the \mbox{$2 \times 2 $} matrix 
Lax pair, $L_n$ and $V_n$, 
%to a ``normalized'' form, i.e., 
so that \mbox{$\det L_n$} and \mbox{$\det V_n$} 
become equal to $1$. 
%\mbox{$\det L_n=\det V_n=1$}. 
Indeed, this is easily achieved by dividing $L_n$ and $V_n$ 
by \mbox{$\sqrt{\det L_n}$} and \mbox{$\sqrt{\det V_n}$}, 
respectively (cf.\ (\ref{fd-Lax}) and (\ref{fd-cons1})). 
However, the normalized
%``normalized'' 
Lax pair 
%without 
not involving 
the auxiliary variables appears to be 
%not more convenient 
rather cumbersome
%unhandy 
and we 
do not present it here. 
%neither present nor use it. 
\\
\\
%(2) 
$\bullet$ {\it The case \mbox{$c=d=0$}}. 

In this case, the two auxiliary variables $c_n$ and $d_n$ 
%disappear
vanish. 
Thus, 
the discrete-time system (\ref{fd-AL1}) reduces to 
\begin{equation}
\left\{
\hspace{2pt}
\begin{split}
& \frac{1}{h}\left(\widetilde{q}_n -q_n \right)  - a q_{n+1} 
 + b \widetilde{q}_{n-1} 
 - a (\widetilde{q}_n \widetilde{r}_{n-1} -q_{n+1} r_n)q_n  
- \alpha_{n} q_n 
+ \widetilde{q}_{n} \beta_n 
%- \alpha_{n+1} q_n + \widetilde{q}_{n} \beta_n 
 =0, 
\\[1mm]
& \frac{1}{h}\left(\widetilde{r}_n -r_n \right)  - b r_{n+1} 
 + a \widetilde{r}_{n-1} 
 - b (\widetilde{r}_n \widetilde{q}_{n-1} -r_{n+1} q_n)r_n 
- \beta_{n} r_n + \widetilde{r}_{n} \alpha_n 
%- \beta_{n+1} r_n + \widetilde{r}_{n} \alpha_n 
 =0, 
\\[1mm] &
\begin{split}
 & \alpha_{n+1} - \alpha_n  = 
	a (\widetilde{q}_n \widetilde{r}_{n-1} -q_{n+1} r_n), 
%\hspace{5mm} \lim_{n \to \pm \infty} \alpha_n =\alpha,
\\[1mm]
 & \beta_{n+1} - \beta_n = 
	b (\widetilde{r}_n \widetilde{q}_{n-1} -r_{n+1} q_n), 
%\hspace{5mm} 
%\lim_{n \to \pm \infty} \beta_n =\beta, 
%\lim_{n \to -\infty} \left( \alpha_n, \beta_n \right)
% = \left( \alpha, \beta \right) 
%\;\, \mathrm{or} \;\, 
%\lim_{n \to + \infty} \left( \alpha_n, \beta_n \right)
% = \left( \alpha, \beta \right),
\end{split}
%\label{}
\end{split}
\right.
\label{fd-AL2}
\end{equation}
where \mbox{$\lim_{n \to -\infty} \left( \alpha_n, \beta_n \right)
 = \left( \alpha, \beta \right)$} or \mbox{$
\lim_{n \to + \infty} \left( \alpha_n, \beta_n \right)
 = \left( \alpha, \beta \right)$}.
%;
%and 
The algebraic system (\ref{ALalg1}) simplifies
%boils down 
to 
\begin{equation}
\left\{
\hspace{2pt}
\begin{split}
&  1  - h a q_n \widetilde{r}_{n-1} 
 +h \beta_n 
= \left( 1 + h \beta \right) \Lambda_n \;\; \textrm{if} \;\; a \neq 0,
%\textrm{or} \;\; a=0,
\\[1mm]
&  1 - h b \widetilde{q}_{n-1} r_n 
 +h \alpha_n 
= \left( 1 + h \alpha \right) \Lambda_n \;\; \textrm{if} \;\; b \neq 0,
%\textrm{or} \;\; b=0, 
\\[1mm]
& h^2 ab \left( 1 - q_n r_n - \widetilde{q}_{n-1} \widetilde{r}_{n-1}\right)
 + \left( 1 + h \alpha_n \right)
   \left( 1 + h \beta_n \right)
% - h^2 a b q_n r_n - h^2 a b \widetilde{q}_{n-1} \widetilde{r}_{n-1} 
\\[-1mm] & 
= \left[ h^2 ab + (1+h \alpha)(1+h \beta) \right] \Lambda_n.
%\nonumber 
\end{split}
\right.
\label{ALalg2}
\end{equation}
System (\ref{fd-AL2}) is invariant under
the following transformation:
\mbox{$q_n \mapsto \mu^n q_n$}, \mbox{$r_n \mapsto \mu^{-n} r_n$},
\mbox{$ha \mapsto \mu^{-1} ha$}, \mbox{$hb \mapsto \mu hb$},
\mbox{$h\alpha \mapsto h\alpha$}, \mbox{$h\beta \mapsto h\beta$},
where $\mu$ is 
%any 
a nonzero constant.

%
%In fact, 
It can be easily verified 
%checked 
%confirmed 
that 
%in (\ref{ALalg2}) 
the first 
%(second) 
equality in (\ref{ALalg2}) 
%relation 
is 
also valid 
%even 
%also 
for 
%even if 
\mbox{$a=0$}, 
%(\mbox{$b=0$})
and the second equality is valid 
%holds true 
for \mbox{$b=0$}. 
%but 
%; however, this information is not necessary. 
%Thus, 
Using the first and second equalities, the third equality in (\ref{ALalg2}) 
results in a 
%(generally) 
quadratic equation in 
%for 
$\Lambda_n$,
\begin{align}
& \left( 1 + h \alpha \right) \left( 1 + h \beta \right) \Lambda_n^2 
-\left[ 
\left( 1 + h \alpha \right) \left( 1 + h \beta \right) + h^2 ab 
- \left( 1 + h \alpha \right)h a q_{n} \widetilde{r}_{n-1} 
\right. 
\nonumber \\
& \left. 
\mbox{} - \left( 1 + h \beta \right)h b \widetilde{q}_{n-1} r_{n}
\right] \Lambda_n 
+ h^2 ab \left( 1 - q_n r_n\right)
\left( 1- \widetilde{q}_{n-1} \widetilde{r}_{n-1}\right) =0.
\label{quadra1}
\end{align}
%\newpage
In terms of the normalized parameters 
\begin{equation}
%\widehat{a} 
\hat{a} := \frac{ha}{1 + h \beta}
, \hspace{5mm} \hat{b} :=\frac{hb}{1 + h \alpha}, 
\label{abhat}
\end{equation}
(\ref{quadra1}) can be rewritten as 
%is written simply as 
%reads as 
\begin{align}
& \Lambda_n^2 -\left( 1 + 
\hat{a} \hat{b} 
- \hat{a} q_{n} \widetilde{r}_{n-1} 
- \hat{b} \widetilde{q}_{n-1} r_{n}
\right) \Lambda_n 
+ \hat{a} \hat{b} \left( 1 - q_n r_n\right)
\left( 1- \widetilde{q}_{n-1} \widetilde{r}_{n-1}\right) =0. 
\label{quadra2}
%\nonumber
\end{align}
%It can 
%should 
%be recalled 
%noted 
We recall
that the Lax matrix $V_n$ given by (\ref{AL-Vn2}) 
with \mbox{$c_n=d_n=0$} 
is required to allow the asymptotic expansion 
%exhibit the asymptotic behavior 
(\ref{asymp-h}) for small $h$. 
This implies that the auxiliary variables 
%quantities 
$\alpha_n$ and $\beta_n$ 
determined by (\ref{ALalg2}) 
%and (\ref{quadra2}) 
are 
at most of order \mbox{$O(1)$}
%, 
and do not involve $1/h$. 
Thus, one of the two solutions of 
the quadratic equation (\ref{quadra2}) 
%should 
%can 
%be 
is rejected, 
%%taking into account 
%%recalling the prescribed 
%asymptotic behavior (\ref{asymp-h})
%of %$\Lambda_n$ as \mbox{$h \to 0$} (or \mbox{$n \to \pm \infty$}), 
and 
we obtain its 
%the 
proper 
%unique 
solution 
%of 
%%this 
%the quadratic equation (\ref{quadra2}), 
as 
\begin{equation}
%\frac{1}{\Lambda_n} = 
\Lambda_n = \frac{1 + A_n + \sqrt{\left( 1 + A_n \right)^2 
-4 B_n}}{2}, 
\label{Lambda1}
\end{equation}
with 
\begin{equation}
\begin{split}
A_n & := \hat{a} \hat{b} - \hat{a} q_{n} \widetilde{r}_{n-1} 
 - \hat{b} \hspace{1pt}\widetilde{q}_{n-1} r_{n}, 
\\
B_n & := \hat{a} \hat{b} \left( 1 - q_n r_n\right)
\left( 1- \widetilde{q}_{n-1} \widetilde{r}_{n-1}\right).
\end{split}
%\nonumber 
\label{AnBn}
\end{equation}
The decaying boundary conditions for $q_n$ and $r_n$ imply that 
\mbox{$\lim_{n \to - \infty} \Lambda_n =\lim_{n \to + \infty} \Lambda_n =1$}, 
and consequently,
\mbox{$\lim_{n \to - \infty} \left( \alpha_n, \beta_n \right)=
\lim_{n \to + \infty} \left( \alpha_n, \beta_n \right)
 = \left( \alpha, \beta \right)$}.
When \mbox{$\hat{a}\hat{b} \in \mathbb{R}$},
%this
the local expression (\ref{Lambda1}) is
valid 
only if 
%as long as 
\mbox{$\hat{a}\hat{b} \le 1$}. 
If \mbox{$\hat{a}\hat{b} > 1$},
the other solution of (\ref{quadra2})
should be employed;
%adopted. 
alternatively,
%otherwise,
one can understand
the right-hand side of (\ref{Lambda1}) as being defined by
%through
%its
%the 
a Taylor series for small $\hat{a}$ and $\hat{b}$,
and
its
%proper
analytic continuation. 
At the ``threshold'' value
\mbox{$\hat{a}\hat{b} =1$},
the discrete-time system (\ref{fd-AL2})
has the
%``trivial''
trivial solution \mbox{$\widetilde{q}_n = \hat{a} q_{n+1}$},
\mbox{$\widetilde{r}_n = \hat{b} r_{n+1}$},
\mbox{$\alpha_n = \alpha$}, \mbox{$\beta_n = \beta$}. 
Thus,
the discriminant
%in the square root
of the quadratic equation (\ref{quadra2})
vanishes at \mbox{$\hat{a}\hat{b}=1$},
and the two solutions indeed intersect. 
Unless \mbox{$\hat{a}\hat{b} =1$},
a unified expression for \mbox{$\Lambda_n$},
\[
\Lambda_n = \frac{1 + A_n + \bigl( 1-\hat{a}\hat{b} \hspace{1pt}\bigr)
\sqrt{\left( \frac{1 + A_n}{1-\hat{a}\hat{b}} \right)^2 
-\frac{4 B_n}{( 1-\hat{a}\hat{b} \hspace{1pt})^2}}}{2}, 
\]
can resolve the sign problem of the square root. 
In addition, 
%the above 
%square root 
this expression allows the Taylor expansion
%for small
with respect to
%perturbative
\mbox{$ \{ q_{n}, r_{n}, \widetilde{q}_{n-1}, \widetilde{r}_{n-1} \}$}.
%but 
However, we avoid the use of
%are not inclined to use
%do not use
%it looks
%%too
this 
unwieldy
%and 
%unattractive 
form. 
%formula.
%for

%
Using the first two relations in (\ref{ALalg2}), 
we can rewrite $\alpha_n$ and $\beta_n$ in the first two 
relations in (\ref{fd-AL2}) 
%to 
in terms of $\Lambda_n$, 
\begin{equation}
\left\{
\hspace{2pt}
\begin{split}
& \frac{1}{h}\left(\widetilde{q}_n -q_n \right)  
 + \beta \widetilde{q}_{n} - \alpha q_n 
 + \frac{1-q_n r_n }{\Lambda_n} 
 \left(- a q_{n+1}  + b \widetilde{q}_{n-1}\right) =0, 
\\[1mm]
& \frac{1}{h}\left(\widetilde{r}_n -r_n \right)  
+ \alpha \widetilde{r}_{n} - \beta r_n 
 + \frac{1-q_n r_n }{\Lambda_n} \left( 
 - b r_{n+1} + a \widetilde{r}_{n-1} \right) =0. 
\end{split}
\right.
%\nonumber 
\label{fdAL1}
\end{equation}
This system is linear in \mbox{$\widetilde{q}_n$} and 
\mbox{$\widetilde{r}_n$}. 
%We can solve 
%so that 
Thus, the 
%{\it explicit} 
%explicit 
%
forward 
time evolution 
%that is 
%{\it 
%explicit 
%under the 
%decaying 
%boundary conditions 
%(\ref{ALbc})
can be expressed 
%written 
%given 
as follows (cf.\ (\ref{abhat})): 
%described by the following mapping 
\begin{equation}
\left\{
\hspace{2pt}
\begin{split}
& \widetilde{q}_n = \frac{1+h \alpha}{1+h \beta}\hspace{1pt}q_n  
 + \frac{1-q_n r_n }{\Lambda_n} 
 \left( \hat{a} q_{n+1} - 
\frac{1+h \alpha}{1+h \beta} \hspace{1pt}\hat{b}\hspace{1pt} 
 \widetilde{q}_{n-1}\right), 
\\[1mm]
& 
\widetilde{r}_n = \frac{1+h \beta}{1+h \alpha}\hspace{1pt}r_n  
 + \frac{1-q_n r_n }{\Lambda_n} 
 \left(  \hat{b} r_{n+1} - 
\frac{1+h \beta}{1+h \alpha} \hspace{1pt} \hat{a} \hspace{1pt}
 \widetilde{r}_{n-1}\right), 
\end{split}
\right.
\label{ALmap1}
\end{equation}
where $\Lambda_n$ is given by 
(\ref{Lambda1}) with (\ref{AnBn}). 

In the simplest case of \mbox{$a=\hat{a}=0$} or \mbox{$b=\hat{b}=0$}, 
%\mbox{$ab= \hat{a} \hat{b}=0$}, 
%of \mbox{$a=0$} or \mbox{$b=0$}, 
the quadratic 
%algebraic 
equation (\ref{quadra2}) in 
%for 
$\Lambda_n$ reduces to a linear equation, 
and 
%consequently, 
the forward time evolution (\ref{ALmap1}) 
is 
%written 
%expressed as 
%becomes 
given by 
a simple rational mapping. 
%the mapping becomes rational 
This fact 
%result 
was 
%simple case recovers the result of S
discovered by Suris~\cite{Suris97',Suris00}; 
actually, 
slightly prior 
to his 
%Suris's 
work, 
%to~\cite{Suris97'}, 
%similar 
relevant results in a preliminary form 
were reported 
%obtained 
by Pempinelli, Boiti, and Leon~\cite{Pemp}. 
%in~\cite{Pemp}. 

%In the case where 
When $h$ is real, \mbox{$\beta= \alpha^\ast$}, and 
\mbox{$b = a^\ast$}, we can impose 
%consider 
%the reality (
the complex conjugacy reduction \mbox{$r_n 
%\pm  
%\propto 
=\sigma q_n^\ast$} with a real constant $\sigma$. 
%is allowed.
%is possible. 
In particular, setting 
%as 
\mbox{$\alpha= -\beta=-\mathrm{i} \gamma /\varDelta^2$}, 
\mbox{$a = -b=\mathrm{i}/\varDelta^2$}, 
and \mbox{$r_n= - \varDelta^2 q_n^\ast$}, we obtain 
%\mbox{$$}
the 
%a 
fully discretized NLS equation
\begin{equation}
\hspace{2pt}
\begin{split}
 \frac{\mathrm{i}}{h}\left(\widetilde{q}_n -q_n \right)  
 & + \frac{2 \left( 1 + \varDelta^2\left| q_n \right|^2 \right)}{
 1 + A_n + \sqrt{\left( 1 + A_n \right)^2 -4 B_n}} 
 \frac{q_{n+1} +\widetilde{q}_{n-1}}{\varDelta^2}
 -\frac{\gamma}{\varDelta^2} \left(\widetilde{q}_{n} + q_n \right)
%\\[-1mm] & \mbox{} 
 =0, 
\end{split}
\label{fdNLS1}
\end{equation}
%with 
where 
\begin{equation}
\begin{split}
%h 
A_n & = \frac{h^2}{
%\left( \frac{\varDelta^2}{h} - \mathrm{i}\gamma \right)
\left(\varDelta^2  + \mathrm{i}\gamma h \right)
\left( \varDelta^2 - \mathrm{i}\gamma h \right)}  
 + \frac{\mathrm{i} \varDelta^2 h 
}{\varDelta^2 + \mathrm{i}\gamma h } 
q_{n} \widetilde{q}_{n-1}^{\hspace{2pt}\ast}
 - \frac{\mathrm{i} \varDelta^2 h}{ \varDelta^2 - \mathrm{i}\gamma h}
\widetilde{q}_{n-1} q_{n}^{\ast}, 
\\
%h^2 
B_n & = \frac{h^2}{
\left(\varDelta^2  + \mathrm{i}\gamma h \right)
\left( \varDelta^2 - \mathrm{i}\gamma h \right)}  
\left( 1 + \varDelta^2 \left| q_n \right|^2 
\right)
\left( 1 + \varDelta^2 \left| \widetilde{q}_{n-1} \right|^2 \right).
\end{split}
%\nonumber 
%\label{AnBn2}
\nonumber
\end{equation}
Thus, 
this is equivalent to the fully discrete NLS equation (\ref{fdNLS2}) 
in the case \mbox{$a=b=0$}, 
up to the space reflection 
%space inversion // space reflection // spatial [spacial] inversion
%the inversion of the spacial variable 
\mbox{$n 
%\mapsto 
\to -n$}; this 
%relationship 
%equivalence 
correspondence 
can 
%also 
be readily 
noticed 
%uncovered 
%unveiled 
%found 
%more easily 
by 
%exploiting 
%looking into 
considering
%using 
the dispersion relation~\cite{AL77}. 
%~\cite{AL77}. 
%(see~\cite{AL77}). 
The choice of 
%case 
\mbox{$\gamma=1$} 
%of 
for the real 
%free 
parameter $\gamma$ 
is the most natural for 
%directly 
taking 
%considering 
the continuum limit, 
while the choice of 
%case 
\mbox{$\gamma=0$} simplifies the equation considerably. 
The aforementioned
condition \mbox{$\hat{a}\hat{b} \le 1$} implies that
%\mbox{$h^2 \le \varDelta^4 + (\gamma h)^2$};
\mbox{$ (1-\gamma^2) h^2 \le \varDelta^4 $};
this is automatically satisfied if \mbox{$\gamma^2 \ge 1$}.
%It is interesting to note 
%%should be noted 
%that, like the continuous NLS equation, 
%this full-discretization is homogeneous under the following weighting scheme: 
%\mbox{weight\hspace{1pt}$(\varDelta)=-1$}, 
%\mbox{weight\hspace{1pt}$(h)=-2$}, 
%\mbox{weight\hspace{1pt}$(q_n)=1$}. 
%%invariant possesses 
%%under the recaling 
%At present, 
%it is unclear 
%%at the present time 
%whether 
%the corresponding one-parameter group of scaling symmetries, 
%\mbox{$(\varDelta, h, q_n) \mapsto 
%(\varDelta/k, h/k^2, 
%%\mu 
%k q_n)$}, 
%can define 
%%a meaningful class of self-similar 
%meaningful ``self-similar'' solutions 
%%of 
%to (\ref{fdNLS1}). 
%
Note that 
%the two quantities 
$\widetilde{q}_{n-1}$ and $q_n$ are 
%not fully independent 
correlated and 
always satisfy the inequality \mbox{$ 1 + A_{n} \ge 2 \sqrt{B_{n}} $} 
if \mbox{$\varDelta^2 >0$}. 

%Note that 
%In order 
To 
%explicitly obtain 
%write out 
express the backward time evolution explicitly, 
%which 
%%is necessary for 
%%should be done for 
%%apears to be especially 
%is highly desired for 
%%in 
%%necessary in 
%the case \mbox{$h < 0$}, 
%(or, equivalently, \mbox{$h < 0$}), 
we have 
%need 
to rewrite $\Lambda_n$ in (\ref{fdAL1}) 
%to 
in terms of \mbox{$\Lambda_{n+1}$} 
using 
%with the help of 
(\ref{Lambda}), 
and then 
%\mbox{$\alpha_{n}$} and \mbox{$\beta_{n}$} 
%to \mbox{$\alpha_{n+1}$} and \mbox{$\beta_{n+1}$} 
%before substituting the local expressions for 
%\mbox{$\alpha_{n}$} and \mbox{$\beta_{n}$}. 
substitute 
%before substituting 
the local expression 
%for $\Lambda_{n}$ 
(\ref{Lambda1}) with (\ref{AnBn}). 
%With this effort, 
By this procedure, 
%Thus, 
the fully discrete NLS equation (\ref{fdNLS1}) 
in the case \mbox{$c=d=0$} 
%is equivalent to 
can also be 
%made coincide 
identified with 
the fully discrete NLS equation (\ref{fdNLS2}) 
in the case \mbox{$a=b=0$} 
%by 
through the time reflection \mbox{$m 
%\mapsto 
\to -m$} 
% and 
%as well as 
%including 
and 
%a 
the 
sign 
%change 
inversion 
of $h$, 
%namely, 
\mbox{$h \to -h$}~\cite{AL77,Suris97',Suris00}. 

%It should be noted 
Note that the auxiliary variables 
appearing in the stage of time discretization 
can be interpreted, after a minor redefinition, 
as giving the 
%conserved 
fluxes 
%in 
of 
the first few conservation laws. 
Indeed, 
(\ref{Lambda}) and (\ref{fd-AL2}) provide the 
following conservation
laws: 
\begin{subequations}
\begin{align}
& \boldsymbol{\Delta}_m^+
\log \left( 1 - q_n r_n \right)
%+ \Delta_n^+
= \boldsymbol{\Delta}_n^+
%\left( \mathrm{tr} \hspace{2pt}M_n \right)
\log \Lambda_n, 
%\left(\right), 
%\label{}
\\[1mm]
& \boldsymbol{\Delta}_m^+
\left(q_n r_{n-1} \right) 
= \boldsymbol{\Delta}_n^+ \left[ 
%\frac{1}{a} 
	\left(\alpha_n -\alpha \right)/a + q_n r_{n-1} \right], 
\label{cons2}
\\[1mm]
& \boldsymbol{\Delta}_m^+ \left(  q_{n-1} r_{n} \right) 
= \boldsymbol{\Delta}_n^+ \left[ \left(\beta_n -\beta \right)/ b 
	+ q_{n-1} r_n \right]. 
\label{cons3}
\end{align}
\end{subequations}
%$a$ and $b$ may depend on the discrete time $m$.
%In the derivation of the for the fluxes, 
The cases \mbox{$a=0$} 
%or 
and 
%the case 
\mbox{$b=0$} 
can be understood as 
%the 
limiting cases; 
%of the above expressions. 
%as 
%and 
the final 
local expressions for the fluxes 
do not contain 
%\mbox{$a^{-1}$} and \mbox{$b^{-1}$}, 
\mbox{$1/a$} and \mbox{$1/b$}, 
and also 
hold true 
for 
%the two 
these 
cases. 
%as well. 
%as they 
%do not contain \mbox{$1/a$} and \mbox{$1/b$}. 
%(The following computations should be checked.)
%Indeed, 
Using (\ref{Lambda}), 
(\ref{ALalg2}), (\ref{abhat}), (\ref{Lambda1}), and (\ref{ALmap1}), 
we can 
%locally 
express 
%write 
%rewrite 
the 
%last two 
fluxes in (\ref{cons2}) and (\ref{cons3})
%locally and 
explicitly 
as 
\begin{align}
& \frac{1}{a}\left(\alpha_n -\alpha \right) + q_n r_{n-1} 
= \frac{1+h \alpha}{ha}\left[
\Lambda_n-1 + \hat{b}\hspace{1pt} \widetilde{q}_{n-1} r_n
 \right] + q_n r_{n-1} 
\nonumber \\
&= \frac{1+h \alpha}{ha}\left[
 - \frac{  1 + A_n - \sqrt{\left( 1 + A_n \right)^2 -4 B_n} 
}{2} 
 + \hat{a} \hat{b} - \hat{a} q_{n} \widetilde{r}_{n-1}
 \right] + q_n r_{n-1} 
\nonumber \\
&= \frac{1+h \alpha}{1+ h\beta}\left[
 - \frac{2 \hat{b} \left( 1 - q_n r_n\right)
 \left( 1- \widetilde{q}_{n-1} \widetilde{r}_{n-1}\right)}
 {1 + A_n +\sqrt{\left( 1 + A_n \right)^2 -4 B_n}}
 + \hat{b} - q_{n} \widetilde{r}_{n-1}
 \right] + q_n r_{n-1}
\nonumber \\
&= \frac{h}{1+ h\beta}\left[
  \frac{2 \left( 1- \widetilde{q}_{n-1} \widetilde{r}_{n-1}\right)
\left( -  b + a q_{n} \widetilde{r}_{n-2}\right)}
 {1 + A_n +\sqrt{\left( 1 + A_n \right)^2 -4 B_n}}
 + b \right] , 
\nonumber 
%\\[3mm]
\end{align}
\vspace{-1mm}
%and 
\begin{align}
& 
 \frac{1}{b}\left(\beta_n -\beta \right) + q_{n-1} r_n 
= \frac{1+h \beta}{hb}\left[
\Lambda_n-1 + \hat{a} q_n \widetilde{r}_{n-1} 
 \right] + q_{n-1} r_n 
\nonumber \\
&= \frac{1+h \beta}{hb}\left[
 - \frac{  1 + A_n - \sqrt{\left( 1 + A_n \right)^2 -4 B_n} 
}{2} 
 + \hat{a} \hat{b} - \hat{b}\hspace{1pt} \widetilde{q}_{n-1} r_{n} 
 \right] + q_{n-1} r_n 
\nonumber \\
&= \frac{1+h \beta}{1+ h\alpha}\left[
 - \frac{2 \hat{a} \left( 1 - q_n r_n\right)
 \left( 1- \widetilde{q}_{n-1} \widetilde{r}_{n-1}\right)}
 {1 + A_n +\sqrt{\left( 1 + A_n \right)^2 -4 B_n}}
 + \hat{a} - \widetilde{q}_{n-1} r_{n} 
 \right] + q_{n-1} r_n 
\nonumber \\
&= \frac{h}{1+ h\alpha}\left[
  \frac{2 \left( 1- \widetilde{q}_{n-1} \widetilde{r}_{n-1}\right)
\left( - a + b \widetilde{q}_{n-2} r_{n} \right)}
 {1 + A_n +\sqrt{\left( 1 + A_n \right)^2 -4 B_n}}
 + a \right], 
\nonumber
\end{align}
where $A_n$ and $B_n$ are given by (\ref{AnBn}). 
Note that using (\ref{Lambda}) and (\ref{ALmap1}), 
we can further 
%still 
rewrite these fluxes 
%to 
in
%to 
many different forms. 
%ways. 
\\
\\
$\bullet$ 
{\it The general case} \mbox{$(a,b) \neq {\boldsymbol{0}}$}
and \mbox{$(c,d) \neq {\boldsymbol{0}}$}. 

%Among 
%Now we 
Let us consider the 
%time discretization (\ref{fd-AL1})of the 
(nonreduced) discrete-time 
%fully discretized 
Ablowitz--Ladik lattice (\ref{fd-AL1}) with (\ref{ALbc})
in the general case of \mbox{$(a,b) \neq {\boldsymbol{0}}$}
and \mbox{$(c,d) \neq {\boldsymbol{0}}$}. 
Before 
%considering 
%discussing 
dealing with 
the general
%generic 
case of the parameters 
$a$, $b$, $c$, 
%and 
$d$, $\alpha$, and $\beta$, 
we first 
%discuss 
consider 
the special case where the conditions 
%of 
\mbox{$c=-a \hspace{1pt}(\neq 0)$}, 
\mbox{$d=-b \hspace{1pt}(\neq 0)$}, 
and \mbox{$\beta=-\alpha$} are satisfied. 
This 
%case 
is a particularly interesting case 
from the 
%point of view of 
perspective of applications such as numerical experiments, 
%as 
because the linear part of the equations 
%of motion 
for $q_n$ and $r_n$ 
%(cf.~(\ref{fd-AL1})) 
%in this case 
has 
%the 
%%time reflection 
symmetry with respect to 
%is invariant under 
the time reflection \mbox{$m \to -m$} 
%and 
%together with 
combined with 
the sign inversion 
%change 
of $h$. 
%, \mbox{$h \to -h$}. 
%Note that in this special case 
Moreover, 
the 
%(anti-)
symmetry
%/antisymmetry 
with respect to the
%is invariant under 
space reflection \mbox{$n \to -n$} 
can 
also 
be achieved 
%in this special case 
by 
%further 
%assuming 
%requiring 
imposing the further condition 
%setting 
\mbox{$b= -a$}~\cite{Suris03,AL77,Taha1,Taha2,Suris00}. 
%or \mbox{$b= a$}~\cite{AL77,Taha1}. 
%$+a$~\cite{AL77,Taha1}. 
%
%It is interesting to investigate the discrete-time 
%flows that are symmetric or anti-symmetric with respect to 
%the inversion of the spatial direction \mbox{$n \to -n$} 
%and/or the temporal 
%direction \mbox{$m \to -m$}~\cite{AL77,Taha1,Taha2,Suris00}.
%
%\begin{equation}
%\left\{
%\hspace{2pt}
%\begin{split}
%&  ha  - h^2 a^2 q_n \widetilde{r}_{n-1} 
% +h^2 a \beta_n 
%+ h^2 a d q_{n-1} \widetilde{r}_n \Lambda_n 
%\\[-1mm] &
%\mbox{}+h c \Lambda_n \left[ h \alpha_n 
%+h a \left(\widetilde{q}_n \widetilde{r}_{n-1} +  q_n r_{n-1} \right)\right] 
%= \left[ ha + h^2 a \beta  + h^2 c \alpha \right] \Lambda_n,
%\\[1mm]
%& hb  - h^2 b^2 \widetilde{q}_{n-1} r_n 
% +h^2 b \alpha_n 
%+ h^2 b c \widetilde{q}_n r_{n-1} \Lambda_n 
%\\[-1mm] &
%\mbox{}+h d \Lambda_n \left[ h \beta_n 
%+h b \left(\widetilde{q}_{n-1} \widetilde{r}_n + q_{n-1} r_n \right)\right] 
%= \left[ hb + h^2 b \alpha  + h^2 d \beta \right] \Lambda_n,
%\\[1mm]
%& h^2 ab + h^2 cd \Lambda_n^{2}  
% + \left( 1+ h c \widetilde{q}_{n} r_{n-1} \Lambda_n + h \alpha_n \right)
%   \left( 1+ h d q_{n-1} \widetilde{r}_{n} \Lambda_n + h \beta_n \right)
%\\[-1mm] 
%& \mbox{}- h^2 \left( a q_n - c \widetilde{q}_{n} \Lambda_n \right) 
%  \left( b r_n - d \widetilde{r}_{n} \Lambda_n \right) 
%  - h^2 \left( b \widetilde{q}_{n-1} - d q_{n-1} \Lambda_n \right) 
%  \left( a \widetilde{r}_{n-1} - c r_{n-1} \Lambda_n \right)
%\\[-1mm] 
%& = \left[ h^2 ab + h^2 c d + (1+h \alpha)(1+h \beta) \right] \Lambda_n.
%%\nonumber 
%\end{split}
%\right.
%\label{ALalg4}
%\end{equation}
%

In the case of 
\mbox{$c=-a \hspace{1pt}(\neq 0)$}, 
\mbox{$d=-b \hspace{1pt}(\neq 0)$}, and 
\mbox{$\beta=-\alpha$}, 
the algebraic system (\ref{ALalg1}) 
can be reformulated
%rewritten 
as 
%reads as 
%is 
%
\begin{equation}
\left\{
\hspace{2pt}
\begin{split}
& \left( 1 - hb q_{n-1} \widetilde{r}_n \Lambda_n 
	+h\beta_n \right)
 - \Lambda_n \left( 1
   - ha \widetilde{q}_n r_{n-1}\Lambda_n+ h\alpha_n \right)
\\[-1mm] &
=  h a q_n \widetilde{r}_{n-1} +
\left[ - 2h\alpha 
%h \beta  - h\alpha 
+h a \left(\widetilde{q}_n \widetilde{r}_{n-1} +  q_n r_{n-1} \right)
%+ hb q_{n-1} \widetilde{r}_n 
\right] \Lambda_n 
+ ha \widetilde{q}_n r_{n-1}\Lambda_n^2,
\\[1mm]
& \left(1 -ha \widetilde{q}_n r_{n-1} \Lambda_n + h\alpha_n \right)
 - \Lambda_n \left( 1 - hb q_{n-1} \widetilde{r}_n \Lambda_n 
	+ h\beta_n \right)
\\[-1mm] &
= h b \widetilde{q}_{n-1} r_n + 
\left[ 2 h \alpha  
%h \alpha  - h\beta 
 +h b \left(\widetilde{q}_{n-1} \widetilde{r}_n + q_{n-1} r_n \right)
% + ha \widetilde{q}_n r_{n-1}
\right] \Lambda_n + hb q_{n-1} \widetilde{r}_n \Lambda_n^2 ,
\\[1mm]
&  \left( 1- h a \widetilde{q}_{n} r_{n-1} \Lambda_n + h \alpha_n \right)
   \left( 1- h b q_{n-1} \widetilde{r}_{n} \Lambda_n + h \beta_n \right)
\\[-1mm] 
& = 
%\mbox{}+ 
- h^2 ab \left( 1 - q_n r_n - \widetilde{q}_{n-1}\widetilde{r}_{n-1} 
\right)
- h^2 ab \left( 1-\widetilde{q}_{n} \widetilde{r}_{n} - q_{n-1} r_{n-1} 
\right) \Lambda_n^{2}  
\\[-1mm] 
& \hphantom{=} 
\mbox{} + \left[ h^2 ab  \left( 2 + q_n \widetilde{r}_{n} 
+\widetilde{q}_{n} r_n +\widetilde{q}_{n-1} r_{n-1} 
 + q_{n-1} \widetilde{r}_{n-1} 
\right) 
+ (1+h \alpha)(1-h \alpha) 
%(1+h \beta) 
\right] \Lambda_n, 
%=0.
%\nonumber 
\end{split}
\right.
\label{ALalg5}
\end{equation}
%
%where 
with \mbox{$c_n = -a \Lambda_n$} and 
\mbox{$d_n = -b \Lambda_n$}. 
%
%By eliminating 
Using the first two relations, 
we can express 
\mbox{$
%\left( 
1- h a \widetilde{q}_{n} r_{n-1} 
\Lambda_n + h \alpha_n 
%\right)
$} and \mbox{$
%\left( 
1- h b q_{n-1} 
\widetilde{r}_{n} \Lambda_n + h \beta_n 
%\right)
$} in terms of $\Lambda_n$. 
%etc.\ 
%from 
%in the third relation 
%equation 
%Thus, 
Substituting these expressions into the third relation, 
we obtain a sixth-degree equation 
%for one unknown 
in only
$\Lambda_n$. 
%Though we have not checked if it 
%%this equation 
%can be factored, 
%but in either case, 
%Unfortunately, 
However, 
this sextic equation 
%appears 
turns out to be too complicated 
for 
%%practical applications of the resulting time discretization. 
exact treatment. 
%to be treated exactly. 
Thus, we need 
%have 
to 
%, but it can be used in combination with 
%However, 
%explore some of the alternatives. 
take 
a 
different route. 
%routes. 
%find some other way
%Thus, we need to 
%propose two alternate use 
%consider alternative 
%take a different route. 
%
%One alternative 
One sensible approach 
%from a practical standpoint [viewpoint, perspective, point of view]
%from a practical application standpoint
%useful for numerical experiments of the time discretization 
is to solve the algebraic system 
(\ref{ALalg5}) approximately; for example, 
%indeed, 
%for three unknowns 
%by 
using perturbative expansions 
%successive approximations
with respect to $h$, 
%in powers of $h$, 
%with respect to powers of $h$, 
%it is easy to obtain an 
we 
%can 
%easily 
obtain 
%an 
%\mbox{$O(h)$} 
%or \mbox{$O(h^2)$} 
%the 
%\mbox{$O(h)$} 
approximate expressions 
%formula
%an \mbox{$O(h)$} or \mbox{$O(h^2)$} approximation 
%of 
for the auxiliary variables as follows: 
\begin{equation}
\left\{
\hspace{1pt}
\begin{split}
&  \alpha_n - a \widetilde{q}_n r_{n-1}\Lambda_n
= \alpha - \frac{a}{2} 
\left( \widetilde{q}_{n} +q_{n} \right)
\left( \widetilde{r}_{n-1} + r_{n-1} \right) +O(h), 
\\
%[1mm] 
&
\beta_n - b q_{n-1} \widetilde{r}_n \Lambda_n 
= -\alpha - \frac{b}{2} \left( \widetilde{q}_{n-1} +q_{n-1} \right)
\left( \widetilde{r}_{n} + r_{n} \right) +O(h),
\\
%[1mm] 
&
 \Lambda_n 
 = 1- \frac{ha}{2} 
\left( \widetilde{q}_{n} +q_{n} \right)
\left( \widetilde{r}_{n-1} + r_{n-1} \right)
- \frac{hb}{2} \left( \widetilde{q}_{n-1} +q_{n-1} \right)
\left( \widetilde{r}_{n} + r_{n} \right)
+O(h^2). 
\end{split}
\right.
\label{appr}
\end{equation}
%proper exact solution. 
%up to \mbox{$O(h)$}. 
%or \mbox{$O(h^2)$}. 
%Plunging 
%Substituting 
Combining (\ref{appr})
%these 
%them 
with 
%into 
(\ref{fd-AL1}) for 
%with 
\mbox{$c=-a$}, \mbox{$d=-b$}, 
and \mbox{$\beta=-\alpha$}, we obtain 
%arrive at 
%obtain 
a 
%rather 
high-precision 
%time discretization  
numerical scheme that is ``almost
%close to 
integrable'' 
(cf.~\cite{AL77,Taha1,Taha2}). 
%expected to be ``nearly integrable''. 
%algebraic system:  approximately integrable 
%we can obtain approximate solution in power series of $h$. 
%
%too complicated, but can be used to obtain an approximate scheme. 
%It is 
%, in principle, 
%straightforward to obtain 
We can also consider
%go to 
the next order 
in $h$
%, 
and include 
%obtain 
%the 
%next order 
higher-order corrections.  
%terms. 
%in the above 
%%asymptotic 
%expansions. 
%%corrections. 

%In 
%Let us consider 
%Now 
We now
discuss 
%study 
the general 
%generic 
case 
where
%in 
no particular 
%special 
%relations among 
conditions on 
the parameters $a$, $b$, $c$, $d$, $\alpha$, and $\beta$ 
are assumed. 
Instead of exploiting 
%studying 
%using 
the complicated algebraic system (\ref{ALalg1}), 
we 
%investigate 
%propose 
%study 
explain 
%a factorization of 
how to 
%factor 
recreate 
%reconstruct
the discrete-time 
%evolution 
Ablowitz--Ladik lattice 
%defined by 
(\ref{fd-AL1}) 
%into 
by 
%as 
the composition 
of a few 
more 
elementary 
%(but asymmetric) 
time evolutions such as 
%; two of them have been 
those studied earlier 
%as 
for the special cases 
%: 
\mbox{$a=b=0$} or \mbox{$c=d=0$}. 
%
%is too complicated to solve directly. 
%Thus, 
To this end, 
%aim, 
we 
%need 
%have 
%to 
%should 
recall some 
%an 
basic 
%important 
properties 
%feature 
of integrable hierarchies. 
That is, under suitable
%proper 
boundary conditions, 
an integrable hierarchy 
generally 
%under appropriate boundary conditions 
%consists of
comprises infinitely
%(infinitely) 
many 
%an infinite number of 
%mutually 
%pairwise 
%commutative 
%{\it 
commuting flows. 
%; 
%and 
Any composition of these flows also 
%again 
belongs 
to the same hierarchy; 
each 
%flow 
of the flows 
%defines an independent time evolution 
%; 
%and 
%the flow defines an independent time evolution, 
%that the importance of 
%the important role of the dispersion relation in an integrable hierarchy 
%of infinitely many commuting flows; that is, 
%the dispersion relation specifies the flow in a unique way. 
%each flow in 
%%a given integrable 
%the hierarchy 
is 
%{\it 
uniquely determined 
%designated 
%
%in a unique way 
%only 
by specifying its linear 
%leading 
part, or equivalently, 
%i.e., 
%the 
its dispersion relation. 
%In the small amplitude limit, 
%%of the dependent variables, 
Under the boundary conditions (\ref{ALbc}) 
or, equivalently, the streamlined version, 
the discrete-time Ablowitz--Ladik lattice (\ref{fd-AL1}) 
%under the boundary conditions (\ref{ALbc}) 
%(or, equivalently, the streamlined version)
has 
%, 
the following linear 
leading 
part: 
%reads as 
%
\begin{equation}
\left\{
\hspace{2pt}
\begin{split}
& \frac{1}{h}\left(\widetilde{q}_n -q_n \right) 
 + c \widetilde{q}_{n+1} + \beta \widetilde{q}_{n} + b \widetilde{q}_{n-1} 
 - a q_{n+1}- \alpha q_n - d q_{n-1}  
\sim 0,
\\
%[1mm]
& \frac{1}{h}\left(\widetilde{r}_n  -r_n \right)
 + d \widetilde{r}_{n+1} 
 +\alpha \widetilde{r}_{n} + a \widetilde{r}_{n-1} 
 - b r_{n+1}
 - \beta r_n - c r_{n-1}  
\sim 0.
\end{split}
\right.
\label{fd-ALlin}
\end{equation}
Substituting the ansatz 
\mbox{$q_n \sim \lambda^{2n}$}, 
\mbox{$\widetilde{q}_n \sim 
\lambda^{2n} \omega $}, 
\mbox{$r_n \sim \lambda^{-2n}$}, 
\mbox{$\widetilde{r}_n \sim \lambda^{-2n} \omega^{-1}$} into (\ref{fd-ALlin}), 
we obtain the corresponding dispersion relation~\cite{AL77}
%is obtained by 
%\[
%q_n \sim z^{2n}, \hspace{5mm} \widetilde{q}_n \sim z^{2n} \omega, 
%\hspace{5mm} r_n \sim z^{-2n}, \hspace{5mm} 
%\widetilde{r}_n \sim \omega^{-1} z^{-2n},
%\]
\begin{equation}
\omega \left( \lambda^2 \right)= 
\frac{1+ h \left( a \lambda^2 
+ \alpha + \frac{d}{\lambda^2}\right)}
{1+ h \left( c \lambda^2 + \beta + \frac{b}{\lambda^2}\right)}.
\label{ALdisp}
\end{equation}
%A probably more interesting and more rigorous 
%realization of this symmetric time evolution 
%is to factor it into a few more elementary (but asymmetric) 
%time evolutions 
%
%It should be noted 
Note that the Lax matrix $V_n$ (\ref{AL-Vn2}) 
approaches a diagonal matrix as 
\mbox{$n \to \pm \infty$} 
%\mbox{$n \to - \infty$} and/or \mbox{$n \to + \infty$} 
(cf.~(\ref{ALbc})), 
%, 
%or the streamlined version), 
and 
\mbox{$\omega \left( \lambda^2 \right)$} is equal to 
the limit value of the ratio between 
%of 
the 
%its 
%two 
diagonal elements~\cite{AL77}; 
%entries. 
this is 
%quite 
natural 
%reasonable 
%as 
because
%since 
the 
%an 
$n$-independent overall factor of $V_n$ 
plays no role in the zero-curvature condition 
(\ref{fd-Lax}). 

For notational convenience, we express the 
forward time evolution 
%time-evolution mapping 
defined by 
%the discrete-time Ablowitz--Ladik lattice 
(\ref{fd-AL1}) 
%under the boundary conditions 
with (\ref{ALbc}) 
as the mapping 
%%\mbox{$f_h (a,b,c,d,\alpha,\beta)$}, 
%\mbox{$f (a,b,c,d, 
%%;
%%h^{-1}+
%\alpha +1/h,
%%\frac{1}{h}
%%h^{-1}+ 
%\beta +1/h)$}, 
%namely, 
\[
%f_h (a,b,c,d,\alpha,\beta
%;h
{\textstyle f ( a,b,c,d, 
%;
\alpha+\frac{1}{h},\beta+\frac{1}{h} )}: 
%\
\hspace{2mm}
(q_n, r_n ) \mapsto 
  \left( \widetilde{q}_n, \widetilde{r}_n \right). 
\]
Note that 
%Actually, 
%Naturally, 
%it is 
%only 
only the ratio of the six 
arguments in $f$ 
%\mbox{$a:b:c:d: (\alpha +1/h):(\beta +1/h)$} 
%that 
%does 
%really 
%actually 
matters 
%specifies 
%determines 
in identifying 
%specifying 
the corresponding 
flow as well as its linear part. 
%so that the flow cannot uniquely determine 
%\mbox{$f_h (a,b,c,d,\alpha,\beta)$}. 
%However, this representation is not unique, but at least 
%$f$ can determine the mapping uniquely, which is sufficient for our purpose. 
Let us discuss how to factorize this mapping into a composition 
%product 
of 
%more 
%elementary 
%``simpler'' 
simpler
mappings 
%in 
belonging to 
the same class. 
%The 
A rigorous and 
%rigorous and complete treatment
%most 
persuasive discussion 
%argument 
can be given 
%done 
on the basis of 
%the inverse scattering method. 
%
%flows are commutative under the proper boundary conditions ..
%Indeed, 
the scattering data 
%appearing 
used in the inverse scattering formalism. 
%~\cite{AL77}. 
%They  
The scattering data 
are 
%$n$
space-independent quantities that 
%and 
have very simple time 
dependences determined by the dispersion relation, typically, 
\mbox{$\widetilde{R}(\lambda) = \omega \left( \lambda^2 \right) R 
(\lambda)$}~\cite{AL77}. Thus, 
%the time evolution of the scattering data 
%is essentially {\it linear} \/in each discrete-time flow 
their time evolution is 
essentially 
%{\it additive} 
%{\it 
linear 
%\/in every flow 
%in each discrete-time flow 
and allows a 
%{\it 
commutative 
%linear 
superposition of 
%for 
different flows. 
%
%The aforementioned basic properties (common sense/understanding)
%of an integrable hierarchy can be confirmed within the inverse scattering 
%formulation
%
%Indeed, 
%Moreover, the interrelations among 
%%apparently different 
%%seemingly 
%different 
Conversely, the decomposition of a complex flow into 
%simpler 
%more 
elementary 
%distinct 
%discrete-time 
flows 
%in the same hierarchy 
can 
%easily 
be 
%achieved 
uncovered 
%unveiled 
%noticed 
by examining
%looking into 
%%observing 
%the corresponding time dependences 
%%bahaviors 
%of the scattering data in the inverse scattering formalism~\cite{AL77}. 
the factorization 
%problem 
of the dispersion relation, 
%\mbox{$\omega \left( \lambda^2 \right)$}, 
%; 
that is, 
%That is, 
whether \mbox{$\omega \left( \lambda^2 \right)$} 
%determined 
%given by (\ref{ALdisp}) 
%may 
can 
be factorized into a product of simpler quantities 
such as 
%rewritten in the form 
%of a product 
\mbox{$
%\omega \left( \lambda^2 \right) = 
\omega_1 \left( \lambda^2 \right) \omega_2 \left( \lambda^2 \right)$} 
or 
\mbox{$
%\omega \left( \lambda^2 \right) = 
\omega_1 \left( \lambda^2 \right) \omega_2 \left( \lambda^2 \right)
  \omega_3 \left( \lambda^2 \right) $}. 
%within the same functional dependence on $\lambda$. 
%
%Thus we can easily discuss if 
%In other words, 
%If we recall that 
%a given 
%%(discrete) 
%%linear 
%dispersion relation 
%%is sufficient 
%uniquely designates the discrete-time flow in 
%%of 
%the hierarchy, 
%and thus we have only to discuss the factorization problem 
%of the dispersion relation. 
%
%To make the paper self-contained, we take a more intuitive way at the
%expense of mathematical rigor 
%
If the reader is unfamiliar with the inverse scattering 
method, the same 
%guiding principle 
guideline 
%is 
can be derived 
%drawn up 
%given 
%provided 
%worked out 
by noting 
%assuming 
the one-to-one 
%``one-to-one'' 
correspondence 
between 
%among 
(the linear part of) a 
%discrete-time 
%Ablowitz--Ladik 
flow 
%, its linear part (\ref{fd-ALlin}), 
and 
%the 
its dispersion relation. 
%(\ref{ALdisp}). 
One can also understand this guideline 
%it 
``honestly'' 
%on 
at the level 
of the associated linear problem 
(cf.\ the second equation in (\ref{line2}) 
and
%/or 
the sentence 
%below 
beneath (\ref{ALdisp})); that is, 
by considering whether the Lax matrix $V_n$ can be 
factorized into a product of simpler 
%$V_n$-
matrices 
such as 
\mbox{$
V_n^{(2)} \left( \lambda \right) V_n^{(1)} \left( \lambda \right) $} 
or 
\mbox{$
V_n^{(3)} \left( \lambda \right) 
V_n^{(2)} \left( \lambda \right) V_n^{(1)} 
\left( \lambda \right)$}~\cite{Suris03,Suris97,Suris97',Suris00}. 

%Anyway, 
%we can 
%the one-to-one correspondence 
%between the discrete-time 
%%Ablowitz--Ladik 
%flow and the dispersion relation (\ref{ALdisp}) 
%motivates us to 
%factor 
By elementary computations, 
the right-hand side of 
%in 
(\ref{ALdisp}) 
%is 
can be factorized into 
%can be written as 
%coincides with 
the product of two or three simpler 
%``elementary'' 
quantities 
(as functions of $\lambda$) 
%that have simpler structures 
as follows: 
%in the following way: 
%
%into a product of quantities 
%
%more intuitive, but still effective approach 
%
%Using the boundary conditions (\ref{ALbc}), 
%we can compute the limit value of 
%the ratio of the \mbox{$(1,1)$}-entry to the 
%\mbox{$(2,2)$}-entry of the matrix $V_n$ (\ref{AL-Vn2}) 
%as \mbox{$n \to \pm \infty$}. The result (their ratio) 
%can be factored into 
%
\begin{subequations}
\begin{align}
%\lim_{n \to \pm \infty} 
%\omega \left( \lambda^2 \right) &= 
%
& \hphantom{=} \hspace{5pt}
\frac{1+ h \left( a \lambda^2 + \alpha + \frac{d}{\lambda^2}\right)}
{1+ h \left(c \lambda^2 + \beta + \frac{b}{\lambda^2}\right)}
\nonumber \\
&= \frac{1+ h \alpha +\sqrt{(1+h\alpha)^2-4h^2 ad}
 + 2ha \lambda^2 }
{1+ h \beta +\sqrt{(1+h\beta)^2-4h^2 bc}
 + \frac{2hb}{\lambda^2}} \times 
%\cdot
\frac{1+\frac{2hd/\lambda^2}{1+ h \alpha +\sqrt{(1+h\alpha)^2-4h^2 ad}}
% \frac{1}{\lambda^2}
}{1+ \frac{2hc  \lambda^2}
{1+ h \beta +\sqrt{(1+h\beta)^2-4h^2 bc}}}
%\nonumber 
%\label{}
\\
&= \frac{1+ \frac{2ha \lambda^2}{1+ h \alpha +\sqrt{(1+h\alpha)^2-4h^2 ad}}}
{1 + \frac{2hb/\lambda^2}{1+ h \beta +\sqrt{(1+h\beta)^2-4h^2 bc}}
%\frac{1}{\lambda^2}
} 
\times 
%\cdot
\frac{1+ h \alpha +\sqrt{(1+h\alpha)^2-4h^2 ad}
+ \frac{2hd}{\lambda^2}}
{1+ h \beta +\sqrt{(1+h\beta)^2-4h^2 bc}+ 2hc \lambda^2}
%\label{}
\\
&= \frac{1+ h \alpha +\sqrt{(1+h\alpha)^2-4h^2 ad}}
{1+ h \beta +\sqrt{(1+h\beta)^2-4h^2 bc}}
\times 
\frac{1+ \frac{2ha \lambda^2}{1+ h \alpha +\sqrt{(1+h\alpha)^2-4h^2 ad}}}
{1 + \frac{2hb/\lambda^2}{1+ h \beta +\sqrt{(1+h\beta)^2-4h^2 bc}}
%\frac{1}{\lambda^2}
} 
\nonumber \\
& \hphantom{=} \hspace{5pt}
\times \frac{1+\frac{2hd/\lambda^2}{1+ h \alpha 
+\sqrt{(1+h\alpha)^2-4h^2 ad}}
% \frac{1}{\lambda^2}
}{1+ \frac{2hc  \lambda^2}
{1+ h \beta +\sqrt{(1+h\beta)^2-4h^2 bc}}}.
%\label{}
\end{align}
\label{o-factor}
\end{subequations}
Thus, 
%according to 
the aforementioned guideline 
%,  we obtain 
implies the corresponding decomposition of the mapping 
\mbox{${\textstyle f ( a,b,c,d,\alpha+\frac{1}{h},\beta+\frac{1}{h} )}$} 
into simpler ones, i.e.,
%%mappings 
%that is, 
%
\begin{subequations}
%\begin{left}
\begin{align}
& \hphantom{=} \hspace{7pt}
{\textstyle f ( a,b,c,d,\alpha+\frac{1}{h},\beta+\frac{1}{h} )} 
\nonumber \\
&= {\textstyle f ( a,b,0,0,\alpha'+\frac{1}{h},\beta'+\frac{1}{h} )} \circ
{\textstyle f ( 0,0,c',d',\frac{1}{h},\frac{1}{h} )} 
\\
&= {\textstyle f ( a',b',0,0,\frac{1}{h},\frac{1}{h} )} \circ
{\textstyle f ( 0,0,c,d,\alpha'+\frac{1}{h},\beta'+\frac{1}{h} )} 
\\
& = {\textstyle f ( 0,0,0,0,\alpha'+\frac{1}{h},\beta'+\frac{1}{h} )} \circ
{\textstyle f ( a',b',0,0,\frac{1}{h},\frac{1}{h} )} 
\circ
{\textstyle f ( 0,0,c',d',\frac{1}{h},\frac{1}{h} )}, 
\label{3products}
%\hspace{15mm}
\end{align}
%\end{left}
\label{f-decom}
\end{subequations}
where 
\begin{align}
%{\textstyle 
& 
%{\textstyle 
a':= 
\frac{2a}{1+ h \alpha +\sqrt{(1+h\alpha)^2-4h^2 ad}}
%} 
, \hspace{5mm} 
%{\textstyle 
b':= \frac{2b}{1+ h \beta +\sqrt{(1+h\beta)^2-4h^2 bc}}
%}
, 
%\hspace{5mm} 
\nonumber \\[2mm]
& 
%{\textstyle 
c':= \frac{2c}{1+ h \beta +\sqrt{(1+h\beta)^2-4h^2 bc}}
%}
, \hspace{5mm} 
%{\textstyle 
d':= 
\frac{2d}{1+ h \alpha +\sqrt{(1+h\alpha)^2-4h^2 ad}}
%}
, 
%\hspace{5mm} 
\nonumber \\[2mm]
& 
%{\textstyle 
\alpha' := 
\frac{\sqrt{(1+h\alpha)^2-4h^2 ad} - 1+ h \alpha}{2h}
%\frac{2 \alpha -2h^2 ad}{1- h \alpha+ \sqrt{(1+h\alpha)^2-4h^2 ad}}
%}
, \hspace{5mm} 
%{\textstyle 
\beta':= 
\frac{\sqrt{(1+h\beta)^2-4h^2 bc} -1+ h \beta}{2h}
%}
.
%\label{}
\nonumber
\end{align}
Note that the order of composition in (\ref{f-decom}) 
does not matter, 
as 
%because 
the order of multiplication in (\ref{o-factor}) does not matter. 
%Thus, 
%It is very interesting that 
Interestingly, 
the 
%original 
time evolution 
%that was a six-point scheme 
defined 
%described 
on the six points 
%with the cordinates 
\mbox{$(n+i, m+j)$}, 
%where 
%with 
\mbox{$i=-1, 0, 1$} and \mbox{$j=0, 1$}, 
%is equivalent to 
is 
%can be 
now written as the composition of two time evolutions, 
each of which is 
%defined 
%described on four points. 
a previously 
%discussed 
described 
%studied 
four-point scheme, and, if any (cf.~(\ref{3products})), 
%the case of (\ref{3products})), 
one trivial 
%ultralocal 
%mapping. 
time evolution. 
%It should be stressed that 
Moreover, 
if 
%when 
%the conditions 
$h$ is real, \mbox{$b=a^\ast$}, 
\mbox{$d = c^\ast$}, 
and \mbox{$\beta= \alpha^\ast$}, 
%then not only 
the original mapping 
\mbox{${\textstyle f ( a,b,c,d,\alpha+\frac{1}{h},\beta+\frac{1}{h} )}$}
%but also 
%the decomposition (\ref{f-decom}) is consistent with 
allows the reduction of the complex conjugate 
\mbox{$r_n=\sigma q_n^\ast$} ($\sigma$:\ real), 
%($\sigma$ is a real constant), 
%we can impose the reality (complex conjugacy) reduction 
and this reduction is 
%maintained 
consistent with 
%under 
%each 
every component 
%decomposed 
mapping 
%appearing 
involved 
in 
%the decomposition 
(\ref{f-decom}). 

Using the dispersion relation (\ref{ALdisp}), 
%as a representation, 
we can readily find the inverse mapping of 
\mbox{${\textstyle f ( a,b,c,d,\alpha+\frac{1}{h},\beta+\frac{1}{h} )}$}. 
Indeed, the trivial identity 
%evident 
\[
\frac{1+ h \left( a \lambda^2
+ \alpha + \frac{d}{\lambda^2}\right)}
{1+ h \left( c \lambda^2 + \beta + \frac{b}{\lambda^2}\right)}
\times 
\frac{1+ h \left( c \lambda^2 + \beta + \frac{b}{\lambda^2}\right)}
{1+ h \left( a \lambda^2
+ \alpha + \frac{d}{\lambda^2}\right)}
=1
%, 
\]
implies the nontrivial formula 
%relation 
%\begin{equation}
\[
{\textstyle f^{-1}( a,b,c,d,\alpha+\frac{1}{h},\beta+\frac{1}{h} )} 
={\textstyle f( c,d,a,b,\beta+\frac{1}{h},\alpha+\frac{1}{h} )}. 
%\label{}
%\end{equation}
\]
In particular, 
this 
%can 
%directly 
%explains 
confirms 
the already mentioned
%above-mentioned 
correspondence 
between 
the case \mbox{$a=b=0$} and the case \mbox{$c=d=0$} 
through the time reflection. 

%Note that 
%In fact, 
%Moreover, 
%The fact that 
Actually, 
the right-hand side of 
(\ref{ALdisp}) is already 
the product of the numerator and 
the inverse of the denominator; 
%%also 
this 
%fact 
implies the decomposition 
%formula 
%\begin{subequations}
\begin{align}
{\textstyle f ( a,b,c,d,\alpha+\frac{1}{h},\beta+\frac{1}{h} )} 
&= 
{\textstyle f ( a,0,0,d,\alpha+\frac{1}{h},\frac{1}{h} )} 
\circ 
{\textstyle f ( 0,b,c,0,\frac{1}{h},\beta+\frac{1}{h} )} 
\nonumber \\
&= 
{\textstyle f ( 0,b,c,0,\frac{1}{h},\beta+\frac{1}{h} )} 
\circ 
{\textstyle f ( a,0,0,d,\alpha+\frac{1}{h},\frac{1}{h} )}. 
\nonumber 
\end{align}
%\end{subequations}
%
Thus, 
using this formula, 
%property, 
%it is possible to 
%we 
one can further 
decompose (\ref{f-decom}) into ``elementary'' 
flows (cf.~\cite{Suris03,Suris97',Suris00}); 
however, 
%this 
such 
%elementary 
flows 
%is 
%each elementary flow 
%generally 
do not 
%allow 
%reduction 
%preserve 
maintain the complex conjugacy 
relation between $q_n$ and $r_n$, 
which may become 
%an unwelcome property 
%which appears to be a 
a serious bottleneck
%cause a severe problem
%serious flaw 
%and we do not think it as a useful way 
in practical applications. 

%
%A more rigorous and complete treatment of 
%this factorization is possible if we resort to the 
%inverse scattering method and look into 
%the time dependence of the scattering data. 
%Every information on the level of the scattering data 
%is essentially linear and we can establish the 
%factorization ...
%
%We use Greek letters to write the auxiliary variables. 

\subsection{The Volterra lattice}

The Volterra lattice 
\begin{equation}
u_{n,t} = u_n (u_{n+1}-u_{n-1})
%,
\label{Volt}
\end{equation} 
%Note that 
%(\ref{Volt}) 
is obtained 
from the Ablowitz--Ladik lattice (\ref{sd-AL}) through the reduction 
\mbox{$a=b=1$}, 
\mbox{$q_n= u_n-1$}, 
and 
\mbox{$r_n= -1$}. 
%\mbox{$a=b=1$}, 
%together with 
%
%In the same way, 
Thus, 
the time discretization of the 
%results on the discrete-time 
Volterra lattice 
%should 
can 
%in principle 
be 
%is 
obtained from the 
%results on the 
discrete-time 
Ablowitz--Ladik lattice 
in subsection~\ref{secAL} 
%by imposing 
through the reduction 
\mbox{$q_n= u_n-1$} and
%, 
\mbox{$r_n= -1$}
%\mbox{$r_n= \textrm{const.}$} 
%as well as 
%under the conditions 
together with 
\mbox{$a=b$}, 
%\mbox{$\alpha=\beta$}, \mbox{$c=d$}
\mbox{$\alpha_{n}=\beta_{n+1}$}, and \mbox{$c_n=d_n$}. 
%and consequently 
%\mbox{$c_n=d_n$}, \mbox{$\alpha_{n}=\beta_{n+1}$}
%Though 
%That this reduction is not consistent with 
The discrepancy between this reduction and the 
decaying boundary conditions (\ref{ALbc}) 
%assumed in 
%%the Ablowitz--Ladik case, 
%%lattice in 
%subsection~\ref{secAL}, 
%but it 
%this 
%not 
%by no means 
%not 
%there exists no 
%there is no essential difference. 
%a serious 
is nonessential 
in this regard. 
%problem. 
However, in the following, 
%but 
we 
%rather 
prefer to consider 
%deal with 
the Volterra lattice independently as an 
%independent 
interesting example. 
%model. 
The Lax pair for the
continuous-time
Volterra lattice (\ref{Volt}) is given by~\cite{Suris03,FT87}
\begin{subequations}
\begin{align}
L_n &= \left[
\begin{array}{cc}
\lambda & u_n \\
-1 & 0 \\
\end{array}
\right], 
\label{LV-Ln}
\\[2mm]
M_n &= \left[
\begin{array}{cc}
\lambda^2 + u_n & \lambda u_n\\
 -\lambda & u_{n-1}\\
\end{array}
\right].
\end{align}
\label{Lax-LV}
\end{subequations}
Indeed, the substitution of (\ref{Lax-LV}) into
the zero-curvature condition (\ref{Lax_eq})
results in (\ref{Volt}).

In the discrete-time case, we assume the Lax matrix $V_n$ 
of the following form: 
%consider the
\begin{equation}
V_n = I + h \left[
\begin{array}{cc}
\lambda^2 a + \alpha_n +\widetilde{u}_n b_n 
	 & \lambda \left( a u_n + \widetilde{u}_n b_n \right)\\
 -\lambda \left( a + b_n\right) 
	&  -\lambda^2 b_n 
%+\beta_n 
	+\alpha_{n-1} + \widetilde{u}_{n-1} b_{n-1} \\
\end{array}
\right]. 
\label{LV-Vn}
\end{equation}
Here,
%where
$\alpha_n$
%, $\beta_n$, 
and $b_n$ are auxiliary variables.
The zero-curvature condition
(\ref{fd-Lax}) for the Lax pair (\ref{LV-Ln}) and (\ref{LV-Vn})
%amounts to 
is equivalent to the following 
%difference-difference 
system of partial difference equations (cf.~\cite{Taha1}):
\begin{equation}
\left\{
\hspace{2pt}
\begin{split}
& \frac{1}{h}\left(\widetilde{u}_n -u_n \right) = \alpha_{n+1} u_n 
	- \widetilde{u}_n \alpha_{n-1}
%\beta_n 
	+ \widetilde{u}_{n+1} b_{n+1} u_n 
	-  \widetilde{u}_{n} \widetilde{u}_{n-1} b_{n-1}, 
\\
%[0mm]
& \alpha_{n+1} - \alpha_n = a \left(u_{n+1}-\widetilde{u}_{n} \right),
\\
%%[1mm]
%& \beta_{n+1} = \alpha_n,
%\\
%[1mm]
& \widetilde{u}_n b_n = b_{n+1} u_n.
\end{split}
\right.
\label{fd-LV1}
\end{equation}
%Actually,
The general form (\ref{fd-LV1})
of the 
%fully 
time-discretized Volterra lattice
is integrable
%even
for
%the case of
matrix-valued dependent
variables, 
%In the following,
%In what follows,
%However,
but in the following,
%in this paper,
%for simplicity,
we consider only
the case of scalar dependent variables.
%In addition to 
In view of (\ref{V+-}), 
%If
we impose the 
%constant 
%rapidly decaying
following boundary conditions for $u_n$, $\alpha_n$, and $b_n$: 
%that is,
\begin{equation}
\lim_{n \to \pm \infty}
% \left(
u_n =1
%c
%1
, \hspace{5mm}
\lim_{n \to \pm \infty} 
%\left( 
\alpha_n=0, \hspace{5mm}
%\beta_n, 
\lim_{n \to \pm \infty} b_n 
%\right)
%\hspace{5mm}
%\lim_{n \to \pm \infty} \beta_n
= 
%\left( 0, 
%\alpha, 
%\beta, 
b. 
%\right).
\label{LVbc}
\end{equation}
%Moreover, we 
%Note that 
%Any nonzero 
The 
%nonzero 
boundary value of $u_n$ 
is normalized 
by 
%re
scaling $a$ and $b$, and 
the boundary value of $\alpha_n$ is set as zero 
%assumed to that \mbox{$\alpha=\beta=0$} 
%without loss of generality 
by redefining 
%rescaling 
$h$. 
To be precise, the boundary conditions
(\ref{LVbc}) contain {\it redundant} \/information. 
Indeed, it can be shown that
the auxiliary variables have the same limit
values 
%as 
for \mbox{$n \to -\infty$} and \mbox{$n \to +\infty$}. 
Therefore, it is sufficient to assume
either \mbox{$\lim_{n \to - \infty}(\alpha_n,b_n)=(0,b)$}
or \mbox{$\lim_{n \to + \infty}(\alpha_n,b_n)=(0,b)$}.
In the continuum
%continuous
limit of time \mbox{$h \to 
%+
0$}, we obtain
%\[
\mbox{$\alpha_n \to a u_n -a$} and 
%\hspace{5mm} 
\mbox{$b_n \to b$}. 
%\]
Thus, 
in this 
%the 
limit, 
%of time
%\mbox{$h \to 0$} 
the time discretization
%full-discrete Ablowitz--Ladik lattice
(\ref{fd-LV1}) 
%certainly 
reduces to 
%the Volterra lattice 
\[
u_{n,t} = (a+b) u_n (u_{n+1}-u_{n-1}),
\]
which 
%coincides, 
%Indeed, 
%if \mbox{$a+b \neq 0$}, with 
%then 
%this 
is, if \mbox{$a+b \neq 0$}, 
equivalent to 
%conincides 
the Volterra lattice (\ref{Volt}), 
up to a 
%scaling 
redefinition of 
%$\partial_t$ 
$t$. 
%or $u_n$. 
%variables. 
%
Note that 
in the case \mbox{$a+b =0$}, (\ref{fd-LV1}) has the trivial 
solution \mbox{$\widetilde{u}_n=u_n$}, \mbox{$\alpha_n=a u_n -a$}, 
\mbox{$b_n=b$}. 

The determinant of the \mbox{$2 \times 2$} 
Lax matrix $L_n$ (\ref{LV-Ln})
%is 
can be immediately computed as \mbox{$ \det L_n = u_n$}. 
Using the recurrence formulas for $\alpha_n$ and $b_n$ 
%fully discretized system 
in (\ref{fd-LV1}), 
we can rewrite the Lax matrix $V_n$ (\ref{LV-Vn}) as
%to 
%the form 
an ultralocal
%``ultralocal'' 
form with respect to 
%in 
the auxiliary variables $\alpha_n$ and $b_n$. 
Thus, its 
%compute the 
\mbox{$2 \times 2$} determinant 
%of the \mbox{$2 \times 2$} matrix $V_n$ (\ref{LV-Vn}) 
is computed as
\begin{align}
& \hphantom{= \mbox{}}
\det V_n (\lambda)
\nonumber \\
&= \left[ 1+ h \left( \lambda^2 a + \alpha_n + \widetilde{u}_n b_n \right) 
  \right] \left[ 1 + h \left(-\lambda^2 b_n 
	+ \alpha_n -a u_n + a\widetilde{u}_{n-1} + b_n u_{n-1} \right) \right]
%\lambda^4 \hspace{2pt}h^2 a c_n
% + \frac{1}{\lambda^4} \hspace{1pt}h^2 b d_n
\nonumber \\
& \hphantom{=\mbox{}} 
+ h^2 \lambda^2 
	\left( a u_n + \widetilde{u}_n b_n \right) \left( a+ b_n \right) 
\nonumber \\
&= 
-\lambda^4 h^2 a b_n +\lambda^2 h \left[ 
%a + h a \left(\alpha_n -a u_n + a\widetilde{u}_{n-1} + b_n u_{n-1} \right) 
%\right.
%\nonumber \\
%& \hphantom{=\mbox{}} 
%\mbox{} - b_n 
%\left. 
%- h b_n \left( \alpha_n + \widetilde{u}_n b_n \right) 
%+ h \left( a u_n + \widetilde{u}_n b_n \right) \left( a+ b_n \right) 
\left( a - b_n \right) \left( 1+ h \alpha_n \right) 
+ h a^2 \hspace{1pt}\widetilde{u}_{n-1} + ha b_n \left( u_{n-1} 
 + \widetilde{u}_n + u_n \right) 
\right]
\nonumber \\
& \hphantom{=\mbox{}} 
+ \left[ 1+ h \left( \alpha_n + \widetilde{u}_n b_n \right) 
  \right] \left[ 1 + h \left(\alpha_n -a u_n + a\widetilde{u}_{n-1} 
	+ b_n u_{n-1} \right) \right].
\nonumber
\end{align}
%Thus, 
Therefore, 
the equality
(\ref{fd-cons1})
%together
combined with the boundary conditions (\ref{LVbc}) 
(or
%, 
the streamlined version as stated above) 
implies the set of relations
\begin{equation}
\left\{
\hspace{2pt}
\begin{split}
& b_n = b \Lambda_n,
\\[1mm]
&  \left( a - b_n \right) \left( 1+ h \alpha_n \right) 
+ h a^2 \hspace{1pt}\widetilde{u}_{n-1} + ha b_n \left( u_{n-1} 
 + \widetilde{u}_n + u_n \right) 
\\[-1mm] &
%\mbox{} 
= \left( a - b + h a^2 + 3 h a b \right) \Lambda_n,
\\[1mm]
& \left[ 1+ h \left( \alpha_n + \widetilde{u}_n b_n \right) 
  \right] \left[ 1 + h \left(\alpha_n -a u_n + a\widetilde{u}_{n-1} 
	+ b_n u_{n-1} \right) \right]
%\\[-1mm] & 
= \left( 1+ h  b \right)^2 \Lambda_n.
\end{split}
\right.
\label{LValg1}
\end{equation}
Here, \mbox{$\Lambda_n$} is the quantity that satisfies
%satisfying
the recurrence formula
%recursive relation
%\mbox{$
\begin{equation}
 \widetilde{u}_n \Lambda_n = u_n \Lambda_{n+1},
\label{LV-Lam}
\end{equation}
and has the {\it normalized}
%``normalized''
\/boundary value
%,
\mbox{$\lim_{n \to \pm \infty} \Lambda_n =1$}. 
More precisely,
%the normalization at the left end
%\mbox{$\lim_{n \to - \infty} \Lambda_n =1$} 
%guarantees
%\mbox{$\lim_{n \to + \infty} \Lambda_n =1$} 
%and vice versa. 
%or, equivalently,
only
one of the two conditions 
\mbox{$\lim_{n \to - \infty} \Lambda_n =1$} and 
\mbox{$\lim_{n \to + \infty} \Lambda_n =1$} is required,
in accordance
%compliance
with the boundary conditions for $\alpha_n$ and $b_n$.
Then, the other condition can be confirmed. 
Thus, $\Lambda_n$ can be written explicitly (but globally) 
%(globally) 
as
\[
\Lambda_n = \prod_{j=-\infty}^{n-1}
        \frac{\widetilde{u}_j}{u_j}
= \prod_{j=n}^{+\infty}
        \frac{{u}_j}{\widetilde{u}_j}.
\]
Note that
the first equality in (\ref{LValg1}) is valid
even if \mbox{$a=0$} (cf.\ (\ref{fd-LV1})). 
The algebraic system (\ref{LValg1})
essentially comprises two nontrivial relations for the
two unknowns
%, 
%:\ 
$\alpha_n$ and $\Lambda_n$.

Because of the nonzero boundary 
%conditions 
value of $u_n$ (\ref{LVbc}), it is rather cumbersome to identify 
%discuss 
the linear leading part 
%of 
%from 
in (\ref{fd-LV1})
%, 
and 
%as well as 
%consequently, 
the corresponding dispersion relation 
% (cite Manakov?), 
%, as we did in 
(cf.\ subsection~\ref{secAL}). 
Instead, 
%following Suris's approach/fashion~\cite{Suris03,Suris97'}, 
%Alternatively, 
we can compute the asymptotic form of the Lax matrix $V_n$ (\ref{LV-Vn}) 
as 
%a function of $\lambda$ in the limit 
%\mbox{$n \to \pm \infty$}, 
\mbox{$n \to - \infty$} or \mbox{$n \to + \infty$}
%, 
and consider 
%discuss 
its factorization 
%problem 
(cf.~\cite{Suris03,Suris97,Suris97',Suris00}). 
The results 
%implies 
imply that the cases \mbox{$a=0$} and 
%the case 
\mbox{$b=0$} form 
%a 
the basis for the general case; that is, 
the 
%discrete-
time evolution 
%in the general case 
%of 
%where 
for 
%with 
general 
%values of 
$a$ and $b$ 
is equivalent to 
%a proper 
the order-independent 
composition of 
%the two time evolutions corresponding to \mbox{$a=0$} and \mbox{$b=0$}, 
%the 
two 
%discrete-
time evolutions 
%for 
corresponding to \mbox{$a=0$} and 
%for 
\mbox{$b=0$}, respectively. 
%Moreover
In addition, these two 
%fundamental 
cases are related 
%with 
to 
each other through 
%a 
the 
space/time reflection, as 
we will see below. 
%
%All 
%These 
%%The same 
%results 
%%conclusion 
%can also be inferred 
%%drawn 
%from the results for
%%on 
%%the decomposition 
%%%factorization 
%%of the composite mapping into simpler ones 
%the discrete-time Ablowitz--Ladik lattice 
%given in subsection~\ref{secAL}. 

Now, we consider the two 
%special
fundamental cases:\
\mbox{$a=0$} or \mbox{$b=0$}.
\\
\\
$\bullet${\it The case \mbox{$a=0$}}, {$b \neq 0$}.

In this case, the 
%two 
auxiliary variable \mbox{$\alpha_n$} 
%is 
%%become 
%$n$-independent; in fact, the boundary conditions (\ref{LVbc}) 
%implies that it 
vanishes. 
%(cf.~(\ref{LVbc})). 
%: 
%i.e., 
%the boundary conditions 
%namely, 
%\mbox{$\alpha_n =0$}. 
Thus, the discrete-time system (\ref{fd-LV1}) reduces to 
\begin{equation}
\left\{
\hspace{2pt}
\begin{split}
& \frac{1}{h}\left(\widetilde{u}_n -u_n \right) = 
 b \Lambda_{n+1} u_n 
%\widetilde{u}_{n} \Lambda_{n} 
\left( \widetilde{u}_{n+1} - u_{n-1} \right),
\\
%%[1mm]
%& \beta_{n+1} = \alpha_n,
%\\
%[1mm]
& \widetilde{u}_n \Lambda_n = u_n \Lambda_{n+1}, 
\hspace{5mm} 
%\lim_{n \to \pm \infty} \Lambda_n =1
\lim_{n \to - \infty} \Lambda_n =1 \;\, {\mathrm{or}} \;\,
\lim_{n \to + \infty} \Lambda_n =1.
\end{split}
\right.
\label{fd-LV2}
\end{equation}
%and 
Note that 
in the 
%``critical'' 
case \mbox{$hb =-1$}, 
(\ref{fd-LV2}) has the trivial solution \mbox{$\widetilde{u}_n = u_{n-1}$}, 
\mbox{$\Lambda_n = 1/u_{n-1}$}. 
The algebraic system (\ref{LValg1}) 
%boils down 
simplifies to
a quadratic equation in 
%for 
$\Lambda_n$,
%\begin{equation}
\[
 \left( 1+ hb \widetilde{u}_n \Lambda_n \right) 
 \left( 1 + h b u_{n-1} \Lambda_n \right)
%\\[-1mm] & 
= \left( 1+ h  b \right)^2 \Lambda_n, 
%\label{LValg1.5}
%\end{equation}
\]
%
%In (\ref{LValg2}), using the first and second equalities, the third equality
%results in a
%%(generally)
%
%
or equivalently, 
\begin{equation}
(h b)^2 \hspace{1pt}\widetilde{u}_n u_{n-1} \Lambda_n^2 
- \left[ \left( 1+ h  b \right)^2 
  -hb \left( \widetilde{u}_n + u_{n-1} \right)  \right] \Lambda_n 
+ 1
%\\[-1mm] & 
= 0.
\label{LValg2}
\end{equation}
%
%Owing to 
The asymptotic behavior (\ref{asymp-h}) 
of the Lax matrix $V_n$ implies that 
%in small $h$
%asymptotic expansion of $V_n$ with respect to $h$
%(\ref{asymp-h}), 
%its 
the proper solution of (\ref{LValg2}) is given by 
%
%{\small 
\begin{equation}
%\Lambda_n = \frac{2}{\left( 1+ h  b \right)^2 
%  -hb \left( \widetilde{u}_n + u_{n-1} \right) + \sqrt{
% \left[\left( 1+ h  b \right)^2 
%  -hb \left( \widetilde{u}_n + u_{n-1} \right)  \right]^2 
%  -4 (h b)^2 \hspace{1pt}\widetilde{u}_n u_{n-1}}}
%
\left( 1+ h  b \right)^2 \Lambda_n = 
\frac{2}{1- \epsilon \left( \widetilde{u}_n + u_{n-1} \right) + \sqrt{
 \left[1 -\epsilon
 \left( \widetilde{u}_n + u_{n-1} \right)  \right]^2 
  -4 \epsilon^2 \hspace{1pt}\widetilde{u}_n u_{n-1}}},
\label{LV-Lam1}
\end{equation}
where \mbox{$ \epsilon := hb/\left( 1+ h  b \right)^2$}. 
Thus, if \mbox{$hb \in \mathbb{R}$}, then \mbox{$\epsilon \le 1/4$}. 
%; in this case, 
%and 
In this case, the local expression (\ref{LV-Lam1}) is valid 
%for 
only if \mbox{$-1 < hb 
%\le
<  1$}, 
which 
%still 
%corresponds to 
covers 
the 
%entire 
%whole 
range 
%of 
\mbox{$\epsilon < 
%\le 
1/4$}. 
%throughout. 
%the range \mbox{$\epsilon \le 1/4$}. 
%(\ref{LVbc}) 
%This 
%To be precise, 
The case \mbox{$hb=1$}, corresponding to 
\mbox{$\epsilon=1/4$}, 
%(\mbox{$\epsilon=1/4$}) 
involves 
%some subtleties
a subtle sign problem of the square root, 
%but 
%and we do not discuss it here. 
which we do not discuss here. 
%this issue. 
%exceptional case. 
%Indeed, 
If \mbox{$(hb)^2 > 1$}, 
%\mbox{$|hb| > 1$}, 
(\ref{LV-Lam1}) is inconsistent 
with the boundary conditions, and the other solution 
of (\ref{LValg2}) should be 
%used. 
adopted. 
%gives the proper expression. 
%employed. 
In any case, 
the boundary conditions for $u_n$ imply that
\mbox{$\lim_{n \to - \infty} \Lambda_n =\lim_{n \to + \infty} \Lambda_n =1$}. 
As $hb$ approaches
%gets closer to 
%approaches 
$1$, 
%the 
more 
%strict 
%%severe 
%conditions 
%restrictions 
restrictive conditions 
have to be imposed on 
%$u_n$'s for 
the $u_n$ to ensure 
%have to approach $1$ 
%the 
a consistent choice of the solution of (\ref{LValg2}). 

Substituting (\ref{LV-Lam1}) into the first equation 
in (\ref{fd-LV2}), we obtain a
%n integrable 
%local 
time discretization 
of the Volterra lattice (\ref{Volt}) in the local form, 
\begin{equation}
%\frac{1}{\epsilon}\left(\widetilde{u}_n -u_n \right) 
\frac{\widetilde{u}_n -u_n}{\epsilon}= 
\frac{2 u_n 
\left( \widetilde{u}_{n+1} - u_{n-1} \right)}
{1- \epsilon \left( \widetilde{u}_{n+1} + u_{n} \right) + \sqrt{
 \left[1 -\epsilon
 \left( \widetilde{u}_{n+1} + u_{n} \right)  \right]^2 
  -4 \epsilon^2 \hspace{1pt}\widetilde{u}_{n+1} u_{n}}}.
\label{loLV1}
\end{equation}
%
%This is an 
%%{\it 
%explicit scheme under the boundary conditions 
%\mbox{$\lim_{n \to \pm \infty} u_n =1$}, 
%%as 
%because 
Thus, the value of $\widetilde{u}_n$ is uniquely determined by 
%from 
%given in terms of 
$u_n$, $\widetilde{u}_{n+1}$, and $u_{n-1}$. 
%(\ref{LVbc}). 
Note that (\ref{loLV1}) can be rewritten 
as 
\begin{equation}
%\frac{1}{\epsilon}\left(\widetilde{u}_n -u_n \right) 
\frac{\widetilde{u}_n -u_n}{\epsilon}= 
\frac{2 
\left( \widetilde{u}_{n+1} \widetilde{u}_n - u_n u_{n-1} \right)}
{1+ \epsilon \left( \widetilde{u}_{n+1} - u_{n} \right) + \sqrt{
 \left[1 +\epsilon
 \left( \widetilde{u}_{n+1} - u_{n} \right)  \right]^2 
  -4 \epsilon \widetilde{u}_{n+1}}}.
\label{loLV1'}
\end{equation}
%Note that 
If \mbox{$hb \in \mathbb{R}$} and 
%$u_n$'s 
the $u_n$ are nonzero 
and real-valued at 
the initial time, 
%\mbox{$m=m_0$}, 
then 
%\mbox{$\epsilon \le 1/4$}. 
(\ref{fd-LV2}) implies that 
%the reality of 
the auxiliary variable $\Lambda_n$ is always real-valued. 
%As the result, 
Consequently, the discriminant of the quadratic equation (\ref{LValg2}) 
%is 
must be nonnegative. 
%non-negative. 
%takes values in $\mathbb{R}$. 
%\mbox{$\Lambda_n \in \mathbb{R}$}. 
%Therefore
Thus, if \mbox{$\epsilon < 
%\le 
1/4$}, 
%and 
%\mbox{$u_n \to 1$} 
\mbox{$u_n$} approaches $1$
sufficiently smoothly
and fast as \mbox{$n \to \pm \infty$}, 
%\mbox{$0 < \epsilon \le 1/4$} or \mbox{$\epsilon < 0$}, 
and the \mbox{$u_n$}
%'s 
%stay in 
%%vicinity
%%the 
%proximity 
are close to 
%in a neighborhood
$1$ at the initial time, 
then the real-valuedness of $u_n$ is 
%maintained 
preserved 
under the time evolution 
of the discrete-time Volterra lattice 
(\ref{loLV1}) or (\ref{loLV1'}). 
%The case \mbox{$\epsilon =0$} is trivially understood 
%as \mbox{$\widetilde{u}_n =u_n$}. 

%
%In order 
To express
%explicitly obtain
%write out 
the backward time evolution explicitly,
we only have to replace \mbox{$\Lambda_{n+1} u_n$} in
the first equation of (\ref{fd-LV2}) 
%by 
with \mbox{$\Lambda_n \widetilde{u}_n$}
%,
and then substitute the local expression (\ref{LV-Lam1}). 
The resulting equation is 
%reads as 
\begin{equation}
%\frac{1}{\epsilon}\left(\widetilde{u}_n -u_n \right) 
\frac{\widetilde{u}_n -u_n}{\epsilon}= 
\frac{2 \widetilde{u}_n 
\left( \widetilde{u}_{n+1} - u_{n-1} \right)}
{1- \epsilon \left( \widetilde{u}_n + u_{n-1} \right) + \sqrt{
 \left[1 -\epsilon
 \left( \widetilde{u}_n + u_{n-1} \right)  \right]^2 
  -4 \epsilon^2 \hspace{1pt}\widetilde{u}_n u_{n-1}}}.
\label{loLV2}
\end{equation}
\\
$\bullet${\it The case \mbox{$b=0$}}, \mbox{$a \neq 0$}.
%, \mbox{$a \neq 0$}.

In this case, the auxiliary variable $b_n$ vanishes. 
%\mbox{$b_n =0$}.
Thus, the discrete-time system (\ref{fd-LV1}) reduces to 
\begin{equation}
\left\{
\hspace{2pt}
\begin{split}
&  \frac{1}{h} 
 \left(\widetilde{u}_n -u_n \right) 
 +\alpha_n  \left(\widetilde{u}_n -u_n \right) 
= a \left(u_{n+1} u_n 
	- \widetilde{u}_n \widetilde{u}_{n-1} \right),
\\
& \alpha_{n+1} - \alpha_n = a \left(u_{n+1}-\widetilde{u}_{n} \right), 
\hspace{5mm} \lim_{n \to - \infty} \alpha_n = 0 \;\, 
\mathrm{or} \;\, \lim_{n \to + \infty} \alpha_n =0. 
\end{split}
\right.
\label{fd-LV3}
\end{equation}
Note that 
in the 
%``critical'' 
case \mbox{$ha =-1$}, 
(\ref{fd-LV3}) has the trivial solution 
%\mbox{$\widetilde{u}_{n-1} = u_{n+1}$}, 
%\mbox{$1+h \alpha_n = \widetilde{u}_{n-1}$},
\mbox{$1+h \alpha_n = {u}_{n+1}$}, 
\mbox{$\widetilde{u}_{n} = u_{n+2}$}. 
Thus, we can assume \mbox{$ha \neq -1$}. 
The algebraic system (\ref{LValg1}) 
%boils down 
simplifies to
\begin{equation}
\left\{
\hspace{2pt}
\begin{split}
&  1 + h  \alpha_n + h a\widetilde{u}_{n-1} 
  = \left( 1 + h a \right) \Lambda_n, 
%\;\;\; \mathrm{if} \;\, a \neq 0,
\\
& \left( 1+ h \alpha_n \right)
 \left( 1 + h \alpha_n - ha u_n + ha\widetilde{u}_{n-1} \right)
%\\[-1mm] & 
= \Lambda_n.
\end{split}
\right.
\label{LValg3}
\end{equation}
%The first relation is valid even if \mbox{$a=0$}, 
%though this case is trivial 
%%case 
%as \mbox{$\widetilde{u}_n =u_n$}. 
By eliminating $\Lambda_n$, 
(\ref{LValg3}) 
reduces to 
%the 
a quadratic equation 
%for 
in \mbox{$1+ h \alpha_n $}, 
\[
\left( 1 + h a \right) \left( 1+ h \alpha_n \right)^2 
- \left[ 1 + ha \left( 1+ ha \right) 
\left( u_n - \widetilde{u}_{n-1} \right)
 \right] \left( 1 + h \alpha_n \right) 
- h a\widetilde{u}_{n-1} =0.
\]
The proper solution of this quadratic equation is given by 
\begin{equation}
1+ h \alpha_n = \frac{1 + F_n -G_n 
+ \sqrt{\left( 1 + F_n -G_n \right)^2 + 4 G_n }}
{2\left( 1 + h a \right)}, 
\label{LV-alp}
\end{equation}
with 
\[
%\begin{equation}
\begin{split}
F_n &:= ha \left( 1+ ha \right) u_n,
\\
G_n &:= ha \left( 1+ ha \right) \widetilde{u}_{n-1}.
\end{split}
%\label{FnGn}
%\end{equation}
\nonumber
\]
When \mbox{$ha \in \mathbb{R}$}, 
the local expression (\ref{LV-alp}) is valid 
only if \mbox{$ha > 
%\ge 
-1/2$}; 
%(cf.~(\ref{LVbc})); 
%we exclude 
the borderline
%delicate 
%subtle 
%exceptional 
case \mbox{$ha = -1/2$} is excluded 
from our consideration. 
%and 
%otherwise, 
If \mbox{$ha < -1/2$}, 
the other solution of the quadratic equation 
should be employed. 
Unless \mbox{$ha = -1/2$}, 
a unified expression for \mbox{$1+ h \alpha_n$}, 
\[
1+ h \alpha_n = \frac{1 + F_n -G_n 
+ (1+2ha) \sqrt{
%\displaystyle 
\left( \frac{1 + F_n -G_n}{1+2ha} \right)^2 
+ \frac{4 G_n}{(1+2ha)^2 }}}
{2\left( 1 + h a \right)}, 
\]
can resolve the sign problem of the square root, 
but it 
%looks 
%too 
is unwieldy and looks unattractive. 
%for 
%In any case, 
%It should be noted 
Note that 
the boundary conditions for $u_n$ imply that 
\mbox{$\lim_{n \to - \infty} \alpha_n = \lim_{n \to + \infty} \alpha_n =0$},
and consequently,
\mbox{$\lim_{n \to - \infty} \Lambda_n =\lim_{n \to + \infty} \Lambda_n =1$}.

%The substitution of 
Substituting the local expression 
(\ref{LV-alp}) 
for \mbox{$1+ h \alpha_n$} 
%into \mbox{$1+ h \alpha_n$} 
into the first equation 
%relation 
%in 
of (\ref{fd-LV3}), 
%results in ... 
we obtain a 
%local 
time discretization of the Volterra lattice (\ref{Volt}). 
%When $ha$ is a constant independent of the discrete time $m$, 
%i.e., a constant, 
In terms of the new 
%single 
parameter 
\mbox{$\delta := ha \left( 1+ ha \right) 
%=:\delta
$},
we can write this time discretization as 
\begin{equation}
\frac{\widetilde{u}_n -u_n}{\delta} 
 = \frac{2 \left(u_{n+1} u_n - \widetilde{u}_n \widetilde{u}_{n-1} \right)}
{1 + \delta \left( u_n - \widetilde{u}_{n-1} \right)
 + \sqrt{\left[ 1+ \delta \left( u_n - \widetilde{u}_{n-1} \right)
  \right]^2 +4 \delta \widetilde{u}_{n-1} }}.
\label{loLV3}
\end{equation}
%Under the boundary conditions \mbox{$\lim_{n \to \pm \infty} u_n =1$}, 
%this is an 
%%{\it 
%explicit scheme 
%%with respect to the forward time evolution 
%%toward 
%in the forward time direction. 
%(\ref{LVbc}).
%as it 
%%can be solved with respect to 
%is linear in \mbox{$\widetilde{u}_n$}. 
%uniquely. 
Thus, 
%the value of 
$\widetilde{u}_n$ is uniquely determined by
$u_n$, $u_{n+1}$, and $\widetilde{u}_{n-1}$. 
Note that 
%\mbox{$ha \in \mathbb{R}$} implies the inequality \mbox{$\delta \ge -1/4$}. 
if 
%\mbox{$ha \in \mathbb{R}$} (or, 
%%and 
%\mbox{$ha \ge -1/2$}), 
\mbox{$ha > -1/2$}, 
then \mbox{$\delta 
%\ge
> -1/4$}. 
If this inequality is satisfied, 
%and 
\mbox{$u_n$} 
%\to 
approaches {$1$} 
sufficiently smoothly
%\mbox{$u_n$} stays a proper proximity of $1$ 
and fast as \mbox{$n \to \pm \infty$}, 
and the \mbox{$u_n$}
%'s stay in 
%the 
%neighborhood
%proximity 
are close to $1$ at the initial time, 
then the real-valuedness of $u_n$ is 
preserved 
%maintained 
under the time evolution
%of the discrete-time Volterra lattice 
(\ref{loLV3}). 
%
%Indeed, if \mbox{$ha \in \mathbb{R}$} and 
%$u_n$'s are 
%%nonzero and 
%real-valued at the initial time,
%then the auxiliary variable $h \alpha_n$ is always real-valued. 

%Indeed, 
%In order 
To 
%write 
obtain 
the backward time evolution explicitly, 
%in the case \mbox{$b=0$} 
%can be given explicitly by 
%can be written by 
we rewrite \mbox{$\alpha_{n}$} in the first equation of (\ref{fd-LV3}) 
%in terms of $\alpha_{n+1}$ using the second equation 
%to 
as \mbox{$\alpha_{n+1} - a \left(u_{n+1}-\widetilde{u}_{n} \right)$}
%, 
and then substitute the local expression (\ref{LV-alp}). 
The resulting equation is 
%reads as 
\begin{equation}
\frac{\widetilde{u}_n -u_n}{\delta} 
 = \frac{2 \left(u_{n+1} u_n - \widetilde{u}_n \widetilde{u}_{n-1} \right)}
{1 - \delta \left( u_{n+1} - \widetilde{u}_{n} \right)
 + \sqrt{\left[ 1+ \delta \left( u_{n+1} - \widetilde{u}_{n} \right)
  \right]^2 +4 \delta \widetilde{u}_{n} }}.
\label{loLV4}
\end{equation}
It is easy to see that 
%This equation 
(\ref{loLV3}) is equivalent to (\ref{loLV1'}) (or (\ref{loLV2})) through 
the space (or time) reflection and the identification 
%\mbox{$\epsilon \leftrightarrow -\delta$}. 
\mbox{$\delta \leftrightarrow -\epsilon$}. 
In the same manner, (\ref{loLV4}) can be identified with (\ref{loLV1'}) 
through the time reflection. 
%is equivalent to 
Thus, the forward/backward time evolution in the case \mbox{$b=0$} 
corresponds to the backward/forward 
time evolution in the case \mbox{$a=0$}, 
%\mbox{$=0$}, 
up to a redefinition of the 
%time step 
parameters. 
%through 
%
%discrete Volterra: \cite{NijCap,HiroTsuji2}
%
\\
\\
{\it Ultradiscretization~{\rm \cite{TakaMatsu95,ToTaMaSa}}.} 
We 
propose 
%discuss 
an 
%ultradiscretization
ultradiscrete analogue
%~\cite{TakaMatsu95,ToTaMaSa} 
of the 
%above 
%discrete-time 
time-discretized 
Volterra lattice in the case \mbox{$b=0$}. 
For the forward time evolution, we first rewrite (\ref{loLV3}) as 
\begin{equation}
%\[
\frac{\widetilde{u}_n}{u_n}
 = \frac{1 + \delta \left( u_n - \widetilde{u}_{n-1} \right) + 2\delta u_{n+1} 
 + \sqrt{1+ 2 \delta \left( u_n + \widetilde{u}_{n-1} \right)
  + \delta^2 \hspace{-1pt}\left( u_n - \widetilde{u}_{n-1} \right)^2 }}
{1 +  \delta \left( u_n + \widetilde{u}_{n-1} \right)
 + \sqrt{ 1+ 2 \delta \left( u_n + \widetilde{u}_{n-1} \right)
  + \delta^2 \hspace{-1pt}\left( u_n - \widetilde{u}_{n-1} \right)^2 }}.
\label{loLV5}
\end{equation}
%\]
%
Then, 
we assume \mbox{$\delta > 0$} so that the positivity of 
the dependent variable, \mbox{$u_n > 0$}, 
%for 
%all 
\mbox{$\forall \hspace{1pt}n \in {\mathbb Z}$}, 
can be 
%is 
preserved 
%maintained
in the time evolution. By 
%expressing 
%re-parametrizing 
reparametrizing 
the parameter and 
the dependent variable as \mbox{$\delta=: 
\exp(-L/\varepsilon)$} 
and \mbox{$u_n =: 
\exp(U_n /\varepsilon)$}, respectively, 
and taking the logarithm, 
the above equation becomes 
%reads as 
%results in 
%
\begin{subequations}
\begin{align}
\widetilde{U}_n - U_n
&= \varepsilon \log \left[ 
1 + \mathrm{e}^{\frac{U_n -L}{\scriptstyle \varepsilon}} 
 - \mathrm{e}^{\frac{\widetilde{U}_{n-1}-L}{\scriptstyle \varepsilon}} 
 + 2\hspace{1pt}\mathrm{e}^{\frac{U_{n+1}-L}{\scriptstyle \varepsilon}} 
 + \sqrt{X_n} \right] 
\nonumber \\
& \hphantom{=} \; \mbox{} 
-\varepsilon \log \left[ 
1 +  \mathrm{e}^{\frac{U_n -L}{\scriptstyle \varepsilon}} 
 + \mathrm{e}^{\frac{\widetilde{U}_{n-1}-L}{\scriptstyle \varepsilon}} 
 + \sqrt{X_n}
\right],
%\nonumber 
\\[1mm]
X_n 
& 
:=1+ 2 \hspace{1pt}\mathrm{e}^{\frac{U_n -L}{\scriptstyle \varepsilon}} 
 + 2 \hspace{1pt}\mathrm{e}^{\frac{\widetilde{U}_{n-1}-L}{\scriptstyle \varepsilon}} 
 + \left( \mathrm{e}^{\frac{U_n -L}{\scriptstyle \varepsilon}} 
 - \mathrm{e}^{\frac{\widetilde{U}_{n-1}-L}{\scriptstyle \varepsilon}} 
 \right)^2.
%\nonumber
\end{align}
\label{udLV0}
\end{subequations}
%
%In (\ref{udLV0}), 
%Note that 
%For simplicity, 
For brevity,
%'s sake,
the $\varepsilon$-dependence of 
%functions 
the $U_n$
%'s 
is suppressed. 
%in (\ref{udLV0}). 
%Thus, 
%In 
When taking the limit \mbox{$\varepsilon \to +0$}, 
%and using 
the well-known formula~\cite{ToTaMaSa,MaSa97,TaMa97}, 
%for taking the ultradiscrete limit 
%\mbox{$
\[
\lim_{\varepsilon \to + 0} \varepsilon \log \left( \sum_{j=1}^M 
 \mathrm{e}^{\frac{A_j ({\scriptstyle \varepsilon})}{\scriptstyle \varepsilon}}
\right) =
\max_{1 \le j \le M} \left( A_j 
%(0^+) 
\right),
%$}
\]
where \mbox{$A_j (\varepsilon)$}
%'s 
are 
%$\varepsilon$-independent 
real functions 
%of $\varepsilon$ 
%that are 
%%assumed to be smooth 
%continuous 
%%at 
%on some interval \mbox{$0\le\varepsilon < \varepsilon_0$}, 
%having a finite limit value 
allowing the 
one-sided limit 
%from the right 
%as \mbox{$\varepsilon \to +0$}, 
\mbox{$\lim_{\varepsilon \to +0} A_j (\varepsilon) =: A_j$}, 
%numbers, 
does {\it not} \/apply directly; however, the 
%main 
core idea 
%in the same spirit 
%forming 
behind
%behind
this 
%the above 
formula is 
still valid. 
%works. 
%and 
%By 
After simple consideration of 
the cases 
%distinctions 
\mbox{$U_n \gtreqless \widetilde{U}_{n-1}
%\gtreqqless
$} when 
%in the case 
%\mbox{$\max \hspace{1pt}(U_n, \,\widetilde{U}_{n-1}) > L$}, 
\mbox{$U_n > L$}, 
%we obtain the ultradiscrete limit of 
(\ref{udLV0}) is 
%seen 
shown to 
reduce, in 
the limit \mbox{$\varepsilon \to +0$}, to 
\begin{align}
\widetilde{U}_n - U_n
&=   \max \left(
% 0, \, 
f(U_n, \hspace{1pt}\widetilde{U}_{n-1}), \, U_{n+1} -L
 %\Theta 
%H (U_n -\widetilde{U}_{n-1}) \times (U_n -1) 
\right) 
%\nonumber \\
%& \hphantom{=} \;\; \mbox{}
 -\max \left(
 0, \, U_n -L, \, \widetilde{U}_{n-1}-L \right).
\label{udLV1}
\end{align}
Here, $U_n :=\lim_{\varepsilon \to +0} U_n (\varepsilon)$ 
is a real variable, 
and the boundary conditions \mbox{$\lim_{n \to \pm \infty} u_n =1$} 
translate into \mbox{$\lim_{n \to \pm \infty} U_n =0$}. 
%\mbox{$H(x)$} is 
The 
%unit 
%Heaviside step 
function \mbox{$f(U_n, \widetilde{U}_{n-1})$} is 
%is 
defined by 
\begin{align}
f(U_n, \hspace{1pt}\widetilde{U}_{n-1})
 := & \hspace{1pt}
\lim_{\varepsilon \to +0} \varepsilon \log \left[ 
1 + \mathrm{e}^{\frac{U_n -L}{\scriptstyle \varepsilon}} 
 - \mathrm{e}^{\frac{\widetilde{U}_{n-1}-L}{\scriptstyle \varepsilon}} 
  + \sqrt{X_n} \right] 
\nonumber \\[1mm]
= & \hspace{1pt}
\begin{cases}
U_n -L, & U_n > \widetilde{U}_{n-1} \;\, \mathrm{and} \;\, U_n > L\\
\widetilde{U}_{n-1} - U_{n-1} +g(\widetilde{U}_{n-2}, \hspace{1pt}U_{n-1})
%\frac{1}{2}
%1/2
, & 
%{\text{if}} 
U_n = \widetilde{U}_{n-1}>L \\
0, & U_n < \widetilde{U}_{n-1} \;\, \mathrm{or} \;\, U_n \le L
\end{cases}
\nonumber \\[1mm]
\ge & \hspace{1pt}\hspace{2pt} 0,
\nonumber 
\end{align}
where 
%the function 
\mbox{$g(\widetilde{U}_{n-2}, \hspace{1pt}U_{n-1})$} is 
defined in a similar manner as 
%similarly as 
%
\begin{align}
g(\widetilde{U}_{n-2}, \hspace{1pt}U_{n-1}) 
 := & \hspace{1pt}
\lim_{\varepsilon \to +0} \varepsilon \log \left[ 
1 +\mathrm{e}^{\frac{\widetilde{U}_{n-2}-L}{\scriptstyle \varepsilon}} 
  - \mathrm{e}^{\frac{U_{n-1} -L}{\scriptstyle \varepsilon}} 
  + \sqrt{X_{n-1}} \right] 
\nonumber \\[1mm]
= & \hspace{1pt}
\begin{cases}
\widetilde{U}_{n-2}-L, & 
\widetilde{U}_{n-2} > U_{n-1} \;\, \mathrm{and} \;\, \widetilde{U}_{n-2} >L\\
f(U_{n-2}, \hspace{1pt}\widetilde{U}_{n-3}), & 
%{\text{if}} 
\widetilde{U}_{n-2} = U_{n-1}>L \\
0, & \widetilde{U}_{n-2} < U_{n-1} \;\, 
	\mathrm{or} \;\, \widetilde{U}_{n-2} \le L
\end{cases}
\nonumber \\[1mm]
\ge & \hspace{1pt}\hspace{2pt} 0.
\nonumber
\end{align}
One might naively think
%/consider 
that 
%\mbox{$g(U_n,\widetilde{U}_{n-1})=(U_n -L)/2$} 
%at \mbox{$U_n =\widetilde{U}_{n-1}>1$}
\mbox{$f(U_n,\widetilde{U}_{n-1})$ at \mbox{$U_n =\widetilde{U}_{n-1}>L$}
is given by
%equal to 
$(U_n -L)/2$}; 
%and \mbox{$g(\widetilde{U}_{n-2}, \hspace{1pt}U_{n-1})$} at 
%\mbox{$\widetilde{U}_{n-2} = U_{n-1}>L$} is given by 
%\mbox{$(\widetilde{U}_{n-2} -L)/2$}
however, this is not true in general, because the equality 
\mbox{$U_n(+0) =\widetilde{U}_{n-1}(+0)$} does not 
imply \mbox{$U_n (\varepsilon) \equiv \widetilde{U}_{n-1} (\varepsilon) $}, 
%\mbox{$\forall \varepsilon$}
%for all 
\mbox{$0<\varepsilon \ll 1$}. 
%, and the same applies for \mbox{$\widetilde{U}_{n-2} = U_{n-1}$}. 
As indicated above, 
the correct 
%value of 
%expressions for 
value of \mbox{$f(U_n,\widetilde{U}_{n-1})$} 
%and \mbox{$g(\widetilde{U}_{n-2}, \hspace{1pt}U_{n-1})$} 
%computation 
%for this case 
%evaluation 
can be computed 
%``
recursively
%'' 
using the 
%two 
formulas 
%as above 
%performed 
%as follows. 
\begin{equation}
f(U_{n+1}, \hspace{1pt}\widetilde{U}_{n})= 
\widetilde{U}_{n} - U_{n} +g(\widetilde{U}_{n-1}, \hspace{1pt}U_{n}), 
\hspace{5mm}
g(\widetilde{U}_{n}, \hspace{1pt}U_{n+1})
=f(U_{n}, \hspace{1pt}\widetilde{U}_{n-1}). 
\label{fg-recur}
\end{equation}
%
%
%Now, we 
We now 
%verify 
briefly explain how these 
%two 
formulas can be 
%are 
derived in the most general case, i.e., 
without any assumptions on the arguments 
such as \mbox{$U_{n+1}=\widetilde{U}_{n}$}. 
%, \mbox{$\hspace{1pt}\widetilde{U}_{n}=U_{n+1}$}, 
%etc. 
Note that (\ref{fg-recur}) can be regarded as 
%giving 
a conservation law. 
Thus, in view of the zero boundary conditions for $U_n$, 
the following global expressions are 
%implied:
%derived: 
valid if \mbox{$L > 0\,$}: 
\begin{align}
f(U_{n+1}, \hspace{1pt}\widetilde{U}_{n}) &= 
\sum_{j=0}^\infty \left( \widetilde{U}_{n-2j} - U_{n-2j} \right)
= \sum_{j=1}^\infty \left( U_{n+2j} - \widetilde{U}_{n+2j} \right), 
\nonumber \\
g(\widetilde{U}_{n}, \hspace{1pt}U_{n+1}) &=
\sum_{j=0}^\infty \left( \widetilde{U}_{n-2j-1} - U_{n-2j-1} \right)
= \sum_{j=1}^\infty \left( U_{n+2j-1} - \widetilde{U}_{n+2j-1} \right).
\nonumber 
\end{align}
To verify (\ref{fg-recur}), 
we compare (\ref{loLV5}) with 
(\ref{LV-Lam}), wherein the expression for 
$\Lambda_n$ is given by the first equation in 
(\ref{LValg3}) 
%and 
with 
(\ref{LV-alp}). 
This results in the 
%a 
nontrivial 
relation 
%equality
\begin{align}
%&\, 
& 1 + \delta \left( \widetilde{u}_{n} - u_{n+1} \right) 
 + \sqrt{1+ 2 \delta \left( u_{n+1} + \widetilde{u}_{n} \right)
  + \delta^2 \hspace{-1pt}\left( u_{n+1} - \widetilde{u}_{n} \right)^2 }
\nonumber \\ &=
%& \, 
1 + \delta \left( u_n - \widetilde{u}_{n-1} \right) 
 + \sqrt{1+ 2 \delta \left( u_n + \widetilde{u}_{n-1} \right)
  + \delta^2 \hspace{-1pt}\left( u_n - \widetilde{u}_{n-1} \right)^2 }. 
\nonumber 
\end{align}
It is an easy 
% simple 
exercise to rewrite this relation 
%equation 
%through rationalization of the 
by rationalizing its 
numerator as 
\begin{align}
%&\, 
& 1 + \delta \left( u_{n+1} - \widetilde{u}_{n}\right) 
 + \sqrt{1+ 2 \delta \left( u_{n+1} + \widetilde{u}_{n} \right)
  + \delta^2 \hspace{-1pt}\left( u_{n+1} - \widetilde{u}_{n} \right)^2 }
\nonumber \\ &= 
\frac{\widetilde{u}_{n}}{u_n}
% \times
\left[
%& \, 
1 + \delta \left( \widetilde{u}_{n-1} - u_n \right) 
 + \sqrt{1+ 2 \delta \left( u_n + \widetilde{u}_{n-1} \right)
  + \delta^2 \hspace{-1pt}\left( u_n - \widetilde{u}_{n-1} \right)^2 } \
\right]. 
\nonumber 
\end{align}
By substituting 
\mbox{$\delta=
\exp(-L/\varepsilon)$}
and \mbox{$u_n =
\exp(U_n /\varepsilon)$}, and 
taking the logarithm, 
the above two 
%relations 
equalities 
reduce, in
the limit \mbox{$\varepsilon \to +0$}, to the second 
and first 
%relations 
equalities 
%equations 
in (\ref{fg-recur}), respectively. 

%
%Note that
The parameter $L$ may
%be time-dependent.
depend on the discrete time coordinate \mbox{$m \in {\mathbb Z}$}. 
%It is to be hoped that the issue of 
%It remains to 
The {\it commutativity} \/of the ultradiscrete
flows defined by (\ref{udLV1})
%with
for
%various
distinct values of $L$ 
%remains to be completely elucidated. 
remains to be established. 
In this connection, 
%In connection with this, 
we conjecture that (\ref{udLV1}) and its time reversal (see below) 
%form 
comprise 
an ultradiscrete analogue of 
the KdV hierarchy, wherein the parameter $L$ labels 
%specifies 
each flow in the hierarchy. 
%will be investigated. 
%
%indeed give commutative flows
%It remains to be clarified how 
It is also beyond the scope of this paper to investigate 
the relationship between 
%properties of 
the ultradiscretized 
%equation 
Volterra lattice 
(\ref{udLV1})
%, in particular, 
%%
%if and how it can be related with 
%%its relationship with 
and the ``box
%--ball 
%$\&$ 
and ball 
system'' 
%proposed in
%by 
of Takahashi
%--
and Satsuma~\cite{TakaSatsu90}. 

For the backward time evolution, 
%(\ref{loLV4}), 
we 
%can 
rewrite (\ref{loLV4}) as 
\[
\frac{u_n}{\widetilde{u}_n}
 = \frac{1 + \delta \left( \widetilde{u}_{n} - u_{n+1} \right) 
 + 2\delta \widetilde{u}_{n-1} 
 + \sqrt{1+ 2 \delta \left( \widetilde{u}_{n} + u_{n+1} \right)
  + \delta^2 \hspace{-1pt}\left( \widetilde{u}_{n} - u_{n+1}\right)^2 }}
{1 +  \delta \left( \widetilde{u}_{n} + u_{n+1} \right)
 + \sqrt{ 1+ 2 \delta \left( \widetilde{u}_{n} + u_{n+1}  \right)
  + \delta^2 \hspace{-1pt}\left( \widetilde{u}_{n} - u_{n+1} \right)^2 }}.
%\label{loLV6}
%\end{equation}
\]
Thus, 
%following the same procedure 
in the same way as that for the forward 
time evolution, we obtain 
%arrive at 
an ultradiscrete analogue 
of (\ref{loLV4}),
\begin{align}
U_n - \widetilde{U}_n
&=   \max \left(
% 0
g (\widetilde{U}_{n}, \hspace{1pt}U_{n+1}), \, \widetilde{U}_{n-1} -L
%, \, 
 %\Theta 
%H ( \widetilde{U}_{n} - U_{n+1}) \times (\widetilde{U}_n -L) 
\right) 
%\nonumber \\
%& \hphantom{=} \;\; \mbox{}
 -\max \left(
 0, \, \widetilde{U}_n -L, \, U_{n+1}-L \right), 
%\label{udLV2}
\nonumber
\end{align}
%
%This equation 
%result is evident 
%can also be from 
which is related to
%with 
(\ref{udLV1}) 
%the 
%%symmetry with respect to 
%correspondence between (\ref{loLV3}) and (\ref{loLV4}) 
through the combined space
%-time 
and time reflection 
(cf.\ 
%correspondence between 
(\ref{loLV4}) and (\ref{loLV3})). 

\subsection{The modified Volterra lattice}

The modified Volterra lattice 
\begin{equation}
q_{n,t} = (1+q_n^{2}) (q_{n+1}-q_{n-1}),
\label{mLV}
\end{equation} 
where $q_n$ is a scalar dependent variable, 
was introduced by Hirota~\cite{Hiro73} 
(also see~\cite{AL0,AL1,Wadati76}); 
%This lattice can be 
it is obtained from 
%the reduction of 
the Ablowitz--Ladik system (\ref{sd-AL}) 
through the reduction \mbox{$a=b=1$}
%, 
and \mbox{$r_n=-q_n$}. 
%~\cite{AL0,AL1}. 
The overall coefficient
%sign 
of the nonlinear terms, including its sign, 
is nonessential
%not essential 
at the level of the equation 
%of motion 
and the associated 
Lax pair, because
%as 
it can be 
changed by 
%the 
%replacement 
rescaling \mbox{$q_n$}.
% \to \mathrm{i} q_n$}. 
%It should be noted 
Note that (\ref{mLV}) is 
invariant under the transformation 
\mbox{$q_n \mapsto (-1)^n q_n$}, \mbox{$t \mapsto -t$}. 
The Lax pair for the 
continuous-time 
modified Volterra lattice (\ref{mLV})
is given by 
%\begin{subequations} 
%\begin{align}
%& \\
%& \\
%\label{}
%\end{align}
%\end{subequations}
%
\begin{subequations}
\begin{align}
L_n &
= \left[
\begin{array}{cc}
 \lambda & q_n \\
 -q_n & \frac{1}{\lambda} \\
\end{array}
\right], 
\label{AL-Ln2}
\\[3mm]
M_n &= 
\left[
\begin{array}{cc}
 \lambda^2  
%- 1
+q_n q_{n-1}
%- (1-q_n q_{n-1}) 
& \lambda q_n + \frac{1}{\lambda} q_{n-1}\\
 -\lambda q_{n-1} - \frac{1}{\lambda} q_{n} & 
%- 1
 q_n q_{n-1} + 	\frac{1}{\lambda^2} 
%- (1- q_n q_{n-1})
\\
\end{array}
\right]. 
\end{align}
\label{Lax-mLV}
\end{subequations}
Indeed, the substitution of (\ref{Lax-mLV}) into 
the zero-curvature condition (\ref{Lax_eq}) 
results in (\ref{mLV}). 

In the discrete-time case, we consider the 
%following 
reduction \mbox{$a=b$}, \mbox{$\alpha_n=\beta_n$}, 
\mbox{$c_n=d_n$}, and \mbox{$r_n=-q_n$} 
of 
%in 
the 
Lax 
matrix $V_n$ (\ref{AL-Vn2}). Thus, 
we obtain 
\begin{equation}
V_n = I + h 
\left[
\begin{array}{c|c}
\begin{array}{l}
\lambda^2 a - \widetilde{q}_n q_{n-1} c_n 
\\ \mbox{}
 + \alpha_n +\frac{1}{\lambda^2} c_n 
\end{array}
& 
\begin{array}{l}
\lambda (a q_n - \widetilde{q}_n c_n) 
\\ \mbox{}
 +\frac{1}{\lambda_{\vphantom \sum}}(a\widetilde{q}_{n-1} -c_n q_{n-1}) 
\end{array}
%\vspace{1pt}
\\
\hline
%\vspace{3pt}
\begin{array}{l}
 \lambda (-a\widetilde{q}_{n-1} +c_n q_{n-1})  
\\ \mbox{}
	+ \frac{1}{\lambda}(-a q_n + \widetilde{q}_n c_n) 
\end{array} 
& 
\begin{array}{l}
\lambda^{2^{\vphantom \sum}} c_n - \widetilde{q}_n q_{n-1} c_n 
\\ \mbox{}
+ \alpha_n +\frac{a}{\lambda^{2}} 
%\\
\end{array}
\end{array}
\right],
\label{AL-Vn4}
\end{equation}
%Here,
where 
$\alpha_n$ and $c_n$ are 
the auxiliary variables. 
%Then, 
The zero-curvature condition 
(\ref{fd-Lax}) 
for the Lax pair (\ref{AL-Ln2}) and (\ref{AL-Vn4})
amounts to the following 
%difference-difference 
system of partial difference equations:
%of difference-difference equations
%
%\begin{subequations}
\begin{equation}
\left\{
\hspace{2pt}
\begin{split}
&
\begin{split}
\frac{1}{h}\left(\widetilde{q}_n -q_n \right) & - a q_{n+1} 
 + a \widetilde{q}_{n-1} - \alpha_{n+1} q_n + \widetilde{q}_{n} \alpha_n 
%\nonumber 
\\[-1mm]
& \mbox{}
 + \widetilde{q}_{n+1} c_{n+1} (1+ q_n^2) 
 -(1+\widetilde{q}_n^{\hspace{2pt}2}) c_n q_{n-1}=0, 
\end{split}
%\right.
\\[1mm] &
%\end{equation}
%\begin{equation}
%\left\{
\begin{split}
 & \alpha_{n+1} - \alpha_n  = 
	a\left(- \widetilde{q}_n \widetilde{q}_{n-1} + q_{n+1} q_n \right), 
\\[1mm]
 & (1 + \widetilde{q}_n^{\hspace{2pt}2}) c_n =(1+{q}_n^2) c_{n+1}.
\end{split}
%\label{}
\end{split}
\right.
\label{fd-mV1}
\end{equation}
%\label{fd-AL1}
%\end{subequations}
%
%In fact, 
The discussion in 
%the previous 
subsection~\ref{secAL} 
implies that 
the general case of 
\mbox{$a \neq 0$} and \mbox{$c_n \neq 0$} 
%can be reproduced 
%reconstructed 
%by taking the composition of the above simple case at 
%as 
is equivalent to 
the composition 
%composite 
of 
%one time evolution 
%two time evolutions 
%in 
%the 
two simpler cases: 
%of 
the case 
\mbox{$a=0$} and \mbox{$\alpha_n=0$} 
%\mbox{$a=\alpha_n=0$} 
and 
%one time evolution 
%two time evolutions 
%in 
the case \mbox{$c_n=0$}. 
%at 
%%corresponding to 
%%mappings 
%%for 
%two different values of $h$. 
Moreover, the latter case 
%\mbox{$c_n=0$} 
can be 
%regarded as the time reflection of 
%identified 
related with 
%inversion of 
the former case through 
%the inversion of time 
%and the sign 
%the time direction 
the time reflection. 
%% of 
%including the sign 
%%change 
%inversion of $h$. 
%\mbox{$a=0$} and \mbox{$\alpha_n=0$}. 
Thus, in this subsection, 
%the following, 
we only consider 
the 
%simple 
elementary case
%: 
%of 
\mbox{$a=0$}, \mbox{$\alpha_n=0$}, and 
\mbox{$c_n=-\Lambda_n$} 
%with 
under the boundary conditions 
%In addition to (\ref{V+-}), 
%%If 
%we impose the boundary conditions 
\begin{equation}
\lim_{n \to \pm \infty}
% \left( 
q_n 
%\right) = \boldsymbol{0}, 
=0, \hspace{5mm}
\lim_{n \to \pm \infty} \Lambda_n 
%\left( \alpha_n, \beta_n, c_n, d_n \right)
%\hspace{5mm}
%\lim_{n \to \pm \infty} \beta_n 
= 1. 
\label{ALbc2}
\end{equation}
Note that the boundary value of $c_n$ is set as \mbox{$-1$} 
without any 
loss of generality; it can be changed to any nonzero value 
by rescaling the 
%``time 
%spacing 
%interval 
``step size'' parameter $h$. 
To be precise, the boundary conditions
(\ref{ALbc2}) contain {\it redundant} \/information. 
Indeed, it can be shown that
%the auxiliary variable 
$\Lambda_n$ 
has the same limit
value 
%as 
for \mbox{$n \to -\infty$} and \mbox{$n \to +\infty$}. 
Therefore, it is sufficient to assume
either
\mbox{$\lim_{n \to -\infty} \Lambda_n = 1$}
or
\mbox{$\lim_{n \to +\infty} \Lambda_n = 1$}. 
%As is clear from 
%
%In this simple case, 
Now, the general system (\ref{fd-mV1}) 
%now 
reduces to the simpler system 
\begin{equation}
\left\{
\hspace{2pt}
\begin{split}
&
\frac{1}{h}\left(\widetilde{q}_n -q_n \right) 
 = (1+ q_n^2) \Lambda_{n+1} \left(\widetilde{q}_{n+1}-q_{n-1} \right), 
%\right.
\\[1mm] 
&
 (1 + \widetilde{q}_n^{\hspace{2pt}2}) \Lambda_n =(1+{q}_n^2) \Lambda_{n+1}.
%\label{}
\end{split}
\right.
\label{fd-mV2}
\end{equation}
The second 
%last 
%equalities 
relation 
in (\ref{fd-mV2}) 
%, together with the boundary conditions (\ref{ALbc2}), 
implies that the 
%{\it 
global
%} \/
expressions for 
%of 
the auxiliary variable 
$\Lambda_n$ in terms of 
%\mbox{$q_n\textrm{'s}$}:
the \mbox{$q_n$} are
%~\cite{AL77,Suris03}: 
%
\begin{equation}
 \Lambda_n = \prod_{j=-\infty}^{n-1} 
	\frac{1+\widetilde{q}_j^{\hspace{2pt}2}}{1+{q}_j^{\hspace{1pt}2}}
= \prod_{j=n}^{+\infty} 
	\frac{1+{q}_j^{\hspace{1pt}2}}
	{1+\widetilde{q}_j^{\hspace{2pt}2}}. 
%\end{split}
\nonumber
%\label{}
\end{equation}
%
%The equivalence of the two expressions 
%%for the auxiliary variable 
%is guaranteed by the boundary conditions 
%%(\ref{V+-}) and 
%(\ref{ALbc2}). 
The substitution of each 
%one 
%this 
expression 
into the first relation in (\ref{fd-mV2}) provides 
%the 
a global-in-space  
time discretization of the 
modified Volterra lattice (\ref{mLV}). 
%~\cite{AL76,AL77}. 
%
%compatibility 
%In the continuum 
%%continuous 
%limit of time \mbox{$h \to 0$}, we obtain 
%\[
%c_n \to c, \hspace{5mm} d_n \to d. 
%\]
%Thus, with an additional 
%%the 
%but unessential constraint on 
%%condition among 
%the parameters, 
%%and a proper redefinition of the parameters 
%%\mbox{$$}, 
%the time discretization 
%%full-discrete Ablowitz--Ladik lattice 
%(\ref{fd-AL1}) reduces to 
%the Ablowitz--Ladik lattice (\ref{sd-AL}) 
%in the 
%%continuous 
%limit 
%%of time 
%\mbox{$h \to 0$}. 
If \mbox{$h \in \mathbb{R}$} and
the $q_n$
%'s 
are real-valued at the initial time,
%\mbox{$m=m_0$},
then
%\mbox{$\epsilon \le 1/4$}.
%(\ref{fd-LV2}) implies that
%%the reality of
%the auxiliary variable $\Lambda_n$ is always real-valued, 
%Consequently, the discriminant of the quadratic equation (\ref{LValg2})
%must be non-negative.
%Thus, 
%and 
the real-valuedness of $q_n$ is 
preserved 
%maintained 
under the 
discrete-time evolution. 
%of the discrete-time modified Volterra lattice. 

Let us resolve this nonlocality. 
The determinant of the \mbox{$2 \times 2$} Lax matrix $L_n$ (\ref{AL-Ln2})
%Note that 
is given by 
\mbox{$ \det L_n = 1 + q_n^2$}, while 
%; 
%We can 
%also 
%compute 
the determinant of the 
%\mbox{$2 \times 2$} 
Lax matrix 
$V_n$ (\ref{AL-Vn4}) with 
\mbox{$a=0$}, \mbox{$\alpha_n=0$}, and
\mbox{$c_n=-\Lambda_n$} 
is computed 
as
\begin{align}
\det V_n (\lambda) 
%\nonumber \\=& 
=& - \left( \lambda^2 +1/\lambda^2
%\frac{1}{\lambda^2} 
\right) 
h \Lambda_n 
%\nonumber \\
%& \mbox{} + h^2 \Lambda_n^2 
% + \left( 1 + h \widetilde{q}_{n} q_{n-1} \Lambda_n \right)
%   \left( 1 + h q_{n-1} \widetilde{q}_{n} \Lambda_n \right)
%  + h^2  \widetilde{q}_{n}^{\hspace{2pt}2} \Lambda_n^2 
%  + h^2  q_{n-1}^{\hspace{1pt}2} \Lambda_n^2 
\nonumber \\ 
& \mbox{} 
+ h^2 \left( 1+ \widetilde{q}_{n}^{\hspace{2pt}2} \right) 
\left( 1 +q_{n-1}^{\hspace{1pt}2} \right)\Lambda_n^2
+ 2h \widetilde{q}_{n} q_{n-1}\Lambda_n + 1.
\nonumber
\end{align}
Thus, the equality 
(\ref{fd-cons1}) 
%together 
combined with the boundary conditions (\ref{ALbc2}) 
(or
%, 
the streamlined version as stated above) 
%implies 
leads to 
%the relation 
%%set of relations 
%%
%\begin{equation}
%\hspace{2pt}
%%\begin{split}
%%& 
%h^2 \Lambda_n^2  
% + \left( 1+ h \widetilde{q}_{n} q_{n-1} \Lambda_n \right)
%   \left( 1+ h q_{n-1} \widetilde{q}_{n} \Lambda_n \right)
% + h^2  \widetilde{q}_{n}^{\hspace{2pt}2} \Lambda_n^2 
%  + h^2  q_{n-1}^{\hspace{1pt}2} \Lambda_n^2 
%%\\[-1mm] & 
%= \left( 1+ h^2 \right) \Lambda_n.
%%\nonumber 
%%\end{split}
%\label{ALalg4}
%\end{equation}
%%where 
%%
%The algebraic system boils down to
%\begin{equation}
%\hspace{2pt}
%\begin{split}
%& h^2 \left(1-\widetilde{q}_{n} \widetilde{r}_{n} - q_{n-1}r_{n-1}\right) 
%  c_n d_n  
% + \left( 1+ h \widetilde{q}_{n} r_{n-1}c_n + h \alpha \right)
%   \left( 1+ h q_{n-1} \widetilde{r}_{n} d_n + h \beta \right)
%\\[-1mm] 
%& = \left[ h^2 c d + (1+h \alpha)(1+h \beta) \right] \Lambda_n.
%%\nonumber 
%\end{split}
%\label{ALalg5}
%\end{equation}
%Using the first and second equalities, the third equality
%results in a
%%(generally)
the quadratic equation in
%for 
$\Lambda_n$, 
\begin{align}
& h^2 \left( 1+ \widetilde{q}_{n}^{\hspace{2pt}2} \right) 
\left( 1 +q_{n-1}^{\hspace{1pt}2} \right)\Lambda_n^2
-\left( 1 + h^2 
%-  h q_{n-1} \widetilde{q}_{n} 
- 2h \widetilde{q}_{n} q_{n-1}
 \right) \Lambda_n + 1 =0.
\label{quadra5}
%\nonumber
\end{align}
%Thus,
%taking into account
By 
recalling the prescribed asymptotic behavior 
(\ref{asymp-h}) of the Lax matrix $V_n$ for small $h$, 
%(\ref{AL-Vn4}), 
%of $\Lambda_n$ as \mbox{$h \to 0$} (or \mbox{$n \to \pm \infty$}),
%and
we obtain the proper
%unique
solution of this quadratic equation in 
%for 
$\Lambda_n$,
%\begin{equation}
\begin{align}
%\frac{1}{\Lambda_n} & =
\Lambda_n &= \frac{2}{1 + h^2 - 2h \widetilde{q}_{n} q_{n-1}
 + \sqrt{\left( 1 + h^2 - 2h \widetilde{q}_{n} q_{n-1} \right)^2
-4 h^2 \left( 1+ \widetilde{q}_{n}^{\hspace{2pt}2} \right) 
\left( 1 + q_{n-1}^{\hspace{1pt}2} \right)}}
\nonumber \\
&= \frac{2}{1 + h^2 - 2h \widetilde{q}_{n} q_{n-1}
 + \sqrt{\left( 1 - h^2 \right)^2
 - 4 h \left( \widetilde{q}_{n} + h q_{n-1} 
\right) \left( h \widetilde{q}_{n} +q_{n-1} \right)}}.
\label{Lambda6}
%\end{equation}
\end{align}
%If 
When \mbox{$h \in \mathbb{R}$}, 
%this 
the local expression (\ref{Lambda6}) is 
valid only if \mbox{$-1 \le h \le 1$}. 
%(cf.~(\ref{ALbc2})). 
If \mbox{$h^2 > 1$}, the other solution of (\ref{quadra5}) 
should be 
%adopted. 
used. 
Unless \mbox{$h =\pm 1$},
a unified expression for \mbox{$\Lambda_n$},
\[
\Lambda_n = \frac{2}{1 + h^2 - 2h \widetilde{q}_{n} q_{n-1}
 + \left( 1 - h^2 \right) \sqrt{\hspace{1pt}1
%\left( 1 - h^2 \right)^2
 - \frac{4 h}{\left( 1 - h^2 \right)^2} \left( \widetilde{q}_{n} + h q_{n-1} 
\right) \left( h \widetilde{q}_{n} +q_{n-1} \right)}},
\]
can resolve the sign problem of the square root,
but we 
%are not inclined to use
do not use
%it looks
%%too
this 
%unwieldy
%and 
%unattractive 
form.
%for
In any case,
the decaying boundary conditions for $q_n$ imply that
\mbox{$\lim_{n \to - \infty} \Lambda_n =\lim_{n \to + \infty} \Lambda_n =1$}. 
Thus, 
the 
%compatibility of the 
%redundant 
boundary conditions for $\Lambda_n$ 
%as 
given in 
(\ref{ALbc2}) 
%is 
are compatible. 
%confirmed. 

Substituting (\ref{Lambda6}) into the first equation in (\ref{fd-mV2}), 
we obtain a 
%local 
time discretization of the modified Volterra lattice 
(\ref{mLV}) in the local form, 
%a discrete-time modified Volterra lattice 
\begin{align}
& \frac{1}{h} \left( \widetilde{q}_n - q_n \right)
\nonumber \\ 
& =
%\hspace{1pt}& 
\frac{2 \left( 1 + q_n^2 \right) 
 \left( \widetilde{q}_{n+1} - {q}_{n-1}\right)}
{1 + h^2 - 2h \widetilde{q}_{n+1} q_{n}
 + \sqrt{\left( 1 - h^2 \right)^2
 - 4 h \left( \widetilde{q}_{n+1} + h q_{n} 
\right) \left( h \widetilde{q}_{n+1} +q_{n} \right)}}.
\label{mVmap}
\end{align}
%where $\Lambda_n$ is given by (\ref{Lambda6}). 
%with (\ref{CnDn}). 
%Now, the first 
%%and second equations 
%equation in 
%%system 
Note that
the right-hand side of (\ref{mVmap})
%(\ref{fd-mV2}) 
does not involve 
%contain 
%is linear in 
\mbox{$\widetilde{q}_n$}. 
%, and thus 
%%and\mbox{$\widetilde{r}_n$}, respectively. 
%%Thus, 
%the
%%{\it explicit}
%forward
%time evolution 
%%that is
%is 
%%{\it 
%explicit under the decaying boundary conditions. 
%(\ref{ALbc2}). 
Similarly to the continuous-time case, 
(\ref{mVmap}) (or (\ref{fd-mV2}))
is invariant under the transformation 
\mbox{$q_n \mapsto (-1)^n q_n$}, \mbox{$h \mapsto -h$}. 
If \mbox{$-1 \le h \le 1$}, 
%\mbox{$h \in \mathbb{R}$}, 
%and $q_n$'s are real-valued at the initial time,
the real-valuedness of $q_n$ is 
preserved 
%maintained 
under the
discrete-time evolution (\ref{mVmap}); 
%time discretization of the modified Volterra lattice 
%In particular, 
the discriminant 
%quantity
%inside
in the square root is 
%non-negative 
nonnegative as long as 
we start with 
sufficiently small 
real-valued $q_n$
%'s 
at the initial time. 
%In order 
To express
%explicitly obtain
%write out 
the backward time evolution explicitly,
%
%which
%%is necessary for
%%should be done for
%%apears to be especially
%is highly desired for
%%in
%%necessary in
%the case \mbox{$h < 0$},
we only have
%need
to replace 
%rewrite 
\mbox{$(1+q_n^2) \Lambda_{n+1}$} in 
the first 
%two 
%equality 
equation 
of (\ref{fd-mV2}) 
%by 
with \mbox{$\left( 1+\widetilde{q}_n^{\hspace{2pt}2} \right) \Lambda_n$} 
%$\Lambda_n$ in (\ref{fd-AL3}) to \mbox{$\Lambda_{n+1}$}
%using 
(cf.\ the second equation)
%, 
%with the help of (\ref{Lambda}),
and then
substitute
the local expression
(\ref{Lambda6}).

%It should be noted that 
In terms of 
%a 
the new parameter $\delta$ 
given by \mbox{$\delta := h/(1+h^2)$}, 
(\ref{mVmap}) can be rewritten in a slightly more 
compact form, 
%as 
\begin{align}
%& 
\frac{1}{\delta} \left( \widetilde{q}_n - q_n \right)
%\nonumber \\ 
%& 
=
%\hspace{1pt}& 
\frac{2 \left( 1 + q_n^2 \right) 
 \left( \widetilde{q}_{n+1} - {q}_{n-1}\right)}
{1 - 2 \delta \widetilde{q}_{n+1} q_{n}
 + \sqrt{\displaystyle 1 - 4\delta \widetilde{q}_{n+1} q_{n} -4\delta^2
  \! \left( 1 + \widetilde{q}_{n+1}^{\hspace{2pt}2} 
+ q_{n}^2 \right) }}.
\label{mVmap2}
\end{align}
Note that 
if \mbox{$h \in \mathbb{R}$} (or
%, 
\mbox{$-1 \le h \le 1$}), then 
%\mbox{$-\frac{1}{2} \le \delta \le \frac{1}{2}$}.
\mbox{$-1/2 \le \delta \le 1/2$}.

%Note that in 
%With the ``unusual'' 
%%``anti-continuum'' 
%%``singular'' 
%%limit \mbox{$h^2 \to 1$}, 
%%case 
%choice 
At the ``threshold'' values of $h$, 
\mbox{$h =
%\to 
\pm 1$},
%\mbox{$h =\pm 1$},
%\mbox{$h^2 \to 1$},
corresponding to \mbox{$\delta=\pm 1/2$}, 
we can extract the square root in (\ref{mVmap}) 
to obtain a rational mapping. 
This is connected 
%related 
with the fact 
%can be understood by noting 
that 
%this is 
in the cases \mbox{$h =\pm 1$}, 
the discrete-time system 
%modified Volterra 
(\ref{fd-mV2}) 
under the boundary conditions (\ref{ALbc2}) 
%allows 
has the 
%``trivial'' 
trivial solution \mbox{$\widetilde{q}_n = \mp q_{n-1}$}, 
\mbox{$\Lambda_n = 1/(1+q_{n-1}^{\hspace{1pt}2})$} (cf.\ 
%section 
\S 4.6 
%of 
in \cite{Suris03}). 
%However, 
Thus, 
the discriminant 
%in the square root 
of the quadratic equation (\ref{quadra5}) 
vanishes at \mbox{$h =\pm 1$}, 
%at \mbox{$h =\pm 1$}, 
and the two solutions 
intersect. 
%
%intersection
%crossing
To 
%crystallize 
%turn 
%refine 
%this observation into a nontrivial 
%n interesting 
%meaningful 
obtain a nontrivial mapping from this observation, 
%result, 
%to see this,  
we 
%first 
set 
%as 
\mbox{$h =+1$} in (\ref{fd-mV2}) 
and replace $q_n$ 
%by 
with
$\mathrm{i} w_n$, namely, 
%i.e., 
\begin{equation}
\left\{
\hspace{2pt}
\begin{split}
&
\widetilde{w}_n -w_n 
 = (1- w_n^2) \Lambda_{n+1} \left(\widetilde{w}_{n+1}-w_{n-1} \right), 
%\right.
\\[1mm] 
&
 (1 - \widetilde{w}_n^{\hspace{2pt}2}) \Lambda_n =(1-{w}_n^2) \Lambda_{n+1}.
%\label{}
\end{split}
\right.
\label{fd-mV3}
\end{equation}
Moreover, 
%Second, 
we 
%need 
%%have 
%to 
%relax 
generalize the boundary conditions (\ref{ALbc2}) 
%to 
as 
%modify the boundary 
%\begin{equation}
\[
\lim_{n \to \pm \infty}
% \left( 
w_n 
%\right) = \boldsymbol{0}, 
=\gamma, \hspace{5mm}
\lim_{n \to \pm \infty} \Lambda_n 
%\left( \alpha_n, \beta_n, c_n, d_n \right)
%\hspace{5mm}
%\lim_{n \to \pm \infty} \beta_n 
= \frac{1}{(1+\gamma)^2}, 
\hspace{5mm} \gamma \neq -1,
%\label{ALbc3}
%\end{equation}
\]
and assume that \mbox{$\left| w_n -\gamma \right|$} is sufficiently small. 
%enough. 
Thus, the additional condition \mbox{$\gamma \neq 0$} 
excludes 
%the possibility of 
%the 
the trivial time evolution 
\mbox{$\widetilde{w}_n = - w_{n-1}$}. 
%\mbox{$\Lambda_n = 1/(1+q_{n-1}^2)$}. 
%following the above procedure for general $h$ 
%under the above boundary conditions 
%Then, 
Following the same 
%above 
procedure 
%for general $h$ 
%in the same way as above, 
as above, 
we obtain 
%a 
the quadratic equation in 
%for 
$\Lambda_n$ that can 
be factorized as 
\[
\left[ \left( 1 + \widetilde{w}_n\right) 
	\left( 1 + w_{n-1} \right) \Lambda_n -1 \right] 
\left[ \left( 1 - \widetilde{w}_n\right) \left( 1 - w_{n-1} \right) \Lambda_n
-1 \right] =0.
\]
Substituting its 
%the 
proper solution 
\mbox{$ \Lambda_n = 
1/\left[ \left( 1 + \widetilde{w}_n\right)
\left( 1 + w_{n-1} \right)\right]
$} into (\ref{fd-mV3}), 
we obtain a single equation 
\mbox{$(\widetilde{w}_n-1) (\widetilde{w}_{n+1} +1) 
= (w_n -1) ({w}_{n-1} +1 )$}. 
By a trivial change of the dependent variable 
%variables, 
\mbox{$w_n =: 1+ 2
%\delta 
\nu y_n  \hspace{3pt}(\nu \neq 0)$}, we obtain
%arrive at 
the well-known 
%``discrete-time Volterra'' lattice~\cite{
``discrete Volterra equation''~\cite{
%Suris03,
NijCap,HiroTsuji2}, 
\begin{equation}
\widetilde{y}_n (1+ 
%\delta 
\nu \widetilde{y}_{n+1}) 
 = y_n ( 1+ 
%\delta 
\nu {y}_{n-1} ). 
%{\mathcal{\delta}} {\mathfrak{\delta}} {\mathrm{\delta}} \cal{\delta}
%\textmu 
%\textdelta
%$\mbox{\rm \delta}$
\label{dLV}
\end{equation}
%values of $q_n$ and $\Lambda_n$ as \mbox{$n \to \pm \infty$}. 
%In particular, the case \mbox{$h \to +1$} 
%%can be related with 
%is closely related with 
%which coincides with 
%the well-known discrete-time 
%Volterra lattice~\cite{NijCap,HiroTsuji1,HiroTsuji2} 
%in (cf.\ section x.x of Suris). 
Therefore, 
%we have confirmed that 
the 
%discrete-time Volterra 
%lattice 
discrete-time 
%system 
equation (\ref{dLV}) belongs to the modified Volterra 
hierarchy and {\it not} \/the original 
Volterra hierarchy~\cite{Suris03};
%%, and it 
this corresponds to the 
%as a 
special case where 
the quadratic equation for the auxiliary variable 
$\Lambda_n$
%can be 
is factorized 
%and has 
%allows 
to provide 
%give 
a rational 
%expression. 
solution. 
\\
\\
{\it Ultradiscretization~{{\rm \cite{TakaMatsu95,ToTaMaSa}}}.} 
We 
%derive 
%propose
%discuss
present an intuitively
%-
plausible derivation of an 
%ultradiscretization
ultradiscrete analogue 
of the
%above
%discrete-time
time-discretized
modified Volterra lattice, 
%; 
although 
this 
%does not appear to be 
may not be 
%not 
%the 
a unique 
%one. 
ultradiscretization. 
%in the case \mbox{$b=0$}.
For the forward time evolution, we 
%first 
rewrite (\ref{mVmap2}) as
%\begin{equation}
\begin{align}
\widetilde{q}_n 
& = \frac{\displaystyle  
 q_n \left\{ 1 
%-2\delta 
+ \epsilon q_{n} q_{n-1} 
 + \sqrt{Y_n
%1 -4\delta \widetilde{q}_{n+1} q_n - 4\delta^2 \hspace{-1pt}\left( 
%1 + \widetilde{q}_{n+1}^{\hspace{2pt}2} + q_n^2 \right)
} \right\} 
%+2\delta 
-\epsilon \left( \widetilde{q}_{n+1} -q_{n-1}\right)}
{\displaystyle 
1 
%-2\delta 
+ \epsilon \widetilde{q}_{n+1} q_n 
 + \sqrt{Y_n
%1 -4\delta \widetilde{q}_{n+1} q_n - 4\delta^2 \hspace{-1pt}\left( 
%1 +  \widetilde{q}_{n+1}^{\hspace{2pt}2}  + q_n^2 \right)
}} 
\nonumber \\
&= \frac{\displaystyle  
 q_n \left\{ 1 + \epsilon \left( q_{n} + 1/q_n \right) q_{n-1} 
 + \sqrt{Y_n} -\epsilon\widetilde{q}_{n+1}/q_n \right\}}
{\displaystyle 1 + \epsilon \widetilde{q}_{n+1} q_n 
 + \sqrt{Y_n}} ,
\label{UDmV0}
%\end{equation}
\end{align}
where \mbox{$\epsilon:= -2\delta$} and
\mbox{$Y_n := 1 + 2\epsilon \widetilde{q}_{n+1} q_n - \epsilon^2 
%1 -4\delta \widetilde{q}_{n+1} q_n - 4\delta^2 
\hspace{-1pt}\left( 1 +\widetilde{q}_{n+1}^{\hspace{2pt}2}  + q_n^2 \right)$}. 
%In the case \mbox{$\epsilon=0$}, 
Note that the 
%resulting 
equation \mbox{$\widetilde{q}_n=q_n$} 
%in 
%obtained 
in the trivial 
case \mbox{$\epsilon=0$} 
preserves the sign
%/value
%magnitude 
%positivity 
of the dependent variable on each lattice 
site \mbox{$n \in {\mathbb Z}$}. 
%, \mbox{$q_n > 0$},
%%for
%%all
%\mbox{$\forall \hspace{1pt}n \in {\mathbb Z}$}. 
Thus, 
%it is natural to 
we 
%first 
consider a class of solutions in the limit \mbox{$\epsilon \to  +0$} 
%Then, 
%we consider 
%take 
%the limit \mbox{$\epsilon \to  +0$} assuming 
%in such a way 
%as 
such that 
%both the nontrivial time evolution and 
the positivity 
of
%the time evolution is nontrivial, but the positivity of
%the dependent variable is still preserved. 
the dependent variable, \mbox{$q_n > 0$},
%for
%all
\mbox{$\forall \hspace{1pt}n \in {\mathbb Z}$}, 
is preserved in the 
%under a 
%the 
nontrivial time evolution. 
We introduce the parametrization 
\mbox{$\epsilon=:
\exp(-L/\varepsilon)$}, \mbox{$L>0$} 
and \mbox{$q_{n,m} =: \rho \hspace{1pt}
%_{n-m} 
\exp(Q_{n,m}/\varepsilon)$}, \mbox{$\rho>0$}, 
%respectively, 
where \mbox{$m \in {\mathbb Z}$} is the discrete time coordinate, which 
%that 
is usually suppressed. 
The scaling 
%normalization 
parameter $\rho$ 
%$c_{n-m}$ 
can 
%is allowed to weakly 
depend weakly on 
%both 
%$n-m$ 
$n$, $m$, and $\varepsilon$, but 
%in the following, 
%for simplicity, 
for brevity,
%'s sake, 
we treat it as a constant. 
%for simplicity. 
%, but the dependence on them is assumed to be weak enough. 
%Moreover, 
%Although 
%After 
%With the substitution of 
Once this parametrization is substituted 
into 
(\ref{UDmV0}), the 
%time evolution 
solution \mbox{$\{ Q_{n,m} \}$} of the initial-value 
problem depends on $\varepsilon$ 
%through 
via the time evolution. 
%that is, $Q_n$ has the $\varepsilon$-dependence. 
%gives 
%In the following, 
We hypothesize 
%for the sake of simplicity 
that by choosing $\rho$ appropriately, 
the $\varepsilon$-dependence of $Q_{n}$ in 
%a class 
the considered class of solutions 
% on $\varepsilon$ 
%is made 
becomes
negligible
%negligibly small 
%weak
as \mbox{$\varepsilon \to +0$}, and 
%it 
$Q_n$ is 
%non-negative, 
nonnegative, \mbox{$ Q_n \ge 0$}. 
%One can also 
%A simple 
Probably, 
the simplest way to 
%understand/
justify the latter condition is 
%can also be understood by slightly modifying 
%as 
%through 
%by 
%a modification of 
%modifying the zero 
to modify the zero boundary conditions 
%\mbox{$\lim_{n \to \pm \infty} q_n =0$} 
%value 
(cf.~(\ref{ALbc2})) to 
%a nonzero value 
%the 
nonzero boundary conditions,
%boundary conditions 
%ones 
\mbox{$\lim_{n \to \pm \infty} q_n =\rho$}, 
%, 
and to assume 
%assuming 
\mbox{$q_n \ge \rho$}, \mbox{$\forall \hspace{1pt}n \in {\mathbb Z}$}. 
%In particular, 
%Anyway, 
We are 
%{\it 
not interested in tracing
%relatively 
infinitely long 
%small 
tails of solitons that decay exponentially as \mbox{$n \to \pm \infty$}; 
rather, we prefer to extract solitons 
with 
%on 
compact support in the limit \mbox{$\varepsilon \to  +0$}. 
%by assuming 
%, and assume 
%Thus, 
This is why 
we impose 
%assume 
the 
%``
zero
%'' 
boundary conditions \mbox{$\lim_{n \to \pm \infty} Q_n =0$} 
%in 
on the new variable $Q_n$. 

The condition \mbox{$Y_n \ge 0$}, 
%which is equivalent to 
guaranteed by the 
nonnegativity of the discriminant of (\ref{quadra5}), 
%constraint 
%condition 
%indicates 
requires that 
\[
\epsilon q_n \le \widetilde{q}_{n+1} + \sqrt{
\left( 1 - \epsilon^2 \right) 
\left( 1+\widetilde{q}_{n+1}^{\hspace{2pt}2} \right)}
%{\epsilon}
\;\; \mathrm{and} \;\; 
\epsilon \widetilde{q}_{n+1} \le q_n + \sqrt{
\left( 1 - \epsilon^2 \right) 
\left( 1+q_{n}^{2} \right)} \hspace{2pt}.
\]
Using the parametrization 
\mbox{$\epsilon=\exp(-L/\varepsilon)$} and 
\mbox{$q_{n} = \rho \hspace{1pt}\exp(Q_{n}/\varepsilon)$} 
with \mbox{$Q_n \ge0$}, 
we can interpret 
these conditions 
%constraints 
in the limit \mbox{$\varepsilon \to  +0$} as 
%\mbox{$\left| \widetilde{Q}_{n+1} - Q_n \right| \le L$}.
\mbox{$-L \le \widetilde{Q}_{n+1} - Q_n \le L$}. 
%In particular, 
%\mbox{$2\epsilon \left( \widetilde{q}_{n+1} -q_{n-1}\right)$} 
%should be 
%%always 
%negligible compared with $q_n$. 
%%$q_n$'s may go $+\infty$, but 
%\mbox{$\sqrt{\epsilon}
%%^{\frac{1}{2}}
%q_n =: Q_n$}
%Then,
%we assume \mbox{$\epsilon > 0$} so that the positivity of
%the dependent variable, \mbox{$u_n > 0$},
%%for
%%all
%\mbox{$\forall \hspace{1pt}n \in {\mathbb Z}$},
%can be
%%is
%%preserved
%%maintained
%in the time evolution. 
%
%By
%%expressing
%re-parametrizing
%the parameter and
%the dependent variable as \mbox{$\epsilon=:
%\exp(-L/\varepsilon)$} and \mbox{$\sqrt{\epsilon} q_n =:
%\exp(Q_n /\varepsilon)$} respectively,
%and 
By taking the logarithm,
%the above equation 
(\ref{UDmV0}) becomes
%reads as
\begin{subequations}
\begin{align}
\widetilde{Q}_n
&= Q_n + \varepsilon \log \left[
1 + 
%\rho^2 \hspace{1pt}
 \mathrm{e}^{\frac{Q_{n}+Q_{n-1}-L}{\scriptstyle \varepsilon}} 
  \left( \rho^2+  \mathrm{e}^{-\frac{2Q_n}{\scriptstyle \varepsilon}} \right)
 + \sqrt{Y_n} 
 - \mathrm{e}^{\frac{\widetilde{Q}_{n+1}-Q_n-L}{\scriptstyle \varepsilon}}
\right]
\nonumber \\
& \hphantom{=} \; \mbox{}
-\varepsilon \log \left[
 1 + \rho^2 \hspace{1pt}\mathrm{e}^{\frac{\widetilde{Q}_{n+1}+Q_n-L}
 {\scriptstyle \varepsilon}} 
 + \sqrt{Y_n}
\right],
%\nonumber
\\[1mm]
Y_n
&
=1-\mathrm{e}^{-\frac{2L}{\scriptstyle \varepsilon}} 
 + \rho^2 \hspace{1pt}
\mathrm{e}^{\frac{\widetilde{Q}_{n+1}+Q_n-L}
{\scriptstyle \varepsilon}} 
 \left( 2
%1+1 
  - \mathrm{e}^{\frac{\widetilde{Q}_{n+1}-Q_n-L}{\scriptstyle \varepsilon}} 
  - \mathrm{e}^{\frac{Q_n -\widetilde{Q}_{n+1}-L}{\scriptstyle \varepsilon}}
\right).
%\nonumber
\end{align}
\label{UDmV1}
\end{subequations}
%
%In deriving the above equations, 
%the parameter $\rho$ is treated as a constant, but 
%%even if it depends weakly on 
%its weak dependence on 
%%both 
%$n-m$ and $\varepsilon$ makes 
%%, there is 
%no 
%%serious 
%%the final result 
%essential difference. 
%For brevity's sake, 
%Recall that the $\varepsilon$-dependence of $Q_n$'s is not considered. 
%
Taking the aforementioned assumptions into account, 
%suppressed. 
%Under the assumption of 
%Thus, 
%the limit \mbox{$\epsilon \to +0$} converts 
(\ref{UDmV1}) reduces in the limit \mbox{$\varepsilon \to +0$} to 
%into 
the following equation: 
\begin{align}
\widetilde{Q}_n
&= Q_n + \max \left(0, \, Q_{n}+Q_{n-1}-L, \,
  \frac{\widetilde{Q}_{n+1}+Q_n-L}{2} \right)
\nonumber \\ & \hphantom{=} \;\; \mbox{} 
-\max \left( 0, \, \widetilde{Q}_{n+1}+Q_n-L \right). 
\label{UDmV}
\end{align}
As mentioned above, 
the boundary conditions 
%\mbox{$\lim_{n \to \pm \infty} q_n =0$}
%translate into 
\mbox{$\lim_{n \to \pm \infty} Q_n =0$} are assumed. 
It is desirable 
%to be confirmed 
to confirm for each solution 
that the 
%additional 
conditions 
\mbox{$Q_n \ge 0$} and \mbox{$\bigl| \widetilde{Q}_{n+1} - Q_n \bigr| \le L$} 
%\mbox{$-L \le \widetilde{Q}_{n+1} - Q_n \le L$} 
%hold true. 
%are 
%indeed 
%preserved 
%certainly 
hold true 
in the time evolution. 
Note that 
the parameter $L$ may 
%be time-dependent. 
depend on the discrete time. 
It is 
%to be 
hoped that the 
issue of the {\it commutativity} \/of the ultradiscrete 
flows defined by (\ref{UDmV}) 
%indeed give commutative flows
%with 
for 
%various 
distinct values of $L$ will be investigated. 
In this regard, 
%connection, 
%In connection with this, 
we expect 
%conjecture 
that (\ref{UDmV}) and its time reversal 
(see below) 
%form 
will comprise 
an ultradiscrete analogue of 
the mKdV hierarchy, wherein the parameter $L$ labels 
%specifies 
each flow in the hierarchy. 
%Note that 
The ultradiscrete equation 
(\ref{UDmV}) is ``linear'' in the sense that it is invariant 
under the rescaling \mbox{$Q_n \mapsto k Q_n $}, 
\mbox{$L \mapsto k L $}, \mbox{$k >0$}. 
Thus, 
%when 
if $L$ is a 
%fixed 
time-independent 
constant, it is possible to fix $L$ at unity. 
%normalize $L$ as \mbox{$L=1$}. 
However, 
%it is often better to 
this normalization 
%rescaling 
often 
%may 
changes 
%a natural number assigned to 
%some 
an integer-valued $Q_n$ 
%integer value of $Q_n$ 
%into 
to a fractional value; 
%number, so 
thus, 
we rather prefer to leave $L$ as a free parameter. 
%the free parameter $L$. 
%as it is useful to 
%
%The 
Our derivation of the ultradiscrete modified Volterra lattice 
(\ref{UDmV}) is more or less intuitive
%, 
%rather than 
and is not mathematically rigorous. 
A more 
%rigorous 
detailed treatment of all the terms 
in the numerator of (\ref{UDmV0}) may 
%could 
lead to a more complicated 
ultradiscrete 
equation, but we prefer the relatively simple equation (\ref{UDmV}). 
Fortunately, 
%as far as we 
%can 
%could check with 
%using 
%some 
for the specific examples that we considered, 
(\ref{UDmV}) allows the ``stable'' propagation of solitons 
and their elastic (but nontrivial) collisions; 
%of solitons; 
thus, 
(at least some of) 
the integrability properties 
appear to be retained in this ultradiscretization. 

The backward time evolution 
of 
%for 
the 
%discrete-time 
time-discretized modified Volterra lattice is obtained from 
(\ref{mVmap2})
through the combined space
%-time
and time reflection 
\mbox{$n
%\mapsto
\to -n$}, \mbox{$m
%\mapsto
\to -m$}.
%\mbox{$(n,m) \to (-n,-m)$}. 
Thus, its ultradiscrete analogue is 
%also 
obtained from 
%immediately 
(\ref{UDmV}) in the same manner.
%correspondence

\subsection{The lattice Heisenberg ferromagnet model}

The lattice Heisenberg ferromagnet model was 
%found 
proposed in 1982 
%independently 
by several different 
authors~\cite{Skl82,Ishi82,Hal82,Fad84}. 
The equation
%s 
of motion for 
%the 
this semi-discrete 
model can be derived almost 
systematically 
%by following 
using either Ishimori's approach~\cite{Ishi82} 
based on a gauge transformation from the Ablowitz--Ladik lattice 
or the $r$-matrix 
formalism based on 
%approach using 
the fundamental Poisson bracket relation~\cite{Skl82,Fad84}.
%TTF83}. 
However, for a concise and easy-to-understand derivation, 
we use a more heuristic approach based on the zero-curvature 
%condition. 
representation. 

%Let us begin 
We start with the 
%semi-discrete 
Lax pair 
%in the semi-discrete case
of the following form in the semi-discrete case: 
\begin{subequations}
\begin{align}
L_n &= I + \lambda S_n,
\label{L-HF} 
\\
%[3mm]
M_n &= \frac{\lambda}{1-\lambda^2} A_n L_n, 
%= \frac{\lambda}{1-\lambda^2} L_{n-1} A_n ,
\end{align}
\label{HF-Lax1}
\end{subequations}
where the 
%two 
%additional 
conditions 
%\mbox{$\mathrm{tr}\hspace{1pt} S_n =0 $} 
%(const.\ not necessary?), 
\mbox{$(S_n)^2 =I$}
%, 
and \mbox{$A_n S_n = S_{n-1} A_n $} are 
%imposed. 
assumed. 
The latter 
%last 
condition 
%implies 
guarantees the useful 
relation \mbox{$A_n L_n = L_{n-1} A_n $}. 
Thus, substituting the Lax pair (\ref{HF-Lax1}) into 
the zero-curvature condition (\ref{Lax_eq}), we obtain 
\[
(L_n^{-1})_t + \frac{\lambda}{1-\lambda^2} (A_{n+1}-A_n) = O. 
%\label{}
\]
%namely, 
%Using 
Noting the identity \mbox{$(I+\lambda S_n) (I-\lambda S_n)=(1-\lambda^2) I$}, 
this results in 
\begin{equation}
S_{n,t} = A_{n+1}-A_n.
\label{S-A}
\end{equation}
Because 
%of 
\mbox{$(S_n)^2 =I$} and \mbox{$A_n S_n = S_{n-1} A_n $}, 
(\ref{S-A}) implies the relation 
%gives 
\[
A_{n+1} (S_{n+1} + S_n) = (S_{n} + S_{n-1}) A_{n}.
\]
%A natural choice 
Thus, we choose 
%set 
$A_n$ as 
\[
A_n = 2\mathrm{i}a S_{n-1}(S_{n} + S_{n-1})^{-1} + 2b (S_{n} + S_{n-1})^{-1}, 
\]
%which automatically satisfies 
so that the above relation is automatically satisfied. 
Here, 
%the 
%scalar 
%coefficients 
$a$ and $b$ are $n$-independent 
scalars, but 
%scalars that 
they may depend on 
the time variable 
$t$. 
Substituting this expression for $A_n$ into (\ref{S-A}), 
we obtain 
%(cf.~(\ref{space-diff}))
\begin{equation}
S_{n,t} = \boldsymbol{\Delta}_n^+ 
\left[ 2\mathrm{i}a S_{n-1}(S_{n} + S_{n-1})^{-1} 
	+ 2 b (S_{n} + S_{n-1})^{-1} \right], 
\label{gHF}
\end{equation}
where $\boldsymbol{\Delta}_n^+$ is
the forward difference operator in 
the discrete space 
%the spatial direction 
(cf.~(\ref{space-diff})). 
Note that the 
%time 
evolution equation 
%defined by 
(\ref{gHF}) 
is consistent with 
the condition \mbox{$(S_n)^2 =I$}
%, where $S_n$ is a general \mbox{$l \times l$} matrix. 
%, 
%and the 
%for an arbitrary dimension 
%size 
%of 
%$S_n$. 
for a general \mbox{$l \times l$} matrix $S_n$. 
%of the sqaure matrix $S_n$ is not fixed. 
%arbitrary. 
%, and the dimension of the matrix $S_n$ is arbitrary 
%Now, we 
We now 
%restrict the dimension of the matrix $S_n$ to two, a
consider the simplest nontrivial case of \mbox{$l=2$}
%, 
%a \mbox{$2 \times 2$} matrix, 
and express 
%it 
$S_n$ in terms of the Pauli matrices as
\begin{align}
S_n & = \sum_{j=1}^3 S_n^{(j)} \sigma_j \hspace{3pt}
 (=: \boldsymbol{\sigma \hspace{-1pt} \cdot \hspace{-1pt}S}_n)
\nonumber \\
&= \left[
\begin{array}{cc}
S_n^{(3)} & S_n^{(1)} - \mathrm{i} S_n^{(2)}\\
S_n^{(1)} + \mathrm{i} S_n^{(2)} & -S_n^{(3)}\\
\end{array}
\right].
%\nonumber 
\label{S-sigma}
\end{align}
%where 
Here, \mbox{$\boldsymbol{S}_n = 
	\bigl( S_n^{(1)}, S_n^{(2)}, S_n^{(3)} \bigr)$} 
is a unit 
%three-component 
vector, i.e., 
\mbox{$\langle \boldsymbol{S}_n, \boldsymbol{S}_{n} \rangle
%| \boldsymbol{S_n} | 
= 1 $}.
Because 
\mbox{$2\hspace{1pt} (S_{n} + S_{n-1})^{-1} = (S_{n} + S_{n-1})/(1+
%\boldsymbol{S_n} \cdot \boldsymbol{S_{n-1}}
\langle \boldsymbol{S}_n, \boldsymbol{S}_{n-1} \rangle)$}~\cite{Ishi82}, 
(\ref{gHF}) reduces to a single vector equation 
involving both 
the scalar 
%inner 
product 
%$\cdot$ 
and the vector product, 
%$\times$ 
%should be distinguished 
\begin{equation}
\boldsymbol{S}_{n,t} = \boldsymbol{\Delta}_n^+ 
\left[ a 
\frac{\boldsymbol{S}_{n} \times \boldsymbol{S}_{n-1}}
{1+ \langle \boldsymbol{S}_n, \boldsymbol{S}_{n-1} \rangle} 
 + b \frac{\boldsymbol{S}_{n} + \boldsymbol{S}_{n-1}} 
{1+ \langle \boldsymbol{S}_n, \boldsymbol{S}_{n-1} \rangle} \right], 
\hspace{5mm}
\langle \boldsymbol{S}_n, \boldsymbol{S}_{n} \rangle= 1.
\label{sdHF}
\end{equation}
The case \mbox{$b=0$} gives 
%corresponds to 
the lattice Heisenberg ferromagnet 
model~\cite{
%Skl82,
Ishi82,Hal82
%,Fad84
}, while the case 
\mbox{$a=0$} corresponds to 
%a higher symmetry of it
its simplest 
higher symmetry~\cite{GIV86}. 

Let us 
%consider the time discretization of (\ref{sdHF}). 
%move to 
examine 
the discrete-time case. 
%For the sake of definiteness, 
We mainly consider the time discretization of the reduced system 
(\ref{sdHF}), and not the general matrix system (\ref{gHF}), 
%as 
because 
the latter problem 
%appears 
is expected to be too complicated. 
%As a discrete-time analogue of the Lax 
%matrix $M_n$ in 
%%pair 
%(\ref{HF-Lax1}), 
We start 
%begin 
with 
%assume 
the Lax matrix $V_n$ of the following form: 
\begin{equation}
V_n = I + h \frac{\lambda}{1-\lambda^2} A_n L_n,
\label{V-HF}
\end{equation}
where the 
%two 
%additional 
conditions 
%\mbox{$\mathrm{tr}\hspace{1pt} S_n =0 $} (const.\ not necessary?), 
\mbox{$(S_n)^2 =I$}
%, 
and \mbox{$A_n S_n = \widetilde{S}_{n-1} A_n $} are 
%imposed. 
assumed. 
The 
%last 
latter condition 
%implies 
guarantees the useful 
relation \mbox{$A_n L_n = \widetilde{L}_{n-1} A_n $}. 
Thus, substituting the Lax pair, (\ref{L-HF}) and (\ref{V-HF}), 
into the zero-curvature condition (\ref{fd-Lax}), we obtain 
\[
\frac{1}{h} \bigl( \hspace{1pt}\widetilde{L}_n^{\hspace{1pt}-1} 
 - L_n^{-1} \bigr) + \frac{\lambda}{1-\lambda^2} (A_{n+1}-A_n) = O, 
%.
%\label{}
\]
or equivalently, 
%which is equivalent to 
%
\begin{equation}
\frac{1}{h}\bigl(\hspace{1pt} \widetilde{S}_{n} -S_{n} \bigr) 
= A_{n+1}-A_n.
\label{S-S-A}
\end{equation}
Under the condition \mbox{$(S_n)^2 =I$}, the relation 
%condition 
\mbox{$A_n S_n = \widetilde{S}_{n-1} A_n $} is automatically satisfied 
%by setting $A_n$ as 
%\[
if $A_n$ 
%is written in 
takes the general 
form \mbox{$A_n = B_n S_n + \widetilde{S}_{n-1} B_n $}. 
%where $B_n$ is an arbitrary matrix. 
%\]
%More specifically, 
%In particular, 
However, in analogy with the semi-discrete case, 
we 
%assume the following form: 
employ a more specific 
%the following 
%choice 
form 
of $A_n$, 
%that is, 
%namely, 
%: 
\[
A_n = 2\mathrm{i}a_n 
%\alpha_n 
\widetilde{S}_{n-1} \bigl( S_n + \widetilde{S}_{n-1} \bigr)^{-1} 
	+ 2 b_n 
%\beta_n 
\bigl( S_n + \widetilde{S}_{n-1} \bigr)^{-1}, 
\]
which also satisfies the relation \mbox{$A_n S_n = \widetilde{S}_{n-1} A_n $}. 
Here, the scalar 
%quantities 
unknowns 
\mbox{$a_n$} and \mbox{$b_n$} 
%\mbox{$\alpha_n$} and \mbox{$\beta_n$} 
are 
%interpreted as 
auxiliary variables. 
Substituting this expression for $A_n$ into (\ref{S-S-A}), we obtain
\begin{equation}
\frac{1}{h}\bigl(\hspace{1pt} \widetilde{S}_{n} -S_{n} \bigr)
 = \boldsymbol{\Delta}_n^+
\left[ 2\mathrm{i}a_n 
%\alpha_n 
\widetilde{S}_{n-1} \bigl( S_n + \widetilde{S}_{n-1} \bigr)^{-1} + 
%\beta_n 
 2b_n \bigl( S_n + \widetilde{S}_{n-1} \bigr)^{-1} \right].
\label{d-gHF}
\end{equation}
It only remains necessary 
%for us 
to fix the auxiliary variables \mbox{$a_n$} and \mbox{$b_n$}. 
%\mbox{$\alpha_n$} and \mbox{$\beta_n$}. 
%so that 
%Note 
%It should be recalled 
To this end, 
we recall 
that 
the
time
evolution 
%equation 
determined by (\ref{d-gHF}) 
%should 
has to be consistent with
the condition \mbox{$(S_n)^2 =I$}. 
In the following, we focus on the case of 
%a \mbox{$2 \times 2$} 
the \mbox{$2 \times 2$} 
matrix $S_n$ 
%written as 
given by (\ref{S-sigma}). 
%Thus, 
The requirement \mbox{$\mathrm{tr}\hspace{1pt} S_n =0 $} 
results in 
%implies 
the $n$-independence of $a_n$, 
%$\alpha_n$, 
%so 
thus 
we set 
%it as 
\mbox{$a_n = a$}. 
%Thus
Therefore, (\ref{d-gHF}) reduces to 
\begin{equation}
\frac{1}{h}\bigl(\hspace{1pt} \widetilde{\boldsymbol{S}}_{n} 
	-\boldsymbol{S}_{n} \bigr) = \boldsymbol{\Delta}_n^+
\left[ a
\frac{\boldsymbol{S}_{n} \times \widetilde{\boldsymbol{S}}_{n-1}}
{1+ \langle \boldsymbol{S}_n, \widetilde{\boldsymbol{S}}_{n-1} \rangle}
 + b_n \frac{\boldsymbol{S}_{n} + \widetilde{\boldsymbol{S}}_{n-1}}
{1+ \langle \boldsymbol{S}_n, \widetilde{\boldsymbol{S}}_{n-1} 
\rangle} \right]. 
%\hspace{5mm} \langle \boldsymbol{S}_n, \boldsymbol{S}_{n} \rangle= 1.
\label{ddHF}
\end{equation}
%In order 
To fix $b_n$, we 
%recall 
%use 
invoke the 
%recipe 
procedure presented in section~\ref{sect2}; 
we assume 
%impose 
the boundary conditions 
\begin{equation}
\lim_{n \to - \infty}
% \left(
%\boldsymbol{S}_{n}  = \boldsymbol{S}, 
\langle \boldsymbol{S}_n, \widetilde{\boldsymbol{S}}_{n-1} \rangle =1, 
\hspace{5mm} \lim_{n \to - \infty} b_n = b. 
\label{HFbc}
\end{equation}
%$\boldsymbol{S}$ is a constant (i.e., time-independent) unit vector. 
Because \mbox{$\langle \boldsymbol{S}_n, \boldsymbol{S}_{n} \rangle= 1 $}, 
we have \mbox{$\det L_n = 1-\lambda^2$} 
%(see 
(cf.\ 
(\ref{L-HF}) and (\ref{S-sigma})). 
Thus, the equality (\ref{Vn:2}) 
derived from (\ref{fd-cons1}) implies that 
%the relation 
the determinant of 
%the Lax matrix 
$V_n$, 
\begin{align}
\det V_n &= 
\det \left( I + h \frac{\lambda}{1-\lambda^2} A_n L_n\right)
\nonumber \\ 
&= \det \left( I-\lambda S_n  + h \lambda A_n \right)
\det \left( \frac{1}{1-\lambda^2} L_n \right)
,
\nonumber 
\end{align}
is an $n$-independent quantity. 
%This is satisfied if 
Therefore, both the trace and the determinant of 
\mbox{$S_n  - h A_n$} must be $n$-independent. 
The $n$-independence of \mbox{$\mathrm{tr} \hspace{1pt}
(S_n  - h A_n)$} 
is already satisfied by 
setting 
%as 
%the $n$-independence of $a_n$, 
\mbox{$a_n = a$}. 
The determinant of \mbox{$S_n  - h A_n$} can be computed as 
\begin{align}
\det \left( S_n  -h A_n \right)
&= \det \left[ S_n  -2\mathrm{i}ha
\widetilde{S}_{n-1} \bigl( S_n + \widetilde{S}_{n-1} \bigr)^{-1} - 2 h b_n
\bigl( S_n + \widetilde{S}_{n-1} \bigr)^{-1} \right]
\nonumber \\
&= \frac{ \det \left[ (1- 2 h b_n )I+ S_n \widetilde{S}_{n-1}  -2\mathrm{i}ha
\widetilde{S}_{n-1} 
%- 2 h b_n I 
\right]}{\det \bigl( S_n + \widetilde{S}_{n-1} \bigr)}
\nonumber \\
&= \frac{\bigl( 1+ \langle \boldsymbol{S}_n, 
	\widetilde{\boldsymbol{S}}_{n-1} \rangle-2h b_n \bigr)^2 
	+1+4(ha)^2 -\langle \boldsymbol{S}_n, 
	\widetilde{\boldsymbol{S}}_{n-1} \rangle^2 }
{-2 \bigl( 1+ \langle \boldsymbol{S}_n, 
	\widetilde{\boldsymbol{S}}_{n-1} \rangle \bigr) },
\nonumber 
\end{align}
%and its value is determined by the boundary conditions (\ref{HFbc}). 
which coincides with its boundary value determined by (\ref{HFbc}). 
This results in 
%Thus, we obtain 
a quadratic equation in 
%for 
$hb_n$, i.e., 
%\begin{multline}
\begin{align}
%& 
2 (h b_n)^2 & -2 \bigl( 1+ \langle \boldsymbol{S}_n, 
	\widetilde{\boldsymbol{S}}_{n-1} \rangle \bigr) h b_n 
\nonumber \\
& 
%\hspace{10mm}	
	\mbox{} +2(ha)^2 + \left[ hb(2-hb) -(ha)^2 \right]
	\bigl( 1+ \langle \boldsymbol{S}_n, 
	\widetilde{\boldsymbol{S}}_{n-1} \rangle \bigr)=0.
\nonumber 
\end{align}
%\end{multline}
For 
%a 
sufficiently small 
%value of 
$h$ (\mbox{$0 < |h| \ll 1$}), 
the proper solution of this quadratic equation is given by 
\[
hb_n = \frac{2(ha)^2 + \left[ hb(2-hb) -(ha)^2 \right]
% \bigl( 1+ \langle \boldsymbol{S}_n,  \widetilde{\boldsymbol{S}}_{n-1} 
% \rangle \bigr)
(1+g_n) }{1+g_n 
%1+ \langle \boldsymbol{S}_n,\widetilde{\boldsymbol{S}}_{n-1} \rangle 
+\sqrt{ ( 1+ g_n)^2 -4(ha)^2 
 -2 \left[ hb(2-hb) -(ha)^2 \right] (1+g_n)}}, 
\]
where \mbox{$g_n := \langle \boldsymbol{S}_n, 
\widetilde{\boldsymbol{S}}_{n-1}  \rangle $}. 
Substituting this local expression for $hb_n$ 
into (\ref{ddHF}), 
we obtain an integrable time discretization of 
%the lattice Heisenberg model 
(\ref{sdHF}); 
%which 
this time discretization 
is essentially equivalent to (B.18) in~\cite{QNCL} 
(see also (26) in~\cite{NCWQ} and (3.10) in~\cite{NijCap}). 
%
%we write the explicit <-  not good as it is not an explicit scheme
We write the equation of motion 
for three 
%relatively simple 
%and 
important 
cases: 
%, that is, 
%That is, 
the case \mbox{$b=0$}, 
\begin{align}
%\frac{1}{h}\bigl(\hspace{1pt} 
& \widetilde{\boldsymbol{S}}_{n}-\boldsymbol{S}_{n} 
%\bigr) 
= \boldsymbol{\Delta}_n^+
\left\{ ha
\frac{\boldsymbol{S}_{n} \times \widetilde{\boldsymbol{S}}_{n-1}}
{1+ \langle \boldsymbol{S}_n, \widetilde{\boldsymbol{S}}_{n-1} \rangle} 
\right.
\nonumber \\
& \left. \mbox{}
 + \left[ 1 - \sqrt{1 - 2(ha)^2 \frac{1 - \langle \boldsymbol{S}_n, 
	\widetilde{\boldsymbol{S}}_{n-1} \rangle}{ \bigl( 
%/
1+ \langle \boldsymbol{S}_n,\widetilde{\boldsymbol{S}}_{n-1} 
 \rangle \bigr)^{2}}}  \hspace{1pt}\right]
%1+ \langle \boldsymbol{S}_n, \widetilde{\boldsymbol{S}}_{n-1}\rangle
\frac{\boldsymbol{S}_{n} + \widetilde{\boldsymbol{S}}_{n-1}}{2}
\right\},
\hspace{5mm} \langle \boldsymbol{S}_n, \boldsymbol{S}_{n} \rangle= 1; 
\nonumber 
\end{align}
the case \mbox{$ hb(2-hb) =(ha)^2$},
\begin{align}
%\frac{1}{h}
 \widetilde{\boldsymbol{S}}_{n}
        -\boldsymbol{S}_{n} & = \boldsymbol{\Delta}_n^+
\left\{ h a
\frac{\boldsymbol{S}_{n} \times \widetilde{\boldsymbol{S}}_{n-1}}
{1+ \langle \boldsymbol{S}_n, \widetilde{\boldsymbol{S}}_{n-1} \rangle}
\right. 
\nonumber \\
& \left. \mbox{}
 + \left[ 1 - \sqrt{1 - \frac{4(ha)^2}{ \bigl( 
%/
1+ \langle \boldsymbol{S}_n,\widetilde{\boldsymbol{S}}_{n-1} 
 \rangle \bigr)^{2}}}  \hspace{1pt}\right]
%1+ \langle \boldsymbol{S}_n, \widetilde{\boldsymbol{S}}_{n-1}\rangle
\frac{\boldsymbol{S}_{n} + \widetilde{\boldsymbol{S}}_{n-1}}{2}
\right\},
\hspace{5mm} \langle \boldsymbol{S}_n, \boldsymbol{S}_{n} \rangle= 1;
\nonumber 
\end{align}
and the case \mbox{$a=0$}, 
%\mbox{$hb (2-hb) =: \delta$}, 
\begin{align}
& \, \frac{1}{\delta}\bigl(\hspace{1pt} \widetilde{\boldsymbol{S}}_{n}
        -\boldsymbol{S}_{n} \bigr) 
\nonumber \\
 = \; & \boldsymbol{\Delta}_n^+
\left[  
 \frac{\boldsymbol{S}_{n} + \widetilde{\boldsymbol{S}}_{n-1}}
{1+ \langle \boldsymbol{S}_n, \widetilde{\boldsymbol{S}}_{n-1}
\rangle 
+ \sqrt{ \bigl( 1+ \langle \boldsymbol{S}_n, \widetilde{\boldsymbol{S}}_{n-1}
 \rangle \bigr)^2 - 2 \delta \bigl( 1+ \langle \boldsymbol{S}_n, 
 \widetilde{\boldsymbol{S}}_{n-1} \rangle \bigr)}} \right],
\nonumber \\[1mm] & 
\hspace{73mm}
%\hfill
\delta := hb (2-hb), 
\hspace{5mm} 
\langle \boldsymbol{S}_n, \boldsymbol{S}_{n} \rangle= 1.
\nonumber 
\end{align}
The first and second cases provide 
%integrable 
time discretizations
%give discrete-time analogues 
of the lattice Heisenberg ferromagnet model
[(\ref{sdHF}) with \mbox{$b=0$}],  
while the third case gives a discrete-time analogue of 
its simplest higher symmetry [(\ref{sdHF}) with \mbox{$a=0$}]. 
Note that 
the first case \mbox{$b=0$} 
%should be 
can be identified with (6.8) in~\cite{QNCL}. 
%Unfortunately, 
%this 
%any of these 
%each of them is an 
All 
these difference 
schemes are seemingly 
%these are 
%on a superficial level 
{\it highly implicit}, 
%\/scheme, 
that is, 
the value $\widetilde{\boldsymbol{S}}_n$ is not 
%determined uniquely ...
written explicitly in a closed form in terms of 
${\boldsymbol{S}}_n$, ${\boldsymbol{S}}_{n+1}$, 
and $\widetilde{\boldsymbol{S}}_{n-1}$, 
%We can compute
%though 
but this is {\it not} \/a serious drawback. 
Indeed, 
we can obtain 
%compute successively 
%recursively 
the power series expansion 
of $\widetilde{\boldsymbol{S}}_n$ in $h$ (or $\delta$) 
to any 
%an arbitrary 
order 
successively. 
%is not 
%
%Alternatively, we give up computing 
%Moreover, instead of trying to compute 
%Alternatively, 
Moreover, we can compute 
the exact value of $\widetilde{\boldsymbol{S}}_n$ 
%in 
by
%two steps. 
the following steps. 
First, we 
%We first 
%derive 
%consider 
take the scalar product between the equation of motion 
and \mbox{$\boldsymbol{S}_{n+1}$} to obtain an equation for 
%and 
\mbox{$\langle \boldsymbol{S}_{n+1}, \widetilde{\boldsymbol{S}}_{n}\rangle$}, 
%; this equation involves 
wherein \mbox{$\widetilde{\boldsymbol{S}}_{n}$} appears 
only through the 
%scalar product 
form 
\mbox{$\langle \boldsymbol{S}_{n+1}, \widetilde{\boldsymbol{S}}_{n}\rangle$}. 
Then, 
the proper 
solution of this equation 
%should be looked for within 
%lying in 
%should lie 
can be found in 
an 
%some 
\mbox{$O(h)$} (or \mbox{$O(\delta)$}) 
%proximity 
neighborhood
%neighbourhood
of \mbox{$\langle \boldsymbol{S}_{n+1},{\boldsymbol{S}}_{n}\rangle$}. 
%; 
%should 
%can be 
%Substituting, ... linear in 
%Plunging 
Substituting it back into the original equation of motion, 
%to yield 
we arrive at 
a linear 
%``linear'' 
equation 
%in 
for \mbox{$\widetilde{\boldsymbol{S}}_{n}$} 
that can be 
%easily 
solved straightforwardly. 
%Thus, 
%we believe that 
%this is not a serious drawback and can be repaired by a 
%%better 
%proper re-parametrization 
%of the vector dependent variable $\boldsymbol{S}_n$; 
%we hope to revisit this problem in the near future. 

In this subsection, 
we 
%derived 
have obtained 
the time discretizations of the first two 
%treat the first flow and the second flow of the hierarchy 
flows of the lattice Heisenberg ferromagnet 
%model 
hierarchy 
in a unified way. 
%However, 
Our derivation is 
original and 
%concise, 
%easy-to-understand
%easy-to-follow, 
easy to follow, 
%straightforward
but not fully systematic, as 
some parts are based on heuristic treatments. 
%Some parts of our derivation are 
%based on heuristic treatments, so this is not 
%a fully systematic derivation
%It would be 
%%an interesting problem 
%interesting 
%It remains to be seen 
%if an extension of Ishimori's approach can provide 
%to extend Ishimori's approach to examine/achieve 
%a (more) systematic derivation of these time discretizations. 
%It is hoped 
%We hope 
It
%seems to
would be 
%very
interesting 
to investigate 
%apply 
%to extend 
%consider 
whether 
these time discretizations 
%in this subsection 
can be derived in a systematic manner 
%using 
%by applying Ishimori's approach~\cite{Ishi82} to 
from the 
%applies 
%discrete-time 
time discretizations of the Ablowitz--Ladik lattice 
obtained 
in 
%the preveous 
subsection~\ref{secAL}, 
using Ishimori's approach~\cite{Ishi82}.
%in a systematic manner. 
%
%
%We can probably derive their time discretizations 
%systematically by applying Ishimori's approach~\cite{Ishi82} 
%to the full-discrete NLS equations obtained 
%in 
%%the preveous 
%subsection~\ref{secAL}. However, this appears 
%to be a long way and we prefer, at the expence of 
%systematic derivation, to obtain the time discretizations 
%more easily and directly. 
%
%
%\begin{subequations}
%\begin{align}
%& \\
%& \\
%&
%\label{}
%\end{align}
%\end{subequations}

\section{Concluding remarks}

%\[
%\mathscr{H}(p)
%\mathfrak{H}, \mathcal{H}, \mathfrak{a}, \mathcal{a}, 
%\mathfrak{b}, \mathcal{b},\mathfrak{c}, \mathcal{c}, 
%\mathfrak{d}, \mathcal{d}, \mbox{\cal{a}}, \mbox{\cal{b}}, 
%\mbox{\cal{c}}, \mbox{\cal{d}},\frak{a}, 
%\]

In this paper, we have 
%proposed and 
developed an 
%highly sophisticated
%new 
%particularly
effective method for obtaining proper time discretizations 
of integrable lattice systems in \mbox{$1+1$} dimensions. 
This method, which is 
based on the zero-curvature condition (\ref{fd-Lax}), 
allows us to obtain local equations of motion 
%for the original physical variables. 
that can determine the time evolution uniquely.  
%and explicitly. 
Using this method, we 
%have 
%obtained 
constructed new time discretizations of 
the Toda lattice, 
%in the Flaschka--Manakov coordinates, 
the Ablowitz--Ladik lattice, 
the Volterra lattice, 
and the modified Volterra lattice, 
while we obtained the same time discretizations of 
the lattice Heisenberg ferromagnet model 
and its symmetry as those in~\cite{QNCL,NCWQ,NijCap}. 
It should be stressed that 
this is a systematic method and 
%is applicable 
also applies to other integrable lattice systems. 
%as well. 
%
%(cf.\ 
%%see, e.g. 
%%those in 
%Part II of~\cite{Suris03}). 
As a bonus, we 
%could 
were able to derive ultradiscrete analogues of 
the Volterra lattice and the modified Volterra lattice 
involving one arbitrary parameter $L$, namely, 
%see 
(\ref{udLV1}) 
and (\ref{UDmV}). 
%We conjecture that 
Each of these ultradiscrete equations, 
%(and its time reversal) 
%and 
as well as its time reversal, 
%their time-reversed versions) 
appears to 
form a hierarchy of mutually commuting flows 
%by varying 
%specified 
labelled by the parameter $L$.
%It should be noted 
Note that the ultradiscrete equations 
such as (\ref{udLV1}) and (\ref{UDmV}) allow 
straightforward ``multicomponent'' generalizations. 
Indeed, for example, if we express
%, {\it e.g.}, 
the dependent variable and the parameter 
in the form \mbox{$U_n = U_n^{(1)} + U_n^{(2)} 
%I 
\mbox{\boldmath$i$}$}, 
\mbox{$U_n^{(j)} \in {\mathbb Z}$} 
and \mbox{$L= L^{(1)} + L^{(2)}\mbox{\boldmath$i$} 
%\hspace{2pt}(>0) 
>0$}, 
\mbox{$L^{(j)} \in {\mathbb Z}$}, 
where $\mbox{\boldmath$i$}$ is an irrational number, and substitute 
%it 
them into 
(\ref{udLV1}), we can uniquely determine the time evolution 
%obtain a 
of the ``two-component'' system 
%variables 
%starting from any initial data. 
%from 
starting from a given initial condition. 
%A 
It 
%seems to 
would be very 
%looks 
%is 
%may be 
%especially 
interesting to investigate 
%consider 
the special case wherein 
the irrational 
\mbox{\boldmath$i$} is arbitrarily 
%unboundedly 
%i-1 is infinitesimally small 
close to $1$. 

%The new product features beautiful images ...
%One of the 
%
A notable 
%%main 
feature
%s 
%The main advantage 
of our method is that 
%the time discretizations 
the time discretization does not modify 
%change 
%belong to the same integrable hierarchies 
%which 
%as the original continuous-time lattice systems. 
%belong. to. 
%do not change 
the integrable hierarchy to which 
the original 
%continuous-time 
lattice system 
%under investigation 
%considered 
belongs. 
Thus, 
%this guarantees the invariance of 
the integrals of motion 
and the functional 
%essential 
form of the solutions 
%the structures of solutions 
%
%formula 
%for the lattice system 
%considered. 
%also 
remain invariant; 
%
%We would like to mention that each local 
%
%We do not claim 
%It should be stressed that 
%The solutions of the time-discretized lattice systems 
%%are new or unknown. 
%%cannot be 
%are by no means essentially new.
%Indeed, 
%Only the time evolution of 
%the soliton parameters corresponding to the angle variables 
%is discretized. 
%%; that is, 
%%%Roughly epeaking, this implies that 
%%%those parameters 
%%the angle variables 
%%%evolution receives 
%%%if 
%%seen in the phase space evolve in a discrete manner. 
%%like a series of snapshots or 
%%%discrete 
%%samplings. 
%%
%%In particular, the solutions of the local full-discrete 
%%NLS equation can be constructed using (or, even contained in) 
%%the results of Ablowitz and Ladik. 
%
%This fact also implies that the {\it symplectic structure}
%\/of the lattice system is completely 
%preserved by the time discretization (cf.\ Suris), 
%but this is beyond the scope of this paper. 
%
%The inverse scattering 
%%transform 
%method (cf.~\cite{AL76,AL77}) implies 
%guarantees 
%that 
%the time discretizations 
%%thus obtained 
%%are 
%%have exactly the same 
%do not change the structures of solutions 
%%as the time-continuous 
%%original 
%of the lattices; 
only the time dependence 
%evolution 
of certain parameters 
%(roughly speaking, angle variables) 
corresponding to the 
%``angle variables'' 
angle variables in the solutions 
%change upon time discretization 
is changed (cf.~\cite{AL76,AL77,Kako}). 
%is discretized. 
%suffer a change. 
%This property possibly suggests 
%the direct 
%%practical 
%applicability (advantages over others) 
%of our results in numerical simulations. 
%%The constructed discrete-time evolutions 
%%are, by construction, 
%%can be 
%%considered 
%regarded as
%% generalized 
%commuting flows, 
%%symmetries, 
%or equivalently, auto-B\"acklund transformations 
%of 
%%for 
%the original continuous-time lattices. 
%%fully discrete 
%
Such a time discretization 
%of an integrable lattice system 
can be identified with 
the spatial part of an auto-B\"acklund transformation 
of 
%for 
the 
%original 
continuous-time lattice system or, from 
%in 
a more unified point 
of view, an auto-B\"acklund transformation 
%for
of the whole hierarchy of commuting 
%continuous-time 
flows. 
%Thus, 
%It is quite 
%natural to expect that 
%the 
%A 
Therefore, any time discretization 
%constructed using 
obtained by 
%with 
our method 
%from a qualitative standpoint 
%generally 
preserves the qualitative properties 
of the original integrable hierarchy
%continuous-time lattice 
%system 
and thus is expected to serve as an excellent 
%numerical 
scheme 
for numerical integration 
%simulation 
of 
%for 
the continuous-time lattice system. 
%stability 
%In contrast, 
This is in contrast with 
other known 
methods 
%for time discretization 
that 
%usually 
generally
%the time discretization 
%{\it do} \/
%have to 
modify 
the integrable hierarchy 
%which the lattice belongs to, 
to find 
%provide 
a 
%local 
time discretization 
%written 
in the local form. 
%written in local equations of motion. 
%that is, 
%%it provides 
%the discrete-time lattice and the continuous-time lattice 
%belong to different hierarchies. 
As is illustrated 
%for 
%with the Ablowitz--Ladik lattice 
in subsection~\ref{secAL}, 
our method can also 
%be used to obtain 
provide 
%high-precision
%time discretization
%local and 
``almost
%close to 
integrable''
numerical schemes that 
%can 
far 
%largely
surpass the 
Ablowitz--Ladik--Taha
local schemes~\cite{AL76,AL77,Taha1,Taha2}
%tested in~\cite{Taha1,Taha2}.
in approximation accuracy.
%accuracy. 
%precision
%
%In the actual numerical computations performed by 
%Taha and Ablowitz, the auxiliary variables are 
%often replaced by constants, typically unity, 
%leading to a nearly integrable scheme. 
%In our paper, we have obtained the {\it exact local} \/expressions 
%of the auxiliary variables. 
%%These 
%They can be used to provide 
%%used to construct 
%a much  better approximation for 
%%more accurate numerical scheme 
%numerical experiments, {\it e.g.}, using 
%the Taylor expansion such as \mbox{$\sqrt{1+f} \sim 1+ f/2 +\cdots$}.

In our approach, 
a 
%every 
local-in-space 
time discretization is 
always 
derived from a 
%the original 
global-in-space 
%scheme
time discretization involving ``nonlocal'' auxiliary 
%dependent 
variables. 
%; 
The obtained
%/resulting 
local equations of motion 
generally 
%can 
determine 
%define 
both the forward and backward 
time evolution 
uniquely 
%and explicitly 
under the 
specified 
%(assumed?)
%assumed 
%suitable
boundary conditions; 
%and 
%is 
%are 
%considered 
thus, 
they indeed give 
a discrete-time analogue of 
%the ``evolutionary system'' 
{\it evolutionary} \/differential-difference equations.
%Then, 
%Thus,
%it would be natural to 
One might 
%think 
consider 
that the original nonlocal 
%global-in-space 
%scheme
time discretization no longer has any use 
%becomes useless wastes 
%product 
%after 
once the local 
%time 
discretization has been 
%was 
%always 
derived from it; 
however, 
%in many cases, 
this is often not the case. 
In actually solving an 
%the 
initial-value problem, 
the nonlocal
%global
%-in-space 
%scheme
time discretizations often provide 
%useful information 
%critical 
%significant
crucial information on 
%yield important information 
the 
%properties 
attributes 
of the 
%its 
solution, 
%of the initial-value problem, 
such as the real-valuedness, positivity, and 
rationality 
with respect to the initial data and the parameters;
%, etc.;
%which 
it appears that 
%on ther other hand, 
the local 
%time 
discretizations are 
%cannot
%usually 
%unable to 
%weak in verifying 
%poor at establishing 
%establish such properties. 
not useful for 
establishing 
%verifying 
such properties directly. 
%immediately 
Let us illustrate this point with one instructive 
%pedagogical
example, namely, 
the discrete-time Ablowitz--Ladik lattice 
in the case \mbox{$a=b=0$} studied in subsection~\ref{secAL}. 
%(cf.\ subsection~\ref{secAL}). 
The global
%-in-space 
scheme
%time discretization 
(\ref{fd-AL3}) 
%(see also
(cf.\ 
(\ref{la-rep})) 
is free from 
%non-
irrational functions; 
%such as the square root 
%indicates immediately that ... 
%and 
consequently, 
under the 
%%rapidly 
decaying boundary conditions (cf.\ (\ref{ALbc})), 
the discrete-time 
updates of the dependent variables 
%thus 
%the time evolution under the 
%%rapidly 
%decaying boundary conditions (cf.\ (\ref{ALbc}))
%is 
are 
%described 
given by a 
%allows a rational expression 
rational mapping. 
%(cf.~(\ref{ALmap2}) with (\ref{Lambda3})). 
On the other hand, the local scheme 
%time discretization 
(\ref{ALmap2}) with (\ref{Lambda3}) 
%(or, (\ref{fdNLS2})) 
involves a 
%the 
square root; thus, 
it is 
extremely 
difficult to foresee that 
%the solution of the initial-value problem
the 
%actual 
time evolution 
%under the 
%%rapidly 
%decaying boundary conditions (cf.\ (\ref{ALbc}))
%is 
%are described 
%is actually 
can be described 
%can be expressed 
%given 
by a 
%allows a rational expression 
rational mapping. 
%and (\ref{CnDn})), 
However, 
%the 
these 
two schemes naturally
%have to 
define 
%one and 
the same time evolution. 
%
%For 
%In (most of) the local time discretizations 
%%considered 
%constructed in this paper, 
%%the solution of 
%%by solving an initial-value problem step by step, 
%%by starting with 
%%from 
%%with 
%%for a given initial value, 
%%we can compute 
%the time evolution 
%%value 
%of the dependent variables 
%%starting 
%from given initial data 
%%\mbox{$(m=0)$} 
%at \mbox{$m=0$} to the next moment 
%%time 
%\mbox{$(m=1)$} can be computed from 
%%one site to another; 
%site to site under the specified boundary conditions. 
%%
%%The property that 
%%the scheme is explicit in 
%Namely, the equations of motion can determine 
%both the forward and backward time evolution 
%%is determined
%%expressed 
%%defined 
%%written 
%uniquely and explicitly; 
%%under the specified boundary conditions; 
%this property is 
%%can be 
%considered a discrete-time analogue of 
%%the ``evolutionary system'' 
%``evolution equations'' in the continious-time case. 
%%; 
%This is quite natural, as 
%before 
%eliminating 
%obtaining the local expressions for the auxiliary variables, 
%It should be recalled that 
%Typically, 
%the original 
%%form of the 
%(global) time discretizations before eliminating  the auxiliary variables 
%%assume 
%have a rational expression 
%%form 
%%do not involve 
%free from 
%%non-
%irrational functions such as 
%the square root
%%, 
%etc. 
%then 
%Thus
Therefore, 
%one finds that 
the quantity inside the square root in (\ref{Lambda3}) 
is always 
%gives 
equal to 
%should give 
the 
%perfect 
square of a rational function 
%polynomial 
%in 
of the dependent variables at the 
%initial/
previous moment 
%\mbox{$m=0$} 
and the parameters; 
%the initial time. 
%\mbox{$q_n(m=0)$}. 
%In computing the forward time evolution, 
%typically, 
%we typically encounter such an 
%
typically, 
%A typical 
%%quantity 
%%form of the square 
%%expression 
%%as 
%%appearing in the square root 
%example acquires the form 
%%assumes the form 
%%in the forward time evolution, 
%\mbox{$\sqrt{
%%f(h)
%( 
%% c 
%f_0 + h f_1 + h^2 f_2 + \cdots 
%%\sum_{j=0}^{j_{max}} h^j f_j 
%)^2}$}, 
%where \mbox{$f_j \, (j \ge 1)$} involve the dependent variables. 
%%As long as \mbox{$f_0>0$}, we wish to identify it as 
%%\mbox{$ f_0 + h f_1 + h^2 f_2 + \cdots$}. 
%%at the present
%%constraints. 
%For example, in the actual 
%%forward 
%time evolution 
%of the discrete-time Ablowitz--Ladik lattice 
%in the case \mbox{$a=b=0$} (cf.~(\ref{ALmap2}) 
%with (\ref{Lambda3})), 
%%and (\ref{CnDn})), 
%one encounters a quantity of 
%it 
%the square root in (\ref{Lambda3}) takes 
one encounters the form 
\[
\sqrt{
%\left
\bigl( 1-\hat{c}\hat{d}
%\right
\hspace{1pt}\bigr)^2
%\left[
%\bigl[
(
%\left\{
1+
%\hat{c}\hat{d}
f_n 
%\left(
%%\hspace{1pt}\mbox{$q_n$, $r_n$,
%\{q_n\}, \{r_n\} \right)
%\right]
%\bigr]
)^2},
\]
where $\hat{c}$ and $\hat{d}$ 
%are defined by 
depend on the 
%``time step'' 
``step size'' 
parameter $h$ through (\ref{cdhat}). 
The rational 
function \mbox{$f_n 
%\left(\{q_n\}, \{r_n\} \right)
$}
%\mbox{$g \left(\{q_n\}, \{r_n\} \right)$} 
involves 
%not only 
%both 
the 
%uncurrent
dependent variables 
%at 
%%\mbox{$m=0$}
%the previous moment 
%but also 
and the parameters appearing in (\ref{ALmap2}), 
and is of order \mbox{$O(h)$}. 
%at most 
%\mbox{$O(h^l)$}, 
%\mbox{$k >0$}. 
%\mbox{$l \ge 1$}. 
%\mbox{$l \in {\mathbb N}$}. 
%Can one 
In addition, 
%Moreover, 
it vanishes when 
%both at 
%\mbox{$h=0$} or, 
%and when 
%\mbox{$q_n=r_n=0$}. 
%$q_n$ and $r_n$ 
the dependent variables 
are 
identically 
zero. 
%for all $n$. 
%Strictly speaking, 
%The condition \mbox{$\hat{c}\hat{d} \le 1$} is 
%%in
%not sufficient 
%In order 
%to 
%naively 
For sufficiently small \mbox{$|h| (\ll 1)$}, 
%the 
this square root is extracted 
as 
\mbox{$
%\bigl
( 1-\hat{c}\hat{d}\hspace{1pt} 
%\bigr
)
%\{ 1+\hat{c}\hat{d}  \left( \mbox{$q_n$, $r_n$} \right)\}
%\left[
(1+f_n 
%\left(\{q_n\}, \{r_n\} \right)
%\right]
)$}. 
%, though 
%%Indeed, 
%for sufficiently 
%small \mbox{$|h| (\ll 1)$} 
%this poses 
%%causes 
%%induces 
%no serious problem.
%but 
However, 
for relatively large $|h|$, 
this is not obvious; 
%evident; 
%%sign/magnitude/
even when \mbox{$\hat{c}\hat{d} < 
%\le 
1$}, 
the values of the dependent variables 
have to satisfy rather restrictive 
conditions. 
%inequalities. <- no reality condition, so we may have (complex number)^2
%
%%it is 
%It is unclear at present
%%at the present time 
%%how 
%whether these conditions are related with 
%%the well-posedness of $\Lambda_n$ (\ref{la-rep}), i.e., 
%the (nonzero and) finiteness conditions 
%of the conserved quantity \mbox{$ \prod_{j=-\infty}^{+\infty} 
%	(1-{q}_j {r}_j)$} (cf.~(\ref{la-rep})) 
%and
%%/or the finiteness conditions of 
%the higher order ones. 
%%conserved quantities. 
%%Probably, the simplest 
One simple way to bypass 
%this 
%the 
this 
sign 
problem 
%in the above case 
%of the square root 
%in the local expressions of the 
%appearing 
%overcome
is to consider 
%understand 
%``define'' 
%consider that 
%\mbox{$g \left(\{q_n\}, \{r_n\} \right)$}
the square root 
%of this 
%%kind 
%type 
as being defined 
%the quantity \mbox{$\sqrt{
%%\left[
%%\bigl[
%%\left\{
%( 1+
%%\hat{c}\hat{d}
%g_n 
%%\left(\{q_n\}, \{r_n\} \right)
%%\right]
%%\bigr]
%)^2}$} 
%in the above expression 
%is defined by the 
by 
%the 
its 
Taylor 
%series 
expansion
%s 
%a Taylor expansion 
%in 
%with respect to 
%in power series of $h$
for small \mbox{$|h|$} 
%and/or 
or small 
%\mbox{$\{q_n\}$}, 
%%and 
%\mbox{$\{r_n\}$}
amplitudes of the dependent variables
%; 
%they are convergent for sufficiently small \mbox{$|h|$},
% (\ll 1)$}, 
%and 
%subsequently, 
and then to 
extend the domain of the definition 
%can be 
%%is 
%extended 
%consider 
by analytic continuation; 
this is 
briefly 
%tersely 
%described 
explained 
%mentioned 
in subsection~\ref{secAL}. 
%with respect to $h$. 
%in the $h$ plane. 
%Therefore, we have a/
%
Thus, this example also illustrates 
%This is 
%one difficult aspect 
the difficulty 
of the sign problem 
%of how to 
%in extracting the square root in the local equations of motion. 
in computing 
%the 
updates of the dependent variables 
using the local equations of motion.
In this paper, 
we 
%have 
mainly considered 
%in this paper 
%the simplest case of 
the ``constant'' boundary conditions
%for $V_n$
at spatial infinity (cf.~(\ref{V+-}))
%, 
and assumed 
%specified 
%chose 
%dealt with 
%the 
%natural and 
simple boundary values of the dependent variables. 
%
%In this paper, we have mostly assumed the zero 
%%decaying 
%boundary conditions at spatial infinity, 
%which 
%%wherein 
%makes the corresponding inverse scattering transform 
%%can be performed in the simplest way. 
%%becomes 
%the simplest one. 
%%as \mbox{$n \to \pm \infty$}. 
%%Thus, 
%%
Note, however, that 
%it should be noted that 
%the validity of 
our method is {\it not} \/sensitive 
to the boundary values of the dependent variables
%, 
and 
%we can 
is also applicable to 
%the case of 
other boundary conditions, 
%as well, 
including 
%some 
periodic, nonvanishing, or 
%``nonconstant'' 
nonconstant 
%ones. 
boundary conditions. 
Indeed, 
as 
%is 
%described 
is 
%obvious 
clear 
from the description 
in subsection~\ref{FCL},
%it suffices that 
%It should be recalled that 
%we 
one can freely modify 
%change 
the boundary conditions 
%of 
for $V_n$ 
%have only to 
%specify 
as long as they 
%can 
determine 
%the 
a definite 
%$n$-independent 
value 
%of 
for the right-hand side of 
(\ref{alge1}) or (\ref{alge2}) that is $n$-independent. 
%, in particular, the specified boundary values of 
%
%%the auxiliary variables 
%%introduced 
%%appearing 
%%in 
%$V_n$, 
%completely 
%determine the proportionality 
%%coefficients/
%factors 
%%between 
%among all the fluxes 
%%associated with 
%corresponding to
%the same ``conserved density'' \mbox{$\log \det L_n$}. 
%%Indeed
Alternatively, 
%otherwise
%before 
%Instead of 
%without first specifying the boundary conditions
%for the dependent variables, 
%For example, instead of specifying the boundary conditions, 
%we 
one can 
first 
%also 
%think of 
%specify 
%treat 
set 
%the 
each 
%$n$-independent 
value of (\ref{alge1}) or 
(\ref{alge2}) as a ``constant'' 
free parameter
%, 
and then 
%discuss 
%deduce 
elicit 
the corresponding 
boundary conditions for the dependent variables. 
%freely choose the proportionality 
%%coefficients/
%factors 
%%between 
%among 
%%%all 
%the fluxes 
%%%associated with 
%%corresponding to the same ``conserved density'' \mbox{$\log \det L_n$}. 
%in the ``lowest-order'' conservation laws 
%(cf.\ (x.x)) 
%instead of specifying the boundary conditions. 
%(\ref{alge1}), (\ref{alge2})
%(This only results in a change of the proportionality factors.) 
In either case, 
%/way, 
one should 
%has to 
%pay much attention in order 
take 
%particular 
care to identify 
%select out 
the proper solution of the resulting 
algebraic system (cf.\ subsection~\ref{sec2.4}). 
%One 
We can 
%be convinced that the boundary conditions do not matter. 
also 
%A more intuitive way to understand that 
understand in a more intuitive way that 
%the integrability of 
the 
%local 
time discretizations 
%equations of motion 
derived in this paper 
%are 
%can be 
are integrable under 
%some 
other 
suitable 
%appropriate
boundary conditions; 
%It is possible to understand that ...  in a more intuitive way; that is, 
%...
%Indeed, 
note that 
the 
%local 
conservation laws 
%can 
should be derived 
%verified 
%verifiable
from the 
local equations of motion 
%by 
using 
only local operations, i.e., 
%that is, 
without referring to 
the boundary conditions. 
%Thus, the integrability by no means heavily restrict 
%be strictly limited // be stringently regulated
%be severely restricted

%Should I write in section 2 
%some comments on 
%%about 
%the normalization (\mbox{$\det =1$}) 
%of the Lax pair in the full-discrete case ?

%In the following, we assume that $h$ is much smaller than 
%any quantity and its inverse appearing in the determining 
%%equation 
%system of the auxiliary variables ...
%overwhelmingly small ??
%Owing to (\ref{asymp-h}), (\ref{V+-}), 
%each auxiliary variable for any $n$ take values in 
%%the 
%%some 
%an $O(h)$ 
%%proximity 
%neighborhood of 
%%the value of 
%the corresponding quantity in $M_n$. 
%For the well-posedness of the auxiliary variables, 
%we assume that the original dependent variables are small enough. 
%In particular, the auxiliary variables do not involve 
%the negative powers of $h$ such as $1/h$ and $1/h^2$, 
%and allow expressions as the Taylor/Maclaurin series in $h$. 
%%expansion 
%\mbox{$h \ll \varDelta^2 \ll 1$}, 
%\mbox{$ q_n r_{n'} = O(\varDelta^2)$}
%
%phase transition at \mbox{$h=1$} ? 
%(the sign in front of the squre root has to be changed). 

Very recently, 
Adler, Bobenko, and Suris~\cite{ABS03,ABS09} 
%gave 
%achieved a classification of 
successfully classified 
%one-field 
discrete integrable 
%two-dimensional 
systems 
%with one 
%for a scalar dependent variable 
on quad-graphs using the notion of 
%the 
three-dimensional consistency~\cite{NijWal}. 
%(cf.~\cite{NijWal}). 
In particular, under some assumptions, 
they presented a short 
but 
complete list of 
one-field integrable equations 
defined by 
%affine-linear
{\it polynomial} \/relations 
%linear 
of degree one 
in 
%with respect to
%all 
each of the four arguments. 
%a 
Their results have been attracting 
%a bunch of 
%many 
a lot of 
%researchers' 
interest from researchers; for example, 
%make a click on a link
click on the NASA ADS
%CiteBase 
link at 
\\
\hspace*{22mm}
%\begin{center}
{\tt http://arxiv.org/abs/nlin/0202024} 
%\quad 
%\, 
%\hspace{2pt}.
%\hspace{1pt}.
.
%\end{center}
\\
%That is
%Indeed, 
The 
%local 
time discretizations of 
%a 
polynomial 
%integrable 
lattice 
%/rational evolutionary 
systems 
%equation 
%constructed using 
obtained by 
%with 
our method generally contain
%, in general, 
%{\it non-polynomial} 
{\it irrational} \/nonlinearity; 
%of non-polynomial
%%/rational 
%type. 
%That is, 
%In many cases, 
%Indeed, 
the nonlinear terms are determined through 
the proper 
solution of 
an ``ultralocal'' 
%nonlinear 
algebraic equation of degree higher than one. 
Thus, 
%the local equations of motion 
%involve 
%cannot be written in any polynomial form, and 
such time discretizations 
essentially lie 
outside the 
%classification of Adler--Bobenko--Suris~\cite{ABS03,ABS09}. 
class considered by 
%Adler--Bobenko--Suris
Adler {\it et al.}~\cite{ABS03,ABS09}, 
%; 
%the 
although their three-dimensional consistency 
%of our time discretizations 
can, in principle, be investigated 
%studied 
using the approach 
%exploited
outlined in the last paragraph of subsection~\ref{nonauto}. 
%Probably, 
%the results of this paper can 
This probably partially explains 
why 
%the Adler--Bobenko--Suris 
their list of one-field integrable equations 
%~\cite{ABS03,ABS09} 
is 
%so 
short. 
%gave 
%
%
%Note that our method does not, in general, maintain 
%the polynomial/rational nature of the equations; 
%that is, the local time discretization of a 
%polynomial/rational evolutionary system 
%%equation 
%%constructed using 
%obtained with our method is not necessarily 
%of polynomial/rational type. 
%
%This is 
%%the 
%a price to pay 
%%be paid in order not to 
%%change 
%for not modifying the underlying integrable hierarchy of the lattice system. 
%Cite the Adler--Bobenko--Suris paper 
%on integrable equations on the quad-graph 
%See~\cite{ABS03} and references citing this paper 
%(for example, click ``CiteBase'' 
%%link 
%at 
%http://arxiv.org/abs/nlin/0202024), or~\cite{ABS09} and 
%references cited therein. 
%
%lattice $\rightarrow$ lattice system (?)
%
%{\bf Acknowledgments} 
\section*{Acknowledgments} 
%We would like to 
The author thanks 
%Prof.\ 
Professor
J.\ Matsukidaira, 
%Prof.\ 
Professor
K.\ Nishinari, 
%Prof.\ 
Professor
S.\ Tsujimoto, 
Professor K.\ Toda, 
Professor T.\ Takenawa, 
Dr.\ F.\ Sugino, 
Dr.\ Z.\ Tsuboi, and Dr.\ K.\ Maruno 
for their useful comments and discussions. 
%and their help. 
%Masahiro Shiroishi for discussions 
%and providing us his unpublished QMC data \cite{Sh06}.
%%%%%%%%%%%%%%%%%%%%%%%%%%%%%%%%%%%%


\addcontentsline{toc}{section}{References}
\begin{thebibliography}{99}
\bibitem{Suris03}
%Yuri 
Y.\ B.\ Suris: 
%, 
{\it The Problem of Integrable Discretization:\ Hamiltonian Approach}
%Series: Progress in Mathematics , Vol. 219 
(Birkh\"auser, Basel, 2003). 

\bibitem{AL76}
M.\ J.\ Ablowitz 
and J.\ F.\ Ladik: 
Stud.\ Appl.\ Math.\ {\bf 55} (1976) 213. 

\bibitem{AL77}
M.\ J.\ Ablowitz and J.\ F.\ Ladik: 
Stud.\ Appl.\ Math.\ {\bf 57} (1977) 1. 

\bibitem{AL1}
M.\ J.\ Ablowitz and J.\ F.\ Ladik: 
J.\ Math.\ Phys.\ {\bf 17} (1976) 1011.

\bibitem{Taha1}
T.\ R.\ Taha and M.\ J.\ Ablowitz: 
J.\ Comp.\ Phys.\ {\bf 55} (1984) 192.

\bibitem{Suris97}
Y.\ B.\ Suris:
Phys.\ Lett.\ A {\bf 234} (1997) 91.

\bibitem{Suris97'}
Y.\ B.\ Suris: 
Inv.\ Probl.\ {\bf 13} (1997) 1121. 

\bibitem{Hirota04}
R.\ Hirota: 
{\it The Direct Method in Soliton Theory} 
(Cambridge Univ.\ Press, Cambridge, 2004) 
edited and translated by A.\ Nagai, J.\ Nimmo and C.\ Gilson. 

\bibitem{Flaschka1}
H.\ Flaschka: 
Phys.\ Rev.\ B {\bf 9} (1974) 1924. 

\bibitem{Flaschka2}
H.\ Flaschka: 
Prog.\ Theor.\ Phys.\
{\bf 51} (1974) 703.

\bibitem{Manakov74}
S.\ V.\ Manakov: 
Sov.\ Phys.\ JETP {\bf 40} (1975) 
%(1974/5) 
269.

\bibitem{TakaMatsu95}
D.\ Takahashi and J.\ Matsukidaira: 
Phys.\ Lett.\ A {\bf 209} (1995) 184. 

\bibitem{ToTaMaSa}
T.\ Tokihiro, D.\ Takahashi, J.\ Matsukidaira
%, 
and J.\ Satsuma: Phys.\ Rev.\ Lett.\ {\bf 76} (1996) 3247.

\bibitem{NQC}
F.\ W.\ Nijhoff, G.\ R.\ W.\ Quispel and H.\ W.\ Capel: 
Phys.\ Lett.\ A {\bf 95} (1983) 273. 

\bibitem{QNCL}
G.\ R.\ W.\ Quispel, F.\ W.\ Nijhoff, H.\ W.\ 
Capel and J.\ van der Linden: 
Physica A
{\bf 125} (1984) 344.

\bibitem{NCWQ}
F.\ W.\ Nijhoff, H.\ W.\ Capel, G.\ L.\ Wiersma 
and G.\ R.\ W.\ Quispel: Phys.\ Lett.\ A {\bf 103} (1984) 293. 

\bibitem{NijCap}
F.\ Nijhoff and H.\ Capel: 
Acta Appl.\ 
%icandae 
Math.\ 
%ematicae
{\bf 39} (1995) 133.

\bibitem{Kako}
F.\ Kako and N. Mugibayashi: Prog.\ Theor.\ Phys.\ {\bf 61} (1979) 776. 

\bibitem{Suris00}
Y.\ B.\ Suris:
Inv.\ Probl.\ {\bf 16} (2000) 1071. 

\bibitem{Orfa1}
S.\ J.\ Orfanidis: 
Phys.\ Rev.\ D {\bf 18} (1978) 3822. 

\bibitem{Ab78}
M.\ J.\ Ablowitz: Stud.\ Appl.\ Math.\ {\bf 58} (1978) 17.

\bibitem{Kaji05}
K.\ Kajiwara and A.\ Mukaihira: 
J.\ Phys.\ A:\ Math.\ Gen.\ {\bf 38} (2005) 6363. 

\bibitem{Orfa2}
S.\ J.\ Orfanidis: 
Phys.\ Rev.\ D {\bf 18} (1978) 3828.

\bibitem{Date1}
E.\ Date, M.\ Jimbo and T.\ Miwa: 
J.\ Phys.\ Soc.\ Jpn.\ {\bf 51} (1982) 4116. 

\bibitem{Sada}
T.\ Sadakane: 
J.\ Phys.\ A:\ Math.\ Gen.\ {\bf 36} (2003) 87.

\bibitem{Lamb}
G.\ L.\ Lamb, Jr.:
Rev.\ Mod.\ Phys.\ {\bf 43} (1971) 99.

\bibitem{Scott}
A.\ C.\ Scott, F.\ Y.\ F.\ Chu and D.\ W.\ McLaughlin: 
Proc.\ IEEE {\bf 61} (1973) 1443.

\bibitem{Miura}
R.\ M.\ Miura (editor): 
{\it B\"acklund Transformations, the Inverse Scattering Method, Solitons, 
and Their Applications} 
(Lect.\ Notes in Math.\ {\bf 515}, Springer, 
Berlin, 1976).

\bibitem{Kono82}
B.\ G.\ Konopelchenko: 
Phys.\ Lett.\ A {\bf 87} (1982) 445.

\bibitem{WahEst}
H.\ D.\ Wahlquist and F.\ B.\ Estabrook: 
Phys.\ Rev.\ Lett.\ {\bf 31} (1973) 1386.

\bibitem{PGR}
V.\ Papageorgiou, B.\ Grammaticos and A.\ Ramani: 
Phys.\ Lett.\ A {\bf 179} (1993) 111. 

\bibitem{STZ}
V.\ P.\ Spiridonov, S.\ Tsujimoto and A.\ S.\ Zhedanov: 
Commun.\ Math.\ Phys.\ {\bf 272} (2007) 139.

\bibitem{KajiOh}
K.\ Kajiwara and Y.\ Ohta:
J.\ Phys.\ Soc.\ Jpn.\ {\bf 77} (2008) 054004. 

\bibitem{Hiro77}
R.\ Hirota: 
J.\ Phys.\ Soc.\ Jpn.\ {\bf 43} (1977) 2079.

\bibitem{NijWal}
F.\ W.\ Nijhoff  and A.\ J.\ Walker: 
Glasgow Math.\ J.\ {\bf 43A} (2001) 109.

\bibitem{ABS03}
V.\ E.\ Adler, A.\ I.\ Bobenko and Y.\ B.\ Suris: 
Commun.\ Math.\ Phys.\ {\bf 233} (2003) 513. 

\bibitem{ABS09}
V.\ E.\ Adler, A.\ I.\ Bobenko and Y.\ B.\ Suris: 
Funct.\ Anal.\ Appl.\ {\bf 43} (2009) 3.
%(to appear) arXiv:\,0705.1663.

\bibitem{Dodd}
R.\ K.\ Dodd: 
J.\ Phys.\ A:\ Math.\ Gen.\ {\bf 11} (1978) 81. 

\bibitem{Suris95}
Y.\ B.\ Suris: 
Phys.\ Lett.\ A {\bf 206} (1995) 153.

\bibitem{Pemp}
F.\ Pempinelli, M.\ Boiti and J.\ Leon: 
%in 
{\it Proceedings of the First Workshop on 
Nonlinear Physics, Theory and Experiment} 
(Gallipoli, 
%(Lecce), 
Italy, 
%June 29--July 7, 
1995) 
%ed.\ 
edited by E.\ Alfinito {\it et al.}
%, World Scientific, 1996, 
(World Scientific, Singapore, 1996) 
pp.\ 261.

\bibitem{Taha2}
T.\ R.\ Taha and M.\ J.\ Ablowitz: 
J.\ Comp.\ Phys.\ {\bf 55} (1984) 203.

\bibitem{FT87}
L.\ D.\ Faddeev and L.\ A.\ Takhtajan: 
{\it Hamiltonian Methods in the Theory of Solitons}
(Springer, 
%-Verlag, 
Berlin, 
%Heidelberg and New York, 
1987) translated by A.\ G.\ Reyman. 

\bibitem{MaSa97}
J.\ Matsukidaira, J.\ Satsuma, D.\ Takahashi, T.\ Tokihiro 
and M.\ Torii: Phys.\ Lett.\ A {\bf 225} (1997) 287. 

\bibitem{TaMa97}
D.\ Takahashi and J.\ Matsukidaira: 
J.\ Phys.\ A:\ Math.\ Gen.\ {\bf 30} (1997) L733. 

\bibitem{TakaSatsu90}
D.\ Takahashi and J.\ Satsuma: 
J.\ Phys.\ Soc.\ Jpn.\ {\bf 59} (1990) 3514.

\bibitem{Hiro73}
R.\ Hirota: J.\ Phys.\ Soc.\ Jpn.\ {\bf 35} (1973) 289. 

\bibitem{Wadati76}
M.\ Wadati: Prog.\ Theor.\ Phys.\ Suppl.\ {\bf 59} (1976) 36. 

%\bibitem{Taha3}
%T. R. Taha and M. J. Ablowitz: J.\ Comp.\ Phys.\ {\bf 55} (1984) 231.

%\bibitem{Taha4}
%Taha and Ablowitz M J: J.\ Comp.\ Phys.\ {\bf 77} (1988) 540.

\bibitem{AL0}
M.\ J.\ Ablowitz and J.\ F.\ Ladik: 
J.\ Math.\ Phys.\ {\bf 16} (1975) 598. 

%\bibitem{Hirota77}
%R.\ Hirota: J.\ Phys.\ Soc.\ Jpn.\ 
%{\bf 43} (1977) 2074. 

%\bibitem{Hirota78}
%R.\ Hirota: 
%J.\ Phys.\ Soc.\ Jpn.\ {\bf 45} (1978) 321. 

%\bibitem{HiroTsuji1}
%R. Hirota, S. Tsujimoto and T. Imai: 
%%Difference Scheme of Soliton equations
%in ``Future Directions of Nonlinear Dynamics in Physical 
%and Biological Systems'' ed.\ by P.\ L.\ Christiansen 
%%J. C. Eilbeck and R. D. Parmentier 
%{\it et al.\ }(Plenum, New York, 1993) p.\ 7. 

\bibitem{HiroTsuji2}
R.\ Hirota and S.\ Tsujimoto: 
J.\ Phys.\ Soc.\ Jpn.\ {\bf 64} (1995) 3125. 

%\bibitem{}
%R.\ Hirota: 
%J.\ Phys.\ Soc.\ Jpn.\ {\bf 66} (1997) 283. 

%\bibitem{}
%M.\ Iwasaki and Y.\ Nakamura: 
%2004 Inverse Problems {\bf 20} 553.

\bibitem{Skl82}
E.\ K.\ Sklyanin: Funct.\ Anal.\ Appl.\ {\bf 16} (1982) 263. 

\bibitem{Ishi82}
Y.\ Ishimori: J.\ Phys.\ Soc.\ Jpn.\ {\bf 51} (1982) 3417. 

\bibitem{Hal82}
F.\ D.\ M.\ Haldane: 
J.\ Phys.\ C:\ Solid State Phys.\ {\bf 15} (1982) L1309. 

\bibitem{Fad84}
L.\ Faddeev: 
{\it Recent Advances in Field Theory and Statistical Mechanics} 
%; 
(Les Houches Summer School, Session XXXIX, 1982)
%, ed.\ 
edited by J.--B.\ Zuber and R.\ Stora
%, 
(North-Holland, Amsterdam, 
%and New York, 
1984)
%, 
pp.\ 561.
%
%Les Houches, France, August 2--September 10 1982, 
%North-Holland, 1984, 
%870 pp. 
%``Recent Advances in Field Theory and Statistical Mechanics

%\bibitem{TTF83}
%V.\ O.\ Tarasov, L.\ A.\ Takhtadzhyan and L.\ D.\ Faddeev: 
%Theor.\ Math.\ Phys.\ {\bf 57} (1983) 1059. 

\bibitem{GIV86}
V.\ S.\ Gerdjikov, M.\ I.\ Ivanov and Y.\ S.\ Vaklev: 
Inv.\ Probl.\ {\bf 2} (1986) 413.

%\bibitem{Date4}
%E.\ Date, M.\ Jimbo 
%and T.\ Miwa: J.\ Phys.\ Soc.\ Jpn.\ {\bf 52} (1983) 761. 

\end{thebibliography}
\end{document}